\documentclass[12pt]{report}
\usepackage[utf8]{inputenc}
\usepackage{hyperref}
\usepackage{authblk}
\usepackage{fullpage}
\usepackage{pdfpages}
\usepackage{enumitem} 
\usepackage[normalem]{ulem}
\usepackage{soul}
\usepackage{upgreek}
\usepackage{blindtext}
\usepackage{boxhandler}
\usepackage[toc]{appendix}
\usepackage{pdfpages}

\usepackage{graphicx}
\graphicspath{ {images/} }
\usepackage{float}
\usepackage{morefloats} 
\usepackage{wrapfig}
\usepackage{caption}
\usepackage{subcaption}
\usepackage{xcolor}

\usepackage{mathtools}
\usepackage{amsthm, amssymb}
\usepackage{mathrsfs}
\usepackage{braket} 
\usepackage{slashed}
\usepackage{cancel}

\newcommand{\av}[1]{\langle #1 \rangle}

\newcommand{\bG}{\mathbb G}

\newcommand{\mink}{\mathbb M}
\newcommand{\bC}{\mathbf C} 
\newcommand{\bK}{{\mathbf K}} 
  
\newcommand{\bL}{\mathbf L}

\newcommand{\bl}{\mathbf l}
\newcommand{\bHH}{\mathbf H}

\newcommand{\tx}{\widetilde x}
\newcommand{\ty}{\widetilde y}
\newcommand{\tO}{\tilde \Omega}
\newcommand{\appropto}{\mathrel{\vcenter{
  \offinterlineskip\halign{\hfil$##$\cr
    \propto\cr\noalign{\kern2pt}\sim\cr\noalign{\kern-2pt}}}}}
\newcommand{\be}{\begin{equation}}
\newcommand{\ee}{\end{equation}}
\newcommand{\bea}{\begin{eqnarray}}
\newcommand{\eea}{\end{eqnarray}}
\newcommand{\bfig}{\begin{figure}}
\newcommand{\efig}{\end{figure}}
\newcommand{\bsfig}{\begin{subfigure}}
\newcommand{\esfig}{\end{subfigure}}

\newcommand{\kgnorm}[1]{( #1 )_{\mathrm{KG}}}
\newcommand{\hD}{\widehat{\Delta}}

\newcommand{\uksj}{s_{\mathbf k}}
\newcommand{\uksjn}{\tilde{s}_{\mathbf k}}
\newcommand{\ukpsj}{s_{\mathbf k'}}

\newcommand{\vkpjp}{\tilde{s}_{\mathbf k}^+}
\newcommand{\vkpjm}{\tilde{s}_{\mathbf k}^-}

\newcommand{\uqkg}{u_{\mathbf q}}
\newcommand{\uqpkg}{u_{\mathbf q'}}

\newcommand{\Aqk}{A_{\mathbf{qk}}}
\newcommand{\Bqk}{B_{\mathbf{qk}}}
\newcommand{\hP}{\widehat{\Phi}}
\newcommand{\hB}{\widehat{\Box}}
\newcommand{\wsj}{W_{\mathrm{SJ}}}
\newcommand{\kr}{\mathrm{Ker}} 
\newcommand{\im}{\mathrm{Im}} 
\newcommand{\ha}{\mathbf a } 
\newcommand{\hak}{\ha_{\mathbf {k}}}

\newcommand{\hb}{\mathbf b } 
\newcommand{\hbk}{\hb_{\mathbf {k}}}
\newcommand{\hbkd}{{\hb_{\mathbf {k}}^\dagger}}

\newcommand{\haq}{\ha_{\mathbf {q}}}
\newcommand{\haqd}{{\ha_{\mathbf {q}}^\dagger}}
\newcommand{\haqp}{\ha_{\mathbf {q'}}}
\newcommand{\haqpd}{{\ha_{\mathbf {q'}}^\dagger}}
\newcommand{\lk}{\lambda_{\mathbf{k}}}

\newcommand{\stkout}[1]{\ifmmode\text{\sout{\ensuremath{#1}}}\else\sout{#1}\fi}
\newcommand{\tre}{\text{Re}}
\newcommand{\cL}{\mathcal{L}}
\newcommand{\cM}{\mathcal{M}}
\newcommand{\Wmink}{W_{\mathrm{mink}}}
\newcommand{\phm}{m_p}
\newcommand{\tm}{m}

\newcommand{\ssee}{{\mathcal{S}}}
\newcommand{\sseec}{{\ssee}^{(c)}}
\newcommand{\diam}{\mathbb D} 
\newcommand{\cH}{\mathcal H}
\newcommand{\cR}{\mathcal R} 
\newcommand{\mx}{\mathrm{max}}
\newcommand{\mn}{\mathrm{min}}

\newcommand{\Ns}{N_O}

\newcommand{\deS}{\mathrm{dS}}
\newcommand{\con}{\alpha}

\begin{document}
	
\begin{titlepage}
 \begin{center}
	\vspace*{1cm}
	
     \Huge
	\textbf{Aspects of Quantum Fields \\on Causal Sets}
	
	\vspace{1.5cm}
	
	\Large
	\textbf{Nomaan X}
	
	\vspace{0.5cm}
	\textit{under the supervision of}
	
	\vspace{0.2cm} 
	
	\textbf{Prof. Sumati Surya}
	
	\vspace{0.2cm}
	
	Department of Theoretical Physics\\
	Raman Research Institute,\\
	Bengaluru - 560080, India
	
	\vspace{0.8cm}
	
	\includegraphics[width=0.13\textwidth]{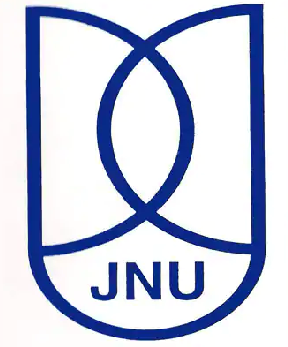} \hspace{5cm} \includegraphics[width=0.09\textwidth]{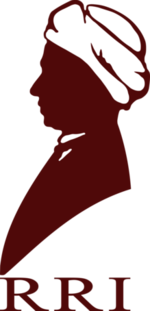}  
	
	\vfill
	
	A thesis submitted for the degree of\\
	Doctor of Philosophy\\
	to
	
	Jawaharlal Nehru University\\
	New Delhi - 110067, India\\
	September, 2020
	
\end{center}
\end{titlepage}
   




\pdfbookmark{Summary}{summary}

\chapter*{Summary}
Quantum gravity has been an outstanding problem in theoretical physics for many decades now \cite{Hooft_1979, Carlip_2008, Schulz:2014gqa, Carlip_2015, Mattingly, rovelli2000notes}. The lack of phenomenology in this area has meant a proliferation of theoretical ideas. These ideas offer different ways of tackling the problem of quantizing the gravitational field. In the past, we have found ways to get glimpses of the final theory by studying quantum fields in curved spacetime \cite{Wald:1995yp, Birrell:1982ix}. This approach has led to the most mathematically rigorous formulations of quantum field theory \cite{Fredenhagen:2015utr}. It has also led to such well known results as Black Hole thermodynamics, particle production in FLRW cosmology, inflation and the information loss paradox. These results have been shown to arise in a variety of scenarios and from different kinds of calculations, further cementing their place as essential features for a final theory despite the lack of phenomenological evidence.   

There are also bottom up approaches that attempt to reach the continuum limit starting from a discrete or quantum construction of the relevant degrees of freedom. Causal set quantum gravity \cite{Surya:2019ndm} is one such approach. It postulates that spacetime is an approximation to an underlying, more fundamental structure - the causal set. This is a set of spacetime events along with information about whether or not they are causally related. Theorems proved by Hawking, King, Mccarthy and Malament \cite{Hawking:1976fe, Malament:1977fe} indicate that this is the only information needed to reproduce (up to conformal equivalence) the continuum spacetime manifold. These ideas in the form of causal set theory were first proposed in 1987 \cite{Bombelli:1987aa}.  

In this thesis we study some kinematical aspects of quantum fields on causal sets. In particular, we are interested in free scalar fields on a fixed background causal set. We present various results building up to the study of the entanglement entropy of de Sitter horizons using causal sets. We begin by obtaining causal set analogs of Green functions for this field. First we construct the retarded Green function in a Riemann normal neighborhood (RNN) of an arbitrary curved spacetime. Then, we show that in de Sitter and patches of anti-de Sitter spacetimes the construction can be done beyond the RNN \cite{NomaanX2017}. This allows us to construct the QFT vacuum on the causal set using the Sorkin-Johnston construction \cite{Afshordi:2012jf}. We calculate the SJ vacuum on a causal set approximated by de Sitter spacetime, using numerical techniques. We find that the causal set SJ vacuum does not correspond to any of the known Mottola-Allen $\alpha$-vacua of de Sitter spacetime. This has potential phenomenological consequences for  early universe physics \cite{Surya:2018byh}. Finally, we study the spacetime entanglement entropy \cite{Sorkin:2012sn} for causal set de Sitter horizons. The entanglement entropy of de Sitter horizons is of particular interest. As in the case of nested causal diamonds in $2d$ Minkowski spacetime, explored in \cite{Saravani:2013nwa}, we find that the causal set naturally gives a volume law of entropy, both for nested causal diamonds in $4d$ Minkowski spacetime as well as $2d$ and $4d$ de Sitter spacetimes. However, as in \cite{Saravani:2013nwa}, an area law emerges when the high frequency modes in the SJ spectrum are truncated. The choice of truncation turns out to be non-trivial and we end with several interesting questions \cite{Surya_2021}.

In chapter 1 we begin with a preliminary discussion of causal set theory and where it falls in the broad spectrum of theories of quantum gravity. We give various definitions associated with causal sets and introduce quantum field theory on a causal set that will be used in the rest of the thesis.

In chapter 2 we discuss why the retarded Green function is important to building a quantum field theory on a causal set \cite{Sorkin:2017fcp}. We examine the validity and scope of Johnston’s models for scalar field retarded Green functions on causal sets in 2 and 4 dimensions \cite{Johnston:2010su}. As in the continuum, the massive Green function can be obtained from the massless one, and hence we first identify the massless Green function. We propose that the 2d model provides a Green function for the massive scalar field on causal sets approximated by any topologically trivial 2-dimensional spacetime. We explicitly demonstrate that this is indeed the case in a Riemann normal neighborhood. In 4d, the model can again be used to provide a Green function for the massive scalar field in a Riemann normal neighborhood which we compare to Bunch and Parker’s continuum Green function \cite{Bunch:1979uk}. We find that the continuum Green function can be reproduced for Ricci flat spacetimes and when $R_{ab}\propto g_{ab}$ i.e., for Einstein spaces. Further, we show that the same prescription can also be used for de Sitter spacetime and the conformally flat patch of anti-de Sitter spacetime. We suggest a generalization of Johnston’s model for the Green function for a causal set approximated by 3-dimensional flat spacetime.

In chapter 3 we present work related to the Sorkin-Johnston (SJ) vacuum in de Sitter spacetime for free scalar field theory. For the massless theory we show that the SJ vacuum can neither be obtained from the $O(4)$ Fock vacuum of Allen and Folacci \cite{Allen:1987tz} nor from the non-Fock de Sitter invariant vacuum of Kirsten and Garriga \cite{Kirsten:1993ug}. Using a causal set discretization of a slab of $2d$ and $4d$ de Sitter spacetime, we show the causal set SJ vacuum for a range of masses $m \geq 0$ of the free scalar field. While our simulations are limited to a finite volume slab of global de Sitter spacetime, they show good convergence as the volume is increased. We find that the 4d causal set SJ vacuum shows a significant departure from the continuum Mottola-Allen $\alpha$-vacua  \cite{Allen:1985ux}. Moreover, the causal set SJ vacuum is well-defined for both the minimally coupled massless $m=0$ and the conformally coupled massless $m=m_c$ cases. This is at odds with earlier work on the continuum de Sitter SJ vacuum where it was argued that the continuum SJ vacuum is ill-defined for these masses \cite{Aslanbeigi:2013fga}. We discuss an important tension between the discrete and continuum behavior of the SJ vacuum in de Sitter and suggest that the former cannot in general be identified with the Mottola-Allen $\alpha$-vacua even for $m >0$.

In chapter 4 we study de Sitter cosmological horizons as they are known to exhibit thermodynamic properties similar to black hole horizons \cite{gibbons}. In particular we study the entanglement entropy of a quantum free scalar field in de Sitter spacetime using Sorkin's spacetime entanglement entropy (SSEE) formula. We use a causal set discretization of de Sitter spacetime and the associated SJ vacuum state to calculate this SSEE numerically in $d=2,4$. We also examine the SSEE in the causal set discretization of $d=4$ Minkowski spacetime for comparison. As in the earlier calculation of the SSEE for 2-dimensional nested causal diamonds \cite{Saravani:2013nwa}, the SSEE on the causal set is seen to follow a volume rather than an area law, unless a truncation scheme is used on the spectrum of the spacetime commutator that enters the calculation. While the 2-dimensional truncation scheme can be motivated quite simply, this is not the case in the examples we study. We propose a few possible truncation schemes and discuss their relative merits vis a vis the area law and complementarity. While the former is satisfied by all the truncation schemes the latter is not.

In chapter 5 we conclude the thesis by tying up the themes and results that have appeared in previous chapters. We emphasize the role of numerical and computational tools in addressing some of these issues. Finally, we point to several open questions that have come up and to potential avenues for further exploration.

\chapter*{Acknowledgements}

First and foremost I thank my supervisor Sumati Surya for providing support and guidance. She has always been available for discussion and has helped me in staying focused. The energy, dedication and outlook she brings towards her work has been inspiring to me as I navigated my own academic life. 

I want to thank my collaborators Fay Dowker, Yasaman Yazdi and Joachim Kambor for enriching discussions and for being critical and thorough in every detail of our work together. I cannot overstate the pleasure of having the opportunity to interact with Rafael Sorkin. His conceptual depth in physics and philosophy has given rise to refreshing ideas in causal set theory. I appreciate discussions with other members of the causal set community - Abhishek Mathur, Ian Jubb, Will Cunningham, David Rideout, Lisa Glaser, Dion Benincasa, Stav Zalel and Niayesh Afshordi. I am thankful that this community has been a friendly and safe space for discussions, talks, criticism and other banter.
I am grateful to members of my academic committee - Joseph Samuel and Justin David for their suggestions on my work.

My stay during the Ph.D. and the logistics around it were supported in various ways by the staff at RRI - hostel, theoretical physics group secretaries, administration and library. I thank all the people involved for making this process simple.       

My non-work life in the last few years has comprised of several activities in RRI, IISc, NCBS, ICTS, JNP and beyond. These experiences have taught me things in many intangible ways. During these activities, I have had the pleasure of meeting amazing people, some of whom have become close friends. The people involved are too many to name and the experiences too diverse to describe. To all of you I say - I see you.

Finally, my parents and family, who were initially hesitant and did not understand my desire to study physics. I am thankful that they decided to trust me and were supportive throughout, I hope that their trust was not in vain.    

 
\vspace*{2cm}

\begin{center}
	
\includegraphics[width=0.5\textwidth]{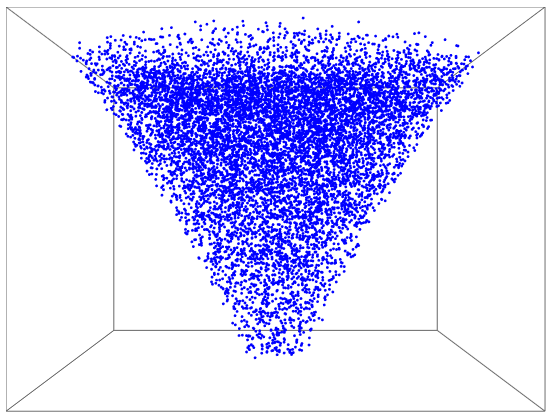}

\vspace{1cm}
	
Recent observations of the microwave background indicate that the universe contains enough matter to cause a time-reversed closed trapped surface. This implies the existence of a singularity in the past, at the beginning of the present epoch of expansion of the universe. This singularity is in principle visible to us. It might be interpreted as the beginning of the universe.\\
\emph{S. W. Hawking and G. F. R. Ellis, The large scale structure of space-time}

\vspace{1.5cm}
	
In the beginning there was nothing, which exploded.\\
\emph{Terry Pratchett, Lords and Ladies}
	
\end{center}

\tableofcontents




\chapter{Introduction}

Fundamental discreteness plays an important role in many theories of physics. In the last century considerable effort has been put into formulating quantum descriptions of the dynamics of matter. Such a shift has been motivated by the discovery of the quantum nature of matter based on considerable experimental evidence. General relativity (GR) is perhaps the epitome of classical field theory. It describes the dynamics of the background on which physics of matter plays out i.e., spacetime itself. However, the need to quantize the gravitational field, the inherent difficulties not withstanding, has been based on attempts to solve inconsistencies arising from GR as well as from working with quantum fields on classical backgrounds. The phenomenological evidence in this direction is scant. Attempts at quantum gravity (QG) have also been motivated by the need for a unifying framework in which the largely algebraic framework of quantum fields and the geometric framework of spacetime can be treated on equal footing \cite{Hooft_1979, Carlip_2008, Schulz:2014gqa, Carlip_2015, Mattingly, rovelli2000notes}.

In the absence of concrete phenomenology, developing a theory requires choices that are motivated by mathematical consistency, by results from other areas of physics or even by aesthetic reasons. Understandably, this leads to several possibilities and various theories of QG are an example of this proliferation of ideas. Basic ontological questions in QG like what should be the degrees of freedom?, what is fundamental in GR - the metric or the causal structure?, what should be the observables? etc. have multiple answers \cite{Sorkin_1997}. Broadly, theories of QG fall into two categories - those that start from the continuum and use some form of quantization and those that start with an underlying fundamental discreteness. Causal set QG falls in the later category - it is a bottom-up approach that replaces the continuum spacetime manifold with a discrete substructure called the causal set. 

The ideas of spacetime discreteness have a rich history\footnote{For details see \cite{Surya:2019ndm} and the introduction in \cite{Johnston:2010su}.} culminating in the seminal work of Bombelli, Lee, Meyer and Sorkin \cite{Bombelli:1987aa} which laid out causal set theory (CST) in its present form. A causal set is a locally finite, partially ordered set (poset) whose elements represent spacetime events along with information about causal ordering between each pair of events. A causal set can be used to replace the spacetime manifold because causality is a building block of Lorentzian geometry - a result based on powerful theorems proved by Hawking, King, McCarthy and Malament \cite{Hawking:1976fe, Malament:1977fe}. These theorems show that there is a bijection between the conformal class of spacetime metrics and the causal ordering (a partially ordered set).

\textbf{\textit{Theorem}}: If a chronological bijection exists between two $d$-dimensional spacetimes which are both future and past distinguishing, then these spacetimes are conformally isometric when $d > 2$.

It was shown by Levichev \cite{levichev1987prescribing} that a causal bijection implies a chronological bijection and hence the above theorem can be generalized by replacing “chronological” with “causal”. Subsequently Parrikar and Surya \cite{Parrikar:2011zn} showed that the causal structure poset $(\mathcal{M},\prec)$ of these spacetimes also contains information about the spacetime dimension. 
In other words, geometric information (barring an overall volume factor) about a spacetime manifold is embedded in the causal ordering of events\footnote{A non-technical discussion of this can be found in Geroch's book \textit{General Relativity from A to B}}.

CST proposes that QG is the quantum theory of causal sets. 
   
\section{Causal Sets}
In this section we define causal sets and a few other order-theoretic constructions which will be used throughout.

A \textit{causal set} $\mathcal{C}$ is a partially ordered set together with an order-relation $\preceq$ that $\forall\,x,y,z\in\mathcal{C}$ satisfies the following conditions:
\begin{enumerate}
    \item \textit{Reflexivity:} $x\preceq x$
    \item \textit{Antisymmetry:} $x\preceq y\preceq x\Rightarrow x=y$
    \item \textit{Transitivity:} $x\preceq y\preceq z\Rightarrow x\preceq z$
    \item \textit{Local finiteness:} $|\{z\in\mathcal{C}|x\preceq z\preceq y\}|<\infty$
\end{enumerate}
Here $|\cdot|$ denotes the cardinality of a set. The elements of $\mathcal{C}$ are spacetime events and the order-relation $\preceq$ denotes the causal order between the events. If $x\preceq y$ we say ``$x$ causally precedes $y$", and we write $x\prec y$ if $x\preceq y$ and $x\neq y$. Causal relations on a Lorentzian manifold (without closed timelike curves) obey conditions 1-3. Condition 4 ensures that there are a finite number of events in any causal interval; this brings in discreteness. 

Two useful ways of characterizing a causal set are the \textit{causal matrix} $C_0$ and the \textit{link matrix} $L_0$. The elements of these matrices are defined as 

  \[  C_0(x,x'):=                                        
\left\{                                                          
	\begin{array}{ll}
		1  & \mbox{if } x' \prec x \\
		0 & \mbox{} \text{otherwise}
	\end{array},
\right.
\quad\quad
L_0(x,x'):=                                        
\left\{                                                          
	\begin{array}{ll}
		1  & \mbox{if } x' \prec x \,\text{and}\, |(x,x')|=0\\
		0 & \mbox{} \text{otherwise}
	\end{array}
\label{cl},
\right.
\]
where $(x,x')$ is the set of events\footnote{Note that there maybe confusion when $(x,x')$ appears in an expression like $C_0(x,x')$, in this case $x,\,x'$ are individual points being used as indices for the matrix $C_0$ and do not represent the interval $(x,x')$. The difference is clear from the context.} that lie in the causal interval between $x$ and $x'$ i.e., $(x,x')=\{a\in\mathcal{C}\,|\,x'\prec a\prec x\}$. Such an interval is called an \textit{Alexandrov set} or \textit{causal diamond}. The relation defined by the extra condition $|(x,x')|=0$ is called a \textit{link}. 

We define a $k$-\textit{chain} of length $k+1$ between $x'$ \textit{and} $x$ in a causal set as a totally ordered subset of $\mathcal{C}$, $\{x_1,x_2,\ldots, x_k \}$ such that $x'\prec x_1\prec x_2\prec....x_{k-1}\prec x_k\prec x$. For $k\ge 1$, define $C_k(x,x')$ to be the number of $k$-chains between $x'$ and $x$ when $x'\prec x$ and zero when $x' \not\prec x$.
The $C_k$'s are powers of the causal matrix:
\be
\label{numberchains}
C_k(x,x') = \underbrace{C_0 \cdot C_0\cdot \ldots C_0}_{k+1} (x,x')\,.
\ee
A $k$-\textit{path} of length $k+1$ between $x'$ and $x$ in $\mathcal{C}$ is a $k$-chain in which each relation is a link. As above, for $k\ge 1$, we define $L_k(x,x')$ to be the number of $k$-paths between $x'$ and $x$ when $x'\prec x$ and zero when $x' \not\prec x$.
The $L_k$'s are powers of the link matrix:
\be
\label{numberchains}
L_k(x,x') = \underbrace{L_0 \cdot L_0\cdot \ldots L_0}_{k+1} (x,x')\,.
\ee 
An \textit{antichain} $A$ is a totally unrelated subset of $\mathcal{C}$ i.e., a subset of $\mathcal{C}$ in which no 2 elements are related to each other. An \textit{inextendible
antichain} is an antichain such that every element $e \in \mathcal{C}\!/\! A$ is related to an element of $A$.

The \textit{nearest neighbours} of an element $x'$ in $\mathcal{C}$ are those that are linked to it i.e., $\{x\in\mathcal{C}\,|\,x' \prec x \,\text{and}\, |(x,x')|=0\}$. This construct can be used to divide a causal set into \textit{layers}. Starting with the \textit{minimal element} (the one with no past nearest neighbours) we can define the following layers
\be
\label{layer}
\mathbb{L}_i=\{x\in\mathcal{C}\,|\,x' \prec x \,\text{and}\, |(x,x')|=i-1\},\quad i\geq 1
\ee 
The same construction can be carried out starting from the \textit{maximal element} (the one with no future nearest neighbours). These constructions are shown in Fig \ref{chain_antichain_layer}.
\bfig[!t]
\bsfig[b]{0.5\textwidth}
\includegraphics[width=\textwidth]{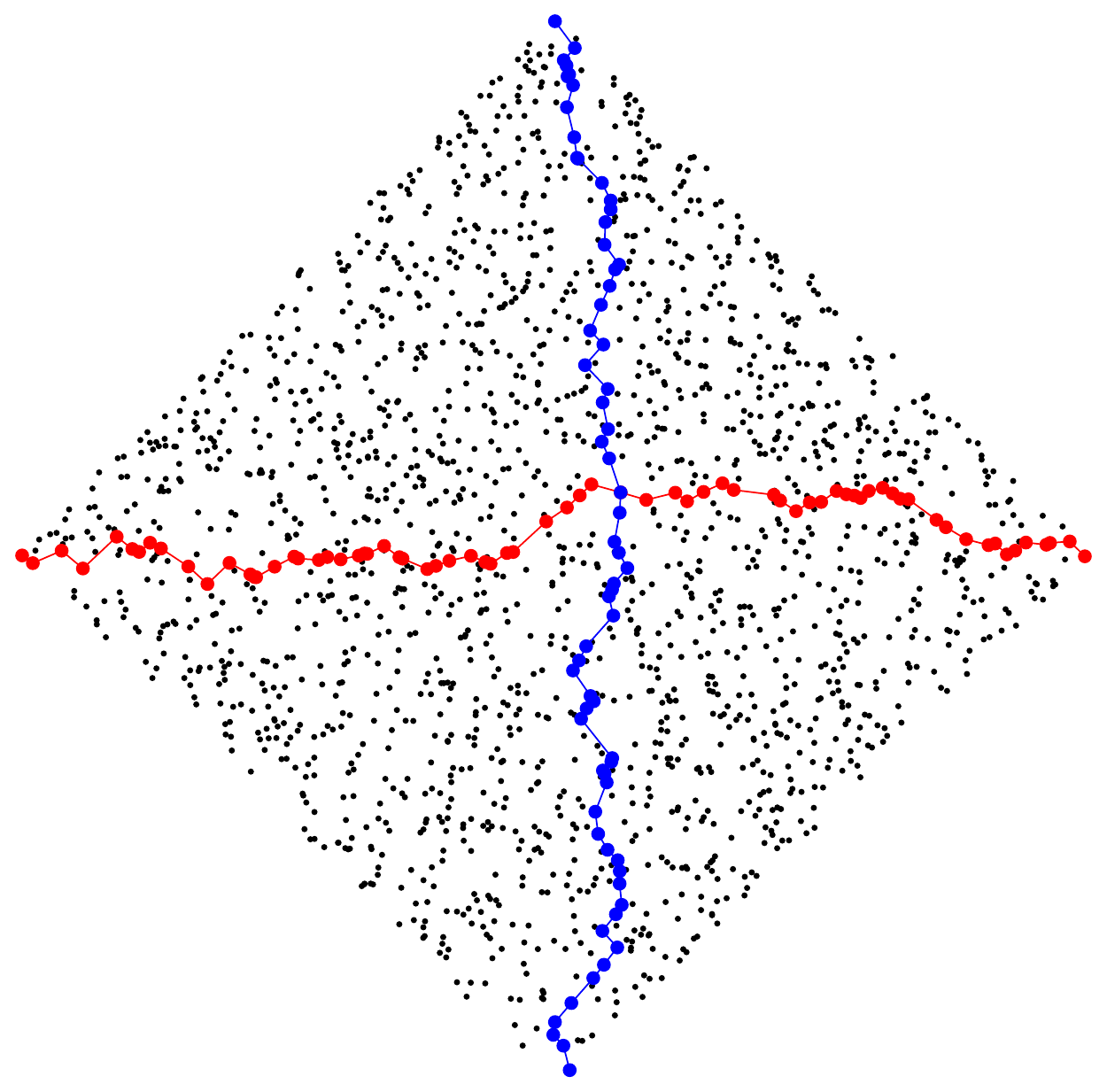}
\caption{}
\esfig
\bsfig[b]{0.5\textwidth}
\includegraphics[width=\textwidth]{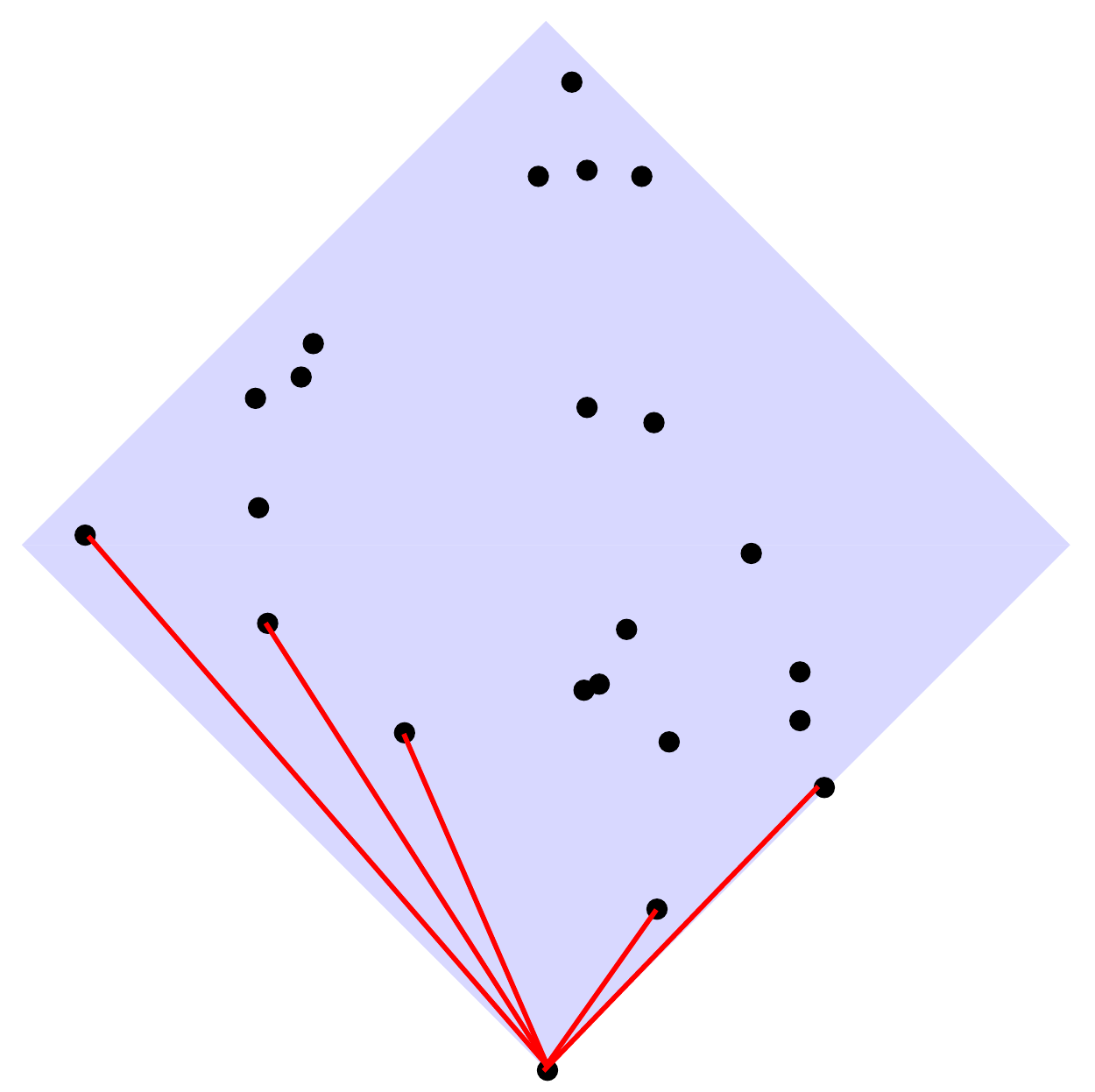}
\caption{}
\esfig
\caption{(a) An example of an antichain in red and a chain in blue is shown. (b) An example of causal set with 20 elements containing 5 elements in the first layer with respect to the minimal element, the corresponding links are also shown.}
\label{chain_antichain_layer}
\efig

\section{Sprinkling}
In thinking about QG, given the action of general relativity, a major hurdle is to evaluate the path integral over all possible Lorentzian metrics (perhaps also over topology). When we replace the manifold structure of general relativity with causal sets, we also replace this with such an evaluation over all possible causal sets i.e.,
\be
Z=\int\mathcal{D}[g]\,e^{iS[g]}\rightarrow\sum_{\{C\}}\,e^{iS[C]}.
\ee
In a full UV complete theory of causal sets, we would need to define an action and evaluate this path integral. The difficulty in evaluating this expression becomes clear when we note that a causal set is a much more general than a spacetime manifold. More precisely, the space of causal sets is dominated by the non-mainfoldlike Kleitmann-Rothschild orders \cite{Kleitmann1975}. Even among manifoldlike causal sets, the dimension of the manifold remains arbitrary which would imply summing over all possible dimensions. 
The most studied proposal for the causal set action is the Benincasa-Dowker-Glaser action~\cite{Benincasa2011,Glaser:2013xha}. Using this, there have been attempts at understanding how non-manifoldlike causal sets can be suppressed~\cite{Loomis2017,mathur2020entropy} and also the behaviour of simple matter models in a restricted class of causal sets~\cite{glaser2018ising}. However, a complete understanding of this expression is still far away.

In this work, we bypass these issues by working with fixed background causal sets i.e., we will setup quantum field theory on a fixed causal set instead of a fixed spacetime manifold. This is a mesoscopic regime between the deep UV and the continuum where spacetime discreteness still plays a role. We expect that work in this regime could be of phenomenological interest. In order to obtain results from CST that have a meaningful interpretation we must work with causal sets that correspond to a spacetime manifold, hence we work with \textit{sprinkled causal sets}.  

Sprinkling is the process of picking points randomly from a region of spacetime $(\mathcal{M},g)$ with a constant density $\rho$. To ensure that such a process is covariant i.e., the points picked are not based on any specific coordinate system we use a random Poisson discretization \cite{Bombelli:2006nm}. The probability of picking $n$ points from a spacetime region of volume $V$, given a fundamental discreteness scale $\rho^{-1}$ is
\be \label{poisson}
P_V(n)=\frac{(\rho V)^n e^{-\rho V}}{n!}
\ee
which also gives us $\langle n \rangle=N=\rho V$.
The causal ordering is inherited from the region's causal ordering restricted to the sprinkled points. The causal sets so obtained are said to approximate $(\mathcal{M},g)$ and we will denote this by $C \sim (\mathcal{M},g)$. Figure \ref{chain_antichain_layer} is an example of a sprinkled causal set in a region of $\mink^2$. 

We refer the reader to \cite{Dowker:2005tz, Surya:2019ndm} for more details on CST.

\section{Quantum Fields on Causal Sets}

Although the eventual goal of CST is to have a fully quantum description of the dynamics of causal sets, it is useful to study the behaviour of quantum fields on fixed background causal sets. This is the analogue of studying quantum fields on fixed spacetime backgrounds. As we will see, this not only allows us to explore a variety of issues but also throws up questions that a theory of discrete spacetime must answer. This section will be an outline and more detail can be found in \cite{Johnston:2010su, Sorkin:2017fcp}. We will also describe the setup and precise analysis relevant to this thesis in subsequent chapters.  

For a \textit{Gaussian} field we can specify a field theory by giving the correlation functions 
$$
\av{\phi(x)}\quad \text{and} \quad W(x,x')=\av{\phi(x)\phi(x')}
$$ 
were we set $\av{\phi(x)}=0$ by convention. The Wightman ``function" $W(x,x')$ is therefore sufficient; any $n$-point function can be determined from it using Wick's rule. An analogous condition for quantum fields can be found in \cite{Sorkin:2011pn}. In defining the theory through the Wightman function, we refer to a Gaussian theory directly instead of having to define a Gaussian state\footnote{A discussion on how to go back and forth between these 2 is given in \cite{Sorkin:2017fcp}.}. This allows us to formulate the notion of a vacuum and its entanglement entropy directly in terms of $W(x,x')$. These constructions applied to regions of de Sitter spacetime in chapters 3, 4 form a large part of this thesis. 

The fact that a Gaussian theory can be defined fully from $W(x,x')$ does not mean that $W$ can be specified freely. It must satisfy the condition
\be
\int dV(x)f(x)W(x,x')f(x')dV(x')\ge0,\quad\quad\forall f\in \text{domain}(W)
\ee 
this condition, called \textit{positive semi-definiteness}, can be taken as an axiom of quantum theory. However, we note that given a state vector $|0\rangle$ in some Hilbert space the above result can be derived as a theorem. It follows immediately from the positivity of $||\psi||^2=\av{\psi|\psi}$ where $|\psi\rangle=\int dV(x)f(x)\,\hat{\phi}\,|0\rangle$. 

We recall that the usual way of obtaining $W$ in quantum field theory is as follows
\be
\Box-m^2\longrightarrow G \longrightarrow \Delta \longrightarrow [\hat{\phi},\hat{\phi}]\longrightarrow a,\,a^\dagger\longrightarrow|0\rangle\longrightarrow W,
\ee
where $G$ is the \textit{retarded (or advanced) Green function} and $\Delta$ is called the \textit{Pauli-Jordan function}. In the intervening step $[\hat{\phi},\hat{\phi}]\longrightarrow a,\,a^\dagger$ we need to make a choice of positive frequency modes and this involves a timelike killing vector in the spacetime region that we consider. The alternate, shorter route that does not involve such a choice and is more suited to causal sets is 
\be
G \longrightarrow \Delta \longrightarrow W.
\label{qftcs}
\ee
In chapter 2 we extend the Minkowski spacetime results of Johnston \cite{Johnston:2010su} to identify appropriate $G$s for causal sets sprinkled in Riemann normal neighbourhoods in an arbitrary curved spacetime as well as in regions of de Sitter and anti de sitter spacetime\footnote{All these results are for $d=2,\,4$.}. We also propose a new Green function for the $3d$ Minkowski case. 

In chapter 3 we use the Sorkin-Johnston (SJ) construction \cite{Afshordi:2012jf} to go from $G$ to $W$ for the Minkowski and de Sitter cases. Through a numerical study on causal sets we find that the vacuum obtained via the SJ construction is distinct from the standard Mottola-Allen $\alpha$-vacua in de Sitter. 

Since the first calculation of entanglement entropy (EE) in a spacetime context, in particular for a black hole horizon \cite{Bombelli:1986rw}, it has been an important part of QFT in curved spacetime and approaches to QG. Historically, it has been customary to define EE using states (or density matrices) on spatial slices but such a non-covariant definition is not adaptable to causal sets. Only recently a covariant definition based entirely on $\Delta,\,W$ has been proposed \cite{Sorkin:2012sn}. Once we obtain a $W$ via \eqref{qftcs} we can find the entanglement entropy on causal sets. While working with black hole horizons using causal sets remains a challenge\footnote{An interesting study of black holes in causal sets is \cite{Asato:2019myn}.}, we can do the next best thing and work with de Sitter horizons which are relevant in a cosmological context. The study of the EE of de Sitter horizons is the subject of chapter 4.

We conclude this chapter with a review of the basics of de Sitter spacetime, mostly following the discussion in \cite{Spradlin:2001pw}.    

\section{de Sitter Spacetime}
\label{dsapp}
 
de Sitter spacetime  $\deS_d$ can be thought of as a surface in $\mathbb{M}^{d+1}$. This surface is characterized by the constraint
\be
-X_0^2+X_1^2+...+X_d^2=\eta_{AB}X^AX^B=\frac{1}{H^2},
\label{hypecoords}
\ee
where $A$ and $B$ run from $0$ to $d$. This is a hyperboloid in $\mathbb{M}^{d+1}$ with ``radius" $l\equiv\frac{1}{H}$. This is also, topologically, $\mathbb{R}\times S^{d-1}$, where the $S^{d-1}$ corresponds to a surface with constant $X_0$. This $(d-1)$-sphere has a radius $\equiv\frac{1}{H^2}+X_0^2$. 

If we assign coordinates $x^a$ on the surface $\deS_d$, then corresponding to each point on the surface we can define vectors $X^A(x)$, in $\mathbb{M}^{d+1}$. Each of these must satisfy (1). We can define another useful quantity as follows: 
\be
Z(x,y)=H^2\eta_{AB}X^A(x)X^B(y)=\cos\theta.
\label{zdef}
\ee
We can think of this as an inner product between two $d+1$-vectors that represent points $x$ and $y$ on the surface $\deS_d$. If there is some angle $\theta$ between these two vectors in $\mathbb{M}^{d+1}$, then the above expression can be written (in exact analogy with the usual ``dot product") in terms of this angle, and the magnitude cancels out with the $H^2$ in front. \\
Now for two points on the surface separated by an angle $\theta$, the geodesic distance (in exact analogy with a sphere) is given by $d(x,y)=\frac{1}{H}\theta$, where $\frac{1}{H}$ plays the role of radius. Therefore we have \cite{Allen:1985ux}
\be
d(x,y)=\frac{1}{H}\cos^{-1}Z(x,y).
\ee
The advantage of this relation is that in general the geodesic distance is given by 
\be
d(x,y)=\int_y^xd\mu\sqrt{\eta_{AB}\frac{dX^A}{d\mu}\frac{dX^B}{d\mu}},
\ee
where $X^A(\mu)$ is a parameterized geodesic between points $x$ and $y$. In general, this integral can be difficult to evaluate. However the closed-form expression of  $Z(x,y)$ allows it to be trivially evaluated once coordinates are assigned to the surface $\deS_d$. The values $Z > 1$, $Z = 1$ and $-1 < Z < 1$ correspond to pairs of points that can be joined by timelike, null, and spacelike geodesics, respectively.

A useful set of coordinates to characterize global de Sitter spacetime are the hyperbolic coordinates. In these, the metric takes the form 
\be
ds^2= -d\tau^2+\frac{1}{H^2}\cosh^2(H\tau)\,d\Omega_{d-1}^2  ,
\label{globmetric}
\ee
where $-\infty<\tau<\infty$ and $\Omega_{d-1}$ are coordinates on $S^{d-1}$. These coordinates are related to those in \eqref{hypecoords} by
\bea
X^0&=&\frac{1}{H}\sinh\tau\\\nonumber
X^i&=&\frac{1}{H} w^i\cosh\tau,\indent i=1,..., d,
\label{globalcoords}
\eea
where $w^i$ are coordinates on the sphere $S^{d-1}$:
\bea
w^1&=&\cos\theta_1,\\\nonumber
w^2&=&\sin\theta_1\cos\theta_2,\\\nonumber
&\vdots&\\\nonumber
w^{d-1}&=&\sin\theta_1...\sin\theta_{d-2}\cos\theta_{d-1},\\\nonumber
w^d&=&\sin\theta_1...\sin\theta_{d-2}\sin\theta_{d-1},
\eea
and where $0\leq\theta_i<\pi$ for $1\leq i\leq d-2$ and $0\leq\theta_{d-1}<2\pi$. $\sum^d_{i=1}\left(w^i\right)^2=1$ and 

\be
d\Omega_{d-1}^2=\sum^d_{i=1}\left(dw^i\right)^2=d\theta_1^2+\sin^2\theta_1d\theta_2^2+...+\sin^2\theta_1...\sin^2\theta_{d-2}d\theta^2_{d-1}
\ee
is the metric on $S^{d-1}$.

Another useful set of coordinates are the conformal/cylindrical coordinates obtained by setting $H\,d\tau/\cosh H\tau=d\tilde{T}$ in the above metric
\be
ds^2=\frac{1}{H^2\,\cos^2\tilde{T}}\left(-d\tilde{T}^2+d\Omega_{d-1}^2\right),
\label{confmetric2}
\ee
where $-\pi/2<\tilde{T}<\pi/2$, which is conformal to the cylinder $S^{d-1}\times [-\pi/2,\pi/2]$. In these coordinates the volume of a region of height $2T$ (i.e., conformal time $\tilde{T}\in[-T,T]$) and radius $l$ is given by
\be
V(T,d)=\frac{2 \pi ^{d/2}l^d}{\Gamma(\frac{d}{2})} \int_{-T}^{T} \sec ^d\tilde{T} \, d\tilde{T} .
\ee
In our cases of interest,
\bea
V(T,2)&=&4\pi l^2\tan T\\
V(T,4)&=&\frac{4}{3}\pi^2 l^4\,\tan T(\cos 2T+2)\sec ^2T.
\eea
The following are some other useful identities relevant to de Sitter spacetime that relate the Ricci scalar $R$ to other commonly used scales -- the cosmological constant ($\Lambda$), the de Sitter radius ($l$) and the Hubble constant ($H$):
\be
R=\frac{2d}{d-2}\Lambda=d\,(d-1)H^2=\frac{d\,(d-1)}{l^2},
\ee   
\be
\text{  where  }\quad\Lambda=\frac{(d-1)(d-2)}{2}H^2.
\ee
The critical mass \footnote{For more details see \cite{Hartong2004}.} is
\be
m_*=\frac{d-1}{2l}.
\ee  
In $d=4$, $R=4\Lambda=12H^2=12/l^2$ and $m_*=\dfrac{3}{2l}$.

Sprinkling into regions of Minkowski spacetime has been discussed elsewhere (see e.g. \cite{Johnston:2010su}). Here we briefly describe the process for de Sitter spacetime. 

\begin{figure}[!h]
	\centering
	\includegraphics[width=0.8\textwidth]{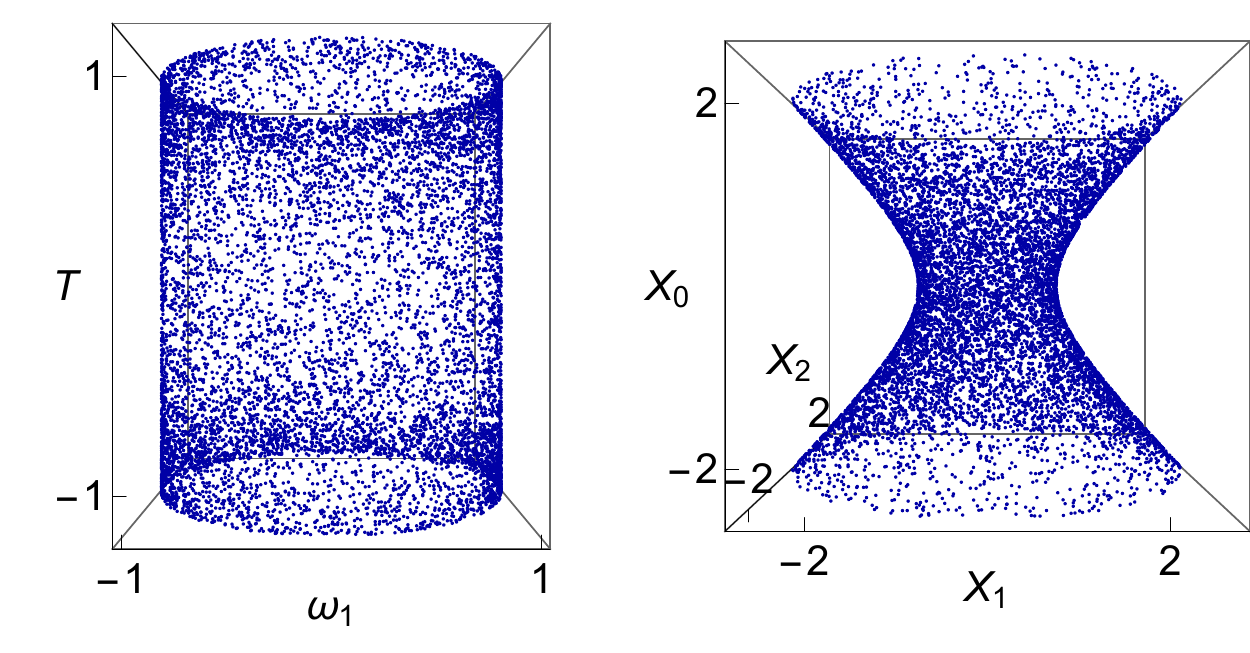}
	\caption{A sprinkling of $N=10000$ elements into the metric of \eqref{confmetric2} for the time interval $-1.2<T<1.2$.}
	\label{dssprinkling}
\end{figure}

A convenient coordinate system in which to do the sprinkling for de Sitter is  the conformal coordinate system of \eqref{confmetric2}. This allows us to work with the simpler conformally related  metric in analyzing the causal structure of de Sitter spacetime.
The sprinkling can be done in two steps. In the first step we pick points randomly on the spatial part, i.e.,, the sphere $S^{d-1}$. One simple way (by no means unique) to do this is to generate normalised $d$-dimensional vectors. These will automatically lie on the surface of $S^{d-1}$. The corresponding spherical coordinates can be obtained by using the standard Cartesian to spherical coordinate transformation.

In the second step we need to obtain the temporal part of the coordinates. As is evident from the  metric, this isn't uniformly distributed but depends on the conformal factor. The effect of the conformal factor can be incorporated by defining a normalised probability distribution with a probability density function equal to $(H\cos T)^{-d}$ in the region of interest. Picking points from this distribution will give us the temporal part of the coordinates. Combining the coordinates from the two steps, we have the required sprinkling. A typical sprinkling is shown in figure \ref{dssprinkling}.

\chapter{Scalar Field Green Functions}

Understanding classical and quantum scalar field propagation on a fixed causal set is an important problem
in causal set quantum gravity~\cite{Johnston2008,Johnston:2010su,Benincasa2011,Aslanbeigi2014}. 
Although ignoring back reaction and the quantum dynamics of the causal set background itself
means that the treatment of scalar field dynamics will be inconsistent in some way, we can hope to learn something about causal set theory by studying this problem.
Recent progress in defining scalar quantum field theory on a causal set puts 
great importance on the retarded Green function for the field on the causal set. Such a Green function can be used to obtain the Feynman propagator, or equivalently the Wightman function, of a distinguished quantum state on a causal set $\mathcal{C}$, the \textit{Sorkin-Johnston
state}~\cite{Johnston:2009fr,Afshordi:2012jf}. Sorkin's related construction of a double path integral form of 
free scalar quantum field theory on a finite casual set is also based on the retarded Green function \cite{Sorkin:2011pn}. 

In \cite{Johnston2008} Johnston found the massive scalar field retarded Green functions, $K_m(x,x')$, for  causal sets approximated by $d=2$ and $d=4$ Minkowski spacetime \cite{Johnston2008,Johnston:2010su}. For each case, he used a ``hop-stop'' ansatz 
in which the Green function equals a sum over appropriately chosen causal trajectories between
the two arguments of the Green function, 
with a weight assigned for every hop between the elements
of the trajectory and another for every stop at an intervening element. Requiring that the
continuum limit of the expectation value of the causal set Green function over multiple sprinklings into Minkowski spacetime equals the continuum retarded Green function  then fixes these weights. Extending the scope of the hop-stop ansatz to a larger class of spacetimes allows us to study causal set quantum field theory further.
  
We begin by describing Johnston's model in Section \ref{sec:model} and explain how it can be motivated by a spacetime treatment. We will see that the key is to identify the appropriate retarded Green function for the \textit{massless} field. This then leads to our proposed extensions of the model in Section \ref{sec:generalisations}.  For $d=2$ we propose that Johnston's definition of $K_m(x,x')$ can be used for a minimally coupled massive scalar field on a causal set approximated by any topologically trivial spacetime. The proposal stems from the fact that the massless, minimally coupled scalar field theory is conformally invariant.  In $d=2$, for non minimal coupling -- \textit{i.e.}, with arbitrary coupling to the Ricci scalar -- we show that the Johnston $K_m(x,x')$ is an appropriate retarded Green function in an approximately flat Riemann Normal Neighbourhood, up to corrections. In $d=4$ we find that it is possible to extend the Minkowski spacetime prescription to a Riemann normal neighbourhood (RNN), as well as de Sitter spacetime and the conformally flat patch of anti de Sitter spacetime. In all cases, the comparison with the continuum fixes the hop-stop weights. Our results are \textit{exact} for de Sitter and the globally hyperbolic patch of  anti de Sitter spacetime, i.e., the limit of the expectation value of the massless causal set Green function is the conformally coupled massless Green function. 
In Section \ref{3dgreen} we use this framework to propose a construction of the retarded Green functions on $d=3$ Minkowski spacetime. In section \ref{discussionch2} we present our conclusions and discuss some open questions.

\section{The Model}
\label{sec:model}

Consider the massless scalar retarded Green function $G_0(x,x')$ on a
globally hyperbolic  $d$ dimensional spacetime $(M, g)$:
\be \label{greenfunc}
\Box_x G_0(x,x')=- \frac{1}{\sqrt{-g(x')} }\delta(x-x')\,.
\ee
The massive retarded Green function, $G_m$, satisfies
\be
(\Box_x - m^2) G_m(x,x')= - \frac{1}{\sqrt{-g(x)} }\delta(x-x')\,,
\ee
and can be written as a formal expansion\footnote{This expression contains IR divergences in each term due to the presence of $G_0$, for a detailed analysis see~\cite{aste2007resummation}. Applying it to causal sets is simpler since we are always working with matrices and finite sums.}
\be
\label{conv}
G_m=G_0 - m^2\,G_0*G_0 + m^4 \,G_0*G_0*G_0+ \ldots = \sum_{k=0}^\infty(-m^2)^k \underbrace{G_0 * G_0* \ldots G_0}_{k+1}
\ee
where
\be
(A\ast B)(x,x')\equiv \int d^dx_1 \sqrt{-g(x_1)} A(x,x_1) B(x_1,x')\,.
\ee
Note that if $G_0(x,x')$ is retarded (\textit{i.e.}, only nonzero if $x'$ is in the causal past of $x$) then so is $G_m(x,x')$. Also note that since $G_0$ is retarded, the convolution integrals are over finite regions of spacetime.
This relation can be reexpressed in the compact form
\be
\label{cconv}
G_m=G_0 - m^2 G_0*G_m\,.
\ee
Conversely  $G_0$ can be obtained  from  $G_m$ via
\be
\label{oppconv}
G_0=\sum_{k=0}^\infty(m^2)^k \underbrace{G_m * G_m* \ldots G_m}_{k+1}
\ee
and
\be
G_0=G_m+m^2 G_m *G_0 = G_m + m^2  G_0*G_m\,.
\ee
Once we have the massless retarded Green function, we can write down a formal series
for the massive retarded Green function.

Now if we have a massless retarded Green function analogue, $K_0(x,x')$, on a causal set which is a sprinkling at density $\rho$ into the $d$-dimensional spacetime, we can immediately propose a massive retarded Green function $K_m(x,x')$ on that causal set via the replacement
\be
\int \sqrt{-g(x)}\,d^d x \rightarrow  \rho^{-1} \sum_{\textrm{causal set elements} }\,,
\ee
leading to
\be
\label{convK}
K_m =  K_0- \frac{m^2}{\rho} K_0*K_0+\frac{m^4}{\rho^2} K_0*K_0*K_0 + \ldots = \sum_{k=0}^\infty\left(-\frac{m^2}{\rho}\right)^k \underbrace{K_0 * K_0* \ldots K_0}_{k+1}
\ee
where now the convolutions have become finite sums over causal set elements in the causal interval $(x,x')$. The series terminates and is well-defined for each pair $x$ and $x'$.

We will now show that Johnston's hop-stop models for the massive retarded Green functions on causal sets approximated by 2 and 4 dimensional Minkowski space are based on natural causal set analogues of the massless Green functions .

\subsection{$d=2$ Minkowski spacetime}

The massless retarded Green  function in $d=2$ Minkowski spacetime $\mink^2$ is
\be
\label{masslesscont2d}
G^{(2)}_0(x,x')=  \frac{1}{2}\theta(x_0-x'_0)\theta(\tau^2(x,x'))
\ee
where $\tau(x,x')$  is defined by
\begin{align}
\label{tau2d}
&\tau(x,x') =   \sqrt{  (x_0-x'_0)^2 - (x_1 - x'_1)^2 } \ \ \textrm{when} \ \ (x_0-x'_0)^2 \ge (x_1 - x'_1)^2 \nonumber \\
&\textrm{and}\nonumber \\
&\tau(x,x') = \,i \, \sqrt{ -  (x_0-x'_0)^2 + (x_1 - x'_1)^2  } \ \ \textrm{when} \ \  (x_0-x'_0)^2 < (x_1 - x'_1)^2  \,.
\end{align}
$\theta$ is the Heaviside step function.

Now consider, on a causal set, the causal matrix $C_0(x,x')$.
The Poisson point process of sprinkling at density $\rho$ in 2 dimensional Minkowski spacetime gives rise to a random variable, $\mathbf{C}_0(x,x')$
for every two points, $x$ and $x'$ via the evaluation of $C_0(x,x')$ on that causal set.
In this case, the random variable takes the same value -- the expectation value -- in each realization. It was observed in \cite{Daughton1993rh} that this value is
\be
\label{expC0}
\av{\bC_0(x,x')} =  2 G^{(2)}_0(x,x')\,.
\ee
This leads to the proposal for a massless retarded Green function, $K^{(2)}_0(x,x')$,
on a $d=2$ flat sprinkled causal set:
\be
\label{massless2d}
K^{(2)}_0(x,x')\equiv  \frac{1}{2}C_0(x,x').
\ee
We define a massive Green function $K^{(2)}_m(x,x')$ on $\mathcal{C}$ using this $K^{(2)}_0(x,x')$ and
(\ref{convK}).

Using Eq.\eqref{convK} and $k$-chains on the causal set then gives
\be
\label{discreteGF2d}
K^{(2)}_m(x,x') =  \sum\limits_{k=0}^\infty  \biggl(- \frac{m^2}{\rho}\biggr)^{k}\biggl(\frac{1}{2}\biggr)^{k+1} C_k(x,x')\,,
\ee
where the sum is written as an infinite sum but terminates for each pair $x$ and $x'$.

For each two points $x$ and $x'$ of $\mink^2$  and each $k$ the random variable
$\bC_k(x,x')$ is $C_k(x,x')$ evaluated on a sprinkled causal set including $x$ and $x'$,
and hence we have the random variable $\bK^{(2)}_m(x,x')$:
 \be
\label{discreteGF2drandom}
\bK^{(2)}_m(x,x') \equiv   \sum\limits_{k=0}^\infty  \biggl(-\frac{m^2}{\rho}\biggr)^{k}\biggl(\frac{1}{2}\biggr)^{k+1} \bC_k(x,x').
\ee
Its expectation value -- for any sprinkling density -- is equal to the continuum Green function since
\be
\label{abundancechains}
 \av{\bC_k(x,x')} = \rho^k (\underbrace{\av{\bC_0}\ast \ldots \ast \av{\bC_0}}_{k+1}) (x,x')
\ee
and so
\begin{align}
\av{\bK^{(2)}_m(x,x')} & = \sum\limits_{k=0}^{\infty} \biggl(-\frac{m^2}{\rho}\biggr)^{k}\biggl(\frac{1}{2}\biggr)^{k+1} \av{\bC_k(x,x')} \label{proof1}\\
& = \sum\limits_{k=0}^{\infty}(-m^2)^{k}\underbrace{G^{(2)}_0 * G^{(2)}_0* \ldots G^{(2)}_0}_{k+1} (x,x')\label{proof2}\\
&= G^{(2)}_m(x,x')\label{proof3}\,.
\end{align}

In \cite{Johnston2008} $K^{(2)}_m(x,x')$ was expressed in terms of the hop and stop weights, $a$ and $b $ respectively:
\be
\label{hopstop}
K^{(2)}_m(x,x') =  \sum\limits_{k=0}^\infty  a^{k+1} b^k C_k(x,x').
\ee
This form was described by Johnston using a particle language as a sum over all chains
between $x$ and $x'$:
for each $k$-chain the hop between two
successive elements is assigned the weight $a$ and the stop at each intervening element between
$x$ and $x'$ is assigned the weight $b$.  Now we see that the weight $a=\frac{1}{2}$ is associated to
 each factor of $K^{(2)}_0$ -- from the relationship between $K^{(2)}_0$ and the
 causal matrix -- and the weight $b=-\frac{m^2}{\rho}$ to each convolution. In
 \cite{Johnston2008} a momentum space calculation was used to find $b$, but as we have just seen the
spacetime formulation is sufficient to read off the value.

\subsection{$d=4$ Minkowski spacetime}

In $d=4$ Minkowski spacetime, $\mink^4$, the retarded Green  function for the massless field only
has support on the light cone:
\be
G^{(4)}_0(x,x')= \frac{1}{2\pi}\theta(x_0-x'_0) \delta( \tau^2(x,x'))\,,
\label{cont4d}
\ee
where
\begin{align}
\label{tau4d}
&\tau(x,x') =   \sqrt{  (x_0-x'_0)^2 - (x_1 - x'_1)^2 -   (x_2 - x'_2)^2 - (x_3 - x'_3)^2} \ \ \textrm{when} \nonumber \\
& \ \ \ \ \ \ \ \ \ \ \ \ (x_0-x'_0)^2 \ge (x_1 - x'_1)^2 + \dots +(x_{3} - x'_{3})^2 \nonumber \\
&\textrm{and}\nonumber \\
&\tau(x,x') = \, i \, \sqrt{ -   (x_0-x'_0)^2 + (x_1 - x'_1)^2 +  (x_2 - x'_2)^2 + (x_3 - x'_3)^2 } \ \ \textrm{when} \nonumber \\
& \ \ \ \ \ \ \ \ \ \ \ \ (x_0-x'_0)^2 < (x_1 - x'_1)^2 + \dots +(x_3 - x'_3)^2  \,.
\end{align}

The causal set analogue is proportional to the link matrix. The expectation value
of the corresponding random variable $\bL_0(x,x')$ in a Poisson sprinkling of density $\rho$ is
\be
\label{linkexp}
\av{\bL_0(x,x')}=\theta(x_0-x'_0) \theta( \tau^2(x,x'))\exp(-\rho V(x,x')),
\ee
where $V(x,x')$ is the volume of the spacetime interval $J^-(x) \cap
J^+(x')$. Here $J^{+}(x)$ and $J^{-}(x)$ denote \footnote{see \cite{Wald_1984} for
  example} the causal future and past of $x$, respectively. In $\mink^4$,
$V(x,x')= \frac{\pi}{24}\tau^4(x,x')$, so that
\begin{align}
\label{linklim}
\lim_{\rho \rightarrow \infty} {\sqrt{\frac{\rho}{6}}\av{\bL_0(x,x')}}&= 2\,\theta(x_0-x_0') \theta(\tau^2)
\delta(\tau^2)\\
& =  \theta(x_0-x_0') \delta(\tau^2)\\
& = 2 \pi G^{(4)}_0(x,x')\,.
\end{align}
This therefore suggests that we take the massless Green function on a flat 4-d causal set to be 
\be
\label{massless4d}
K^{(4)}_0(x,x')= \frac{1}{2 \pi} \sqrt{\frac{\rho}{6}} L_0(x,x')\,.
\ee
The relationship with the continuum Green function is not so direct as
in $d=2$ since here it is only in the continuum limit as $\rho\rightarrow \infty$ that the
expectation value of $K^{(4)}_0$ over sprinklings equals the continuum $G^{(4)}_0$.
We use this $K^{(4)}_0$  to construct a massive Green function $K^{(4)}_m(x,x')$ via
(\ref{convK}) as before
\be
\label{csdiscreteGF4d}
K^{(4)}_m(x,x') =  \sum\limits_{k=0}^\infty  \biggl(-\frac{m^2}{\rho}\biggr)^{k}\biggl(\frac{1}{2\pi}\sqrt{\frac{\rho}{6}}\biggr)^{k+1} L_k(x,x')\,,
\ee
where the sum terminates for each pair $x$ and $x'$.

For each two points $x$ and $x'$ of $\mink^2$  and each $k$, the random variable
$\bL_k(x,x')$ is $L_k(x,x')$ evaluated on a sprinkled causal set including $x$ and $x'$,
and hence we have the random variable $\bK^{(4)}_m(x,x')$:
 \be
\label{discreteGF4d}
\bK^{(4)}_m(x,x') \equiv   \sum\limits_{k=0}^\infty  \biggl(-\frac{m^2}{\rho}\biggr)^{k}\biggl(\frac{1}{2\pi}\sqrt{\frac{\rho}{6}}\biggr)^{k+1} \bL_k(x,x').
\ee
The limit as $\rho\rightarrow \infty$ of its expectation value  is equal to the series for the continuum Green function since
\be
\label{abundancepaths}
 \av{\bL_k(x,x')} = \rho^k (\underbrace{\av{\bL_0}\ast \ldots \ast \av{\bL_0}}_{k+1}) (x,x')
\ee
and so
\begin{align}
\lim_{\rho \rightarrow \infty} \av{\bK^{(4)}_m(x,x')} & = \lim_{\rho \rightarrow \infty} \sum\limits_{k=0}^{\infty} \biggl(-\frac{m^2}{\rho}\biggr)^{k}\biggl(\frac{1}{2\pi}\sqrt{\frac{\rho}{6}}\biggr)^{k+1} \av{\bL_k(x,x')}\\
&= \lim_{\rho \rightarrow \infty}  \sum\limits_{k=0}^{\infty} (-m^2)^{k}\biggl(\frac{1}{2\pi}\sqrt{\frac{\rho}{6}}\biggr)^{k+1}
\underbrace{\av{\bL_0}\ast \ldots \ast \av{\bL_0}}_{k+1} (x,x')\\
& = \sum\limits_{k=0}^{\infty} (-m^2)^{k}\underbrace{G^{(4)}_0 * G^{(4)}_0* \ldots G^{(4)}_0}_{k+1} (x,x')\\
&= G^{(4)}_m(x,x')\,.
\end{align}

Johnston interpreted (\ref{massless4d}) as a sum over paths between $x$ and $x'$. The hop-stop weights can be read off from Eqn (\ref{discreteGF4d}) as
$a=\frac{1}{2 \pi} \sqrt{\frac{\rho}{6}} $ and $b=-\frac{m^2}{\rho}$, respectively.

\bfig[h]
 \centering
  \includegraphics[scale=0.5]{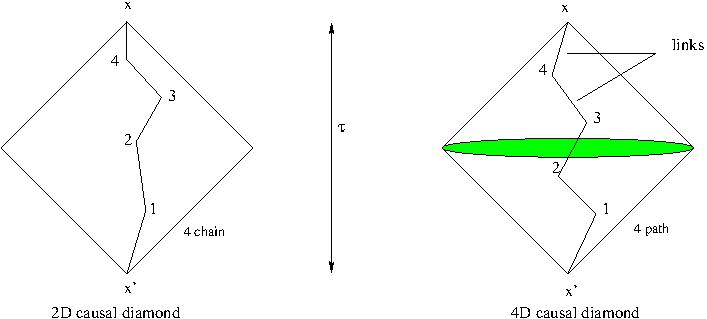}
\caption{The causal trajectories in $d=2$ and $4$ dimensions.}
 \efig

\section{Generalisations}
\label{sec:generalisations}
The key to the above construction of a massive Green function is knowing the massless one. We can repeat it
if we can find the massless retarded Green function for causal sets sprinkled into more general curved spacetimes.

Consider the more general scalar theory with nonminimal coupling with Green function
$G_{m,\xi}(x,x')$ which satisfies
\be
\label{curvedG}
(\Box_g - m^2 -\xi R)G_{m,\xi}(x,x')=\frac{1}{\sqrt{-g(x)} }\delta(x-x')\,.
\ee
$G_{m,\xi}(x,x')$ can be obtained from $G_{0,\xi}(x,x')$ using the same series expansion Eqn(\ref{conv}):
\be
\label{convxi}
G_{m,\xi}= \sum_{k=0}^\infty(-m^2)^k \underbrace{G_{0,\xi} * G_{0,\xi}* \ldots G_{0,\xi}}_{k+1} \,.
\ee

In the special case when the spacetime has constant scalar curvature $R$, then the $\xi R$ term just modifies the mass and  $G_{m,\xi}(x,x')$  can be obtained from
the minimally coupled massless Green  function $G_{0,0}(x,x')$
using a series expansion Eqn(\ref{conv}) with $m^2$ replaced by $m^2+\xi R$.
In general, for constant $R$, we can relate the two Green functions 
\be
\label{cconvxi}
G_{m', \xi'}= \sum_{k=0}^\infty(-{m'}^2 - \xi' R +m^2 + \xi R)^k \underbrace{G_{m,\xi} * G_{m,\xi}*
  \ldots G_{m,\xi}}_{k+1} \, , 
\ee
for any $(m,\xi)$, $(m',\xi')$. 

We seek analogous massive scalar Green functions, $K_{m,\xi}(x,x')$, for causal sets sprinkled into curved spacetimes. We will see that this is possible in special cases.

\subsection{$d=2$}

Every $d=2$  spacetime is locally conformally flat. The conformal coupling in $d=2$ is $\xi=0$, \textit{i.e.},
conformal coupling is minimal coupling. If the spacetime is topologically trivial and
consists of one patch covered by conformally flat
coordinates, then the minimally coupled massless Green function equals the flat spacetime Green function (\ref{masslesscont2d}).

 Therefore, we propose that on
  causal sets sprinkled into such $d=2$ spacetimes,
   the massless minimally coupled causal set Green function,
  $K^{(2)}_{0,0}(x,x')$, is the flat one given by Eqn (\ref{massless2d}) and therefore that
  $K^{(2)}_{m,0}(x,x')$ is the flat one
given by Eqn (\ref{discreteGF2d}):
\be
K^{(2)}_{m,0}(x,x') =  \sum\limits_{k=0}^\infty  \biggl(- \frac{m^2}{\rho}\biggr)^{k}\biggl(\frac{1}{2}\biggr)^{k+1} C_k(x,x')\,,
\ee

The argument that the expectation value over sprinklings of the corresponding random variable will be the
correct continuum Green function proceeds exactly as in the flat case: (\ref{discreteGF2d})--(\ref{proof3}).
However it is formal and we will provide more concrete evidence. We will verify
directly that this $K^{(2)}_{m,0}$ does have the correct expectation value value over sprinklings in an
RNN.

In our calculation below as well as in Section \ref{subsec:4d},   the RNN should be seen as providing an
intermediate scale at which the continuum description is still valid, 
and which is therefore  much larger than the discreteness scale. The
reason to use the RNN is simply that the calculations can be done
explictly to leading order both in the causal set as well as the
continuum.

\subsubsection{ RNN in $d=2$}

Consider the RNN $(O,g)$ with Riemann normal coordinates with origin $x'$.
The metric at $x\in O$ can be expanded to first order about $x'$ in these coordinates as
\be
\label{rnn}
g_{ab}(x)=\eta_{ab}+\dfrac{1}{2!}\partial_{c}\partial_{d}\,g_{ab}(x')x^{c}x^{d}+\mathcal{O}(x^{3}).
\ee
where $\eta_{ab}$ is the metric of Minkowski spacetime in inertial coordinates and $\partial_c g_{ab}(x')=0$. In the RNN, $|R \tau^2(x,x') |<<1$ and we work in an approximation where we drop terms involving derivatives of the curvature or quadratic and higher powers of the curvature.

The $d$ dimensional momentum space Green function in a RNN has been calculated by Bunch and
Parker~\cite{Bunch:1979uk}. To leading order the density
\be
\bG_{m,\xi}(x,x')\equiv (-g(x))^{\frac{1}{4}}G_{m,\xi}(x,x')
\ee
satisfies the  equation
\be
\label{rnnbox}
(\Box_\eta -(m^2+(\xi-\frac{1}{6}) R(x'))\bG_{m,\xi}(x,x')\approx - \delta(x-x')\,,
\ee
where $\Box_\eta = \eta^{ab}\nabla_a\nabla_b$ and acts on the $x$ argument. 
This
has the momentum space solution
\be
\label{ftG}
\bG_{m,\xi}(p) \approx \frac{1}{p^2+m^2} - (\xi -\frac{1}{6}) R(x')\frac{1}{(p^2+m^2)^2}.
\ee
This solution was obtained iteratively using the expansion 
\be 
\bG_{m,\xi}(p)=\bG_{m,\xi,0}(p)+\bG_{m,\xi,1}(p)+\bG_{m,\xi,2}(p) + \ldots 
\ee 
where $\bG_{m,\xi,0}(p)=(k^2+m^2)^{-1}$ is the flat spacetime Green function which is independent of
$\xi$. This expansion is valid when the Compton wavelength of the particle is much smaller than the
curvature scale, i.e., $m^2>> \xi R$, a physically reasonable assumption. 
The spacetime function  can then be expressed as
\be
\bG_{m,\xi}(x,x') \approx G^F_m(x,x')+\frac{1}{2m} (\xi -\frac{1}{6}) R(x')\, \partial_m G^F_m(x,x'),
\ee
where $G_m^F(x,x')$ is the massive minimally coupled Green function in $\mink^d$.
The Green  function is then
\be
\label{rnnG}
G_{m,\xi}(x,x')\approx \biggl(1+\frac{1}{12} R_{ab}(x')x^a x^b\biggr)G^F_m(x,x')+ \frac{1}{2m} (\xi -\frac{1}{6}) R(x')\, \partial_m G^F_m(x,x').
\ee

Now we specialise to $d=2$. Using the $d=2$ Minkowski spacetime solution for the massive retarded solution
\be
 \frac{1}{2} \theta(x_0) \theta(\tau^2) J_0(m\tau)\,,
\ee
where $\tau=\tau(x,x')$ (\ref{tau2d}), the retarded massive Green  function
in $(O,g)$ is given by 
\be
\label{2dcontden}
G^{(2)}_{m,\xi}(x,x') \approx \theta(x_0) \theta(\tau^2)\left[  \dfrac{1}{2}J_{0}(m\tau) +\dfrac{R(x')\tau^2}{48}J_{2}(m\tau)
- \frac{\xi R(x')\tau}{4m}J_1(m\tau)\right]\,.
\ee

We begin by defining a potential causal set Green function motivated from the Minkowski case
\be
\label{2dgeneralab}
{\cal{K}}^{(2)}(a,b)(x,x') \equiv \sum\limits_{k=0}^\infty  a^{k+1} b^k C_k(x,x')
\ee
for arbitrary weights $a$ and $b$. We want to show that the corresponding random variable for sprinklings into a RNN has the correct expectation value, (\ref{2dcontden}) when $a$ and $b$ take their
flat space values $a= \frac{1}{2}$ and $b = -\frac{m^2}{\rho}$.

We can calculate $\av{{\cal{K}}^{(2)}(a,b)(x,x')}$ starting from  Eqn (\ref{2dgeneralab}) if we know
 $\av{\bC_k(x,x')}$ in a small causal diamond. This was
calculated, to first order in curvature, in \cite{Roy:2012uz} for arbitrary $d\geq 2$.  In $d=2$ the expression is
\be
\av{\bC_{k}(x,x')}\approx \av{\bC_{k}(x,x')}_\eta\bigg(1-\dfrac{R(x')\tau^2}{24}\dfrac{k}{k+1}\bigg)
\ee
where $\av{\bC_{k}(x,x')}_\eta= \theta(x_0)\theta(\tau^2)\dfrac{1}{\Gamma(k+1)^2}\biggl(\dfrac{\rho\tau^2}{2}\biggr)^k$
is the expectation value in flat space.
Using the series expansion of the Bessel functions we see that
\bea
\label{2csprop}
\av{{\cal{K}}^{(2)}(a,b)} &\approx& \theta(x_0)\theta(\tau^2) \sum_{k=0}^{\infty}a^{k+1}b^{k}\bigg(\dfrac{\rho \tau^2}{2}\bigg)^{k}\dfrac{1}{(\Gamma(k+1))^2}\bigg(1-\dfrac{R(x')\tau^2}{24}\dfrac{k}{k+1}\bigg)\nonumber\\
&\approx& \theta(x_0)\theta(\tau^2) \left[aI_{0}(\tau\sqrt{2ab\rho}) - \dfrac{aR(x')\tau^2}{24}I_{2}(\tau\sqrt{2ab\rho})\right].
\eea
If we set
$a= \frac{1}{2}$,  $b= - \frac{m^2}{\rho}$ we find
\be
\label{2csprop1}
\av{{\cal{K}}^{(2)}(\frac{1}{2}, - \frac{m^2}{\rho})(x,x')}\approx  \theta(x_0)\theta(\tau^2)\left[\dfrac{1}{2}J_{0}(m\tau) +\dfrac{R(x')\tau^2}{48}J_{2}(m\tau)\right],
\ee
which matches  Eqn (\ref{2dcontden}) for $\xi=0$.

We further note that in the RNN since $R(x') \approx R$,  a constant  to this order of
approximation,  we can use the
observation above that $\xi R$ can be treated as a contribution to the mass.
Putting $a=\frac{1}{2}$ and  $b=-\frac{(m^2+\xi R)}{\rho}$ in (\ref{2csprop})  and using $m^2 >>
\xi R$, we obtain
\begin{align}
& \theta(x_0)\theta(\tau^2)\left[\dfrac{1}{2}J_{0}(\tau\sqrt{m^2+\xi R})+ \dfrac{R\tau^2}{48}J_{2}(\tau\sqrt{m^2+\xi R})\right]\nonumber\\
\approx &\  \theta(x_0)\theta(\tau^2)\left[\frac{1}{2}\sum_{n=0}^{\infty}\frac{(-1)^n}{(n!)^2}\bigg(\frac{\tau}{2}\bigg)^{2n}(m^2+\xi R)^n + \dfrac{R\tau^2}{48}J_{2}(m\tau)\right]\nonumber\\
\approx&\  \theta(x_0)\theta(\tau^2)\left[\dfrac{1}{2}J_{0}(m\tau)  + \dfrac{R\tau^2}{48}J_{2}(m\tau)
 - \frac{\xi R\tau}{4m}J_1(m\tau)\right]
\end{align}
which agrees with Eqn (\ref{2dcontden}).   Thus, for a causal set sprinkled into an approximately flat causal diamond in $d=2$,
Eqn (\ref{2dgeneralab}) with $a=  \frac{1}{2}$ and $b= - \frac{(m^2 + \xi R(x'))}{\rho}$  is approximately the
``right'' massive causal set Green function for general coupling $\xi$.

\subsection{$d=4$}

\subsubsection{RNN  in $d=4$ }
\label{subsec:4d}

The approximate continuum retarded Green  function in the RNN in $d=4$ simplifies to
\begin{align} \label{rnn4dcont}
G^{(4)}_{m,\xi}(x,x') \approx &\ \theta(x_0)\Biggl[  \biggl(\dfrac{1}{2\pi}\delta(\tau^2) -
\theta(\tau^2)\frac{m}{4\pi\tau}J_1(m\tau)\biggr)\biggl (1+
\dfrac{1}{12}R_{ab}(x')x^a x^b\biggr) \Biggr]\\
& - \theta(x_0)\theta(\tau^2)\bigg(\xi-\frac{1}{6}\bigg)\frac{R(x')}{8\pi}J_0(m\tau),
\end{align}
which reduces to the massless Green  function
\begin{align}
\label{rnn4dcont0}
 G^{(4)}_{0,\xi}(x,x') \approx
 \dfrac{1}{2\pi} \theta(x_0)\delta(\tau^2)\bigg(1+\dfrac{1}{12}R_{ab}(x')x^a x^b\bigg) - \theta(x_0)\theta(\tau^2)\bigg(\xi-\frac{1}{6}\bigg)\frac{R(x')}{8\pi}.
\end{align}
Even this simplified expression is formidable to mimic in the causal set since not only does it
require the discrete scalar curvature \cite{Benincasa:2010ac} but also the \textit{components} of the Ricci curvature for
which no expression is known.  However,  for conformal coupling
$\xi=\frac{1}{6}$ and Einstein spaces with Ricci curvature $R_{ab}\propto g_{ab}$,
(\ref{rnn4dcont0}) reduces to the Minkowski spacetime form (\ref{cont4d}). Indeed, we only require
that $R_{ab}(x')\appropto g_{ab}(x') $ upto the order we are considering. 
This suggests that the flat spacetime massless causal set Green function (\ref{massless4d})
may give the right continuum Green function.  Since $R$ is approximately constant
in the RNN (and exactly constant in an Einstein space)  we can
use the series in powers of the massless Green function to propose the
massive one for arbitrary $\xi$.

For the massless field let us calculate the expectation value of the link matrix, given by (\ref{linkexp}).  The
spacetime volume in the RNN has corrections to the Minkowski spacetime volume $V_\eta(x,x')$
\cite{Myrheim:1978ce, Gibbons2007,Khetrapal2013} which in $d=4$ are
\be
V(x,x')\approx V_\eta(x,x')\bigg(1-\dfrac{1}{180}R(x')\tau^2+\dfrac{1}{30}R_{ab}(x')x^ax^b\biggr).
\ee
To leading order then
\bea
\av{\bL_0(x,x')} \approx \theta(x_0) \theta( \tau^2) e^{-\rho V_\eta(x,x')}\bigg(1+\dfrac{\rho V_\eta(x,x')}{180}R(x')\tau^2-\dfrac{\rho V_\eta(x,x')}{30}R_{ab}(x')x^ax^b\bigg).
\eea
Since  $V_\eta(x,x')=\frac{\pi}{24}\tau^4(x,x')$,  $\sqrt{\rho} \av{\bL_0(x,x')}$ contains
terms of the form
\be
h_n(\rho,\tau) \equiv \sqrt{\rho} (c \rho \tau^4)^n \exp(-c\rho\tau^4),
\ee  with $n=0,1$.
  
We now show that given the function 
\bea
h_n(\rho,z) \equiv \sqrt{\rho} (c \rho z^2)^n \exp(-c\rho z^2),\\
\lim_{\rho\to\infty}h_n(\rho, z)=\dfrac{\Gamma(n+1/2)}{\sqrt{ c}}\delta(z).
\label{distid}
\eea
First, we evaluate the integral
\bea 
\int_{-\infty}^{\infty}dz\,\, h_n(\rho, z) &=& 
2\sqrt{\rho}\int_{0}^{\infty}dz\,\, ( c\rho z^2)^ne^{- c\rho z^2} \nonumber \\ 
&=& \dfrac{1}{\sqrt{\pi c}}\int_{0}^{\infty}dt\,\, (t)^{n-1/2}e^{-t}\nonumber\\
&=&\dfrac{\Gamma(n+1/2)}{\sqrt{\pi c}}, 
\eea 
where we made a change of variables $t= c\rho z^2$. This result is independent of $\rho$.
 
Next, we integrate $h_n(\rho, z)$ with an analytic  test function and take the
limit $\rho\rightarrow \infty$.  If $f(z)$ is odd, the integral vanishes (this also happens with the delta function) and we can restrict to even analytic functions 
\be 
f(z)=\sum_{k=0}^{\infty}a_k z^{2k}.  
\ee  
For this, 
\bea
\lim_{\rho\rightarrow\infty}\int_{-\infty}^{\infty}dz\,\,f(z)\,h_n(\rho, z)&=&\lim_{\rho\rightarrow\infty}\sum_{k=0}^{\infty}a_k\int_{-\infty}^{\infty}dz\,\,z^{2k}\,h_n(\rho, z)\nonumber\\
&=&\lim_{\rho\rightarrow\infty}2\sum_{k=0}^{\infty}a_k\int_{0}^{\infty}dz\,\,z^{2k}\sqrt{\rho}( c\rho z^2)^ne^{- c\rho z^2}\nonumber\\
&=&\lim_{\rho\rightarrow\infty}2\sum_{k=0}^{\infty}a_k\dfrac{\sqrt{\rho}}{( c\rho)^k}\int_{0}^{\infty}dz\,\,( c\rho z^2)^{n+k}e^{- c\rho z^2}\nonumber\\
&=&\lim_{\rho\rightarrow\infty}\sum_{k=0}^{\infty}a_k\dfrac{\Gamma(n+k+1/2)}{\sqrt{ c}( c\rho)^k}\nonumber\\
&=&a_0\dfrac{\Gamma(n+1/2)}{\sqrt{\pi c}}=\dfrac{\Gamma(n+1/2)}{\sqrt{ c}}f(0).\nonumber
\eea
Noting that  $n=0$ is the  usual Gaussian integral, and that the behaviour with test functions is one way to define a delta function \cite{deltarefs}, this
proves Eqn (\ref{distid}).

Using this result with $z=\tau^2$, we find
\be
\label{rnncs4d}
\lim_{\rho \rightarrow \infty} \frac{\sqrt{\rho}}{2\pi \sqrt{6}} \av{\bL_0(x,x')} \approx
\dfrac{1}{2\pi}\theta({x_0})\delta(\tau^2)\bigg(1+\dfrac{R(x')\tau^2}{360}-\dfrac{R_{ab}(x')x^ax^b}{60}\bigg).
\ee
The second term vanishes in general and so does the third term when $R_{ab}(x') \propto g_{ab}(x')$
upto this order, and we recover (\ref{cont4d}).  Thus, for sprinklings into a RNN with $R_{ab}(x')
\propto g_{ab}(x')$ to this order, the continuum limit of the expectation value of (\ref{massless4d}) is
approximately the correct value for the Green function of the conformally coupled massless field.

As in the $2d$ case we define 
\be
\label{4dgeneralab}
{\cal{K}}^{(4)}(a,b)(x,x') \equiv \sum\limits_{k=0}^\infty  a^{k+1} b^k L_k(x,x')\,,
\ee
we propose that this is the appropriate causal set Green function for  the massive field and arbitrary coupling $\xi$ in an  RNN with $R_{ab}(x') \propto g_{ab}(x')$ to this order, with $a=\frac{1}{2\pi}\sqrt{\frac{\rho}{6}}$ and $b = -\frac{m^2+ (\xi-\frac{1}{6})R}{\rho}$.

We are unable to verify this directly because there is no known closed form expression for the expectation value of $\bL_k$, the number of $k$-paths for $k\ge 1$, even in an RNN.

\subsubsection{$d=4$ de Sitter and anti de Sitter}
\label{subsec:desitter}

In $d=4$ for  conformally flat spacetimes $g_{ab}=\Omega^2(x)\eta_{ab}$   the conformally
coupled massless Green  function is related to that in $\mink^4$ by  
\be 
\label{confflat}
G_{0,\xi_c}(x,x')=\Omega^{-1}(x) G_0^{F}(x,x')\Omega^{-1}(x'),  
\ee 
where $\xi_c = \frac{1}{6}$ and  $G_0^F(x,x')$ denotes the retarded massless Green  function in $\mink^4$. 
When $g_{ab}$ in
addition has constant scalar curvature the massive Green function for arbitrary $\xi$ can be obtained from
$G_{0,\xi_c}(x,x')$  using Eqn (\ref{convxi}).

An example is the conformally flat patch of de Sitter spacetime 
\be 
ds^2= \frac{1}{(1+H x_0)^2} \biggl( -d x_0^2 + \sum_{i=1}^3 d x_i^2  \biggr), 
\ee 
where $x_0 $ is the conformal time ($-\frac{1}{H}< x_0 <\infty $) and $H=\sqrt{\frac{\Lambda}{3}}$
with $\Lambda$ the cosmological constant.  The conformally coupled massless retarded Green  function is 
\be 
G_{0,\xi_c}(x,x')=\frac{1}{2\pi} \theta(t-t')\delta(\tau^2(x,x')) (1+Hx_0)(1+Hx_0'). 
\ee 
Since de Sitter is homogeneous, one can choose $x'$ to lie at the  convenient location  $x'=(0, \vec 0)$ so that 
\be
 G_{0,\xi_c}(x,x')=\frac{1}{2\pi} \theta(x_0)\delta(\tau^2(x,x')) (1+Hx_0). 
\ee 

Taking our cue from the RNN calculation, we look to the link matrix $L_0(x,x')$ whose expectation 
value is given by Eqn (\ref{linkexp}). Taking $x'=(0,\vec 0)$, we see that
\be 
\label{sqrtv}
\lim_{\rho \rightarrow \infty} \sqrt{\frac{\rho}{\pi}} \av{\bL_0(x,x')}=\theta(x_0)
\theta(\tau^2(x,x'))\delta(\sqrt{V(x,x')}).  
\ee 
In order to evaluate this expression we need to find $V(x,x')$.  In \cite{Gibbons2007} this volume
was calculated for a large interval when $\vec x=\vec x'$. However,  it is the
small volume limit that is relevant to our present  calculation.  When $x$  lies
in an RNN about $x'$, the calculation in the previous section suffices. However, we also need to consider
intervals of small volume that lie outside of the RNN. These ``long-skinny'' intervals hug the future
light cone of $x'$ and it is this contribution to  Eqn (\ref{sqrtv}) that we will now consider. 

In the following light cone coordinates 
\be 
u=\frac{1}{2}(x_0-x_3), v=\frac{1}{2}(x_0+x_3), 
\ee 
let $u(x)=\epsilon$, $v(x)=L$, with $\alpha^2\equiv \frac{\epsilon}{L} <<1$.  Since there is a
spatial rotational symmetry in de Sitter, we can also take $x_1=x_2=0$.  In order to simplify the
calculation of $V(x,x')$, we perform a boost about $x'$ in the $x_0-x_3$ 
plane about $x'$ so that  $\tx=(\tx_0, \vec{0})$. The boost parameter  is then $\beta=\frac{x_3}{x_0} \approx
1-2\alpha$.  In these coordinates the conformal factor at a point $y=(y_0, \vec{y})$ is  
\be 
\tO^2(\ty)\approx \frac{1}{(1+A(\ty_0+\ty_3))^2}
\ee 
where $A=\frac{1}{2}H\alpha$. Further transforming to cylindrical coordinates $(\ty_1,\ty_2,\ty_3)
\rightarrow (r,\phi, \ty_3)$ we can split $V(x,x')$ into two multiple integrals
\bea 
\label{volexprs} 
V_I(x,x')&=& \int_{0}^{-\frac{\tau}{2}} \!\!d\ty_0 \int_{-\ty_0}^{\ty_0} \!\!\!\! d\ty_3\int_0^{\sqrt{\ty_0^2-\ty_3^2}}
\!\! \!\!\!r dr \int_0^{2\pi} d \phi \, \biggl(1+A(\ty_0+\ty_3)\biggr)^{-4}   \\ 
V_{II}(x,x')&=& \int_{\frac{\tau}{2}}^{\tau} \!\! d\ty_0 \int_{-\tau+\ty_0}^{\tau-\ty_0} \!\!\!\! d\ty_3\int_0^{\sqrt{(\tau-\ty_0)^2-\ty_3^2}}
\!\! \!\!\!\!\! \!\!\!r dr \int_0^{2\pi} d \phi \,\, \biggl(1+A(\ty_0+\ty_3)\biggr)^{-4}    
\eea 
with $V(x,x')=V_I(x,x')+V_{II}(x,x')$. Evaluating these expressions using $\tau^2=4 L \epsilon$ we find that 
\be
\label{voldS}
\sqrt{V(x,x')} = \frac{1}{2}\sqrt{\frac{\pi}{6}}\frac{\tau^2}{{(1+A\tau)}}\approx
\frac{1}{2}\sqrt{\frac{\pi}{6}}\biggl(\frac{4 L \epsilon}{1+HL} \biggr), 
\ee 
which substituted into Eqn
(\ref{sqrtv}) gives 
\be 
\lim_{\rho \rightarrow \infty} \sqrt{\frac{\rho}{\pi}} \av{\bL_0(x,x')}=\theta(t-t')
\theta(\tau^2(x,x')) \sqrt{\frac{6}{\pi}}(1+HL)  \delta(4 L \epsilon). 
\ee 
In the small $\alpha$ limit the conformally coupled de Sitter Green  function is 
\be
  G_{0,\xi_c}(x,x') \approx \frac{1}{2\pi} (1+H L) \theta(x_0)\delta(4L\epsilon)  
\ee 
and hence 
\be
\label{masslessdeS}
 \lim_{\rho \rightarrow \infty} \frac{1}{2\pi}\sqrt{\frac{\rho}{6}}
 \av{\bL_0(x,x')}=G_{0,\xi_c}(x,x'). 
\ee 

As in the RNN, defining 
\be
\label{deSgreen4d}
{\cal{K}}^{(4)}(a,b)(x,x') \equiv \sum\limits_{k=0}^\infty  a^{k+1} b^k L_k(x,x')\,,
\ee
we propose that this is the appropriate  causal set Green function for  the massive field and
arbitrary coupling $\xi$  in de Sitter spacetime for  $a=
\frac{1}{2\pi}\sqrt{\frac{\rho}{6}}$ and $b = -\frac{m^2+ (\xi -\frac{1}{6})R}{\rho}$. 

Although our calculation is restricted to the conformally flat patch of de Sitter spacetime, the
result applies to global de Sitter, for the following reason. Let $x' \prec x$ in (global) de Sitter
spacetime. Consider a Lorentz transformation about $x'$ in the 5-dimensional Minkowski spacetime in
which the hyperboloid that is de Sitter spacetime is embedded, which brings $\vec x=\vec x'$.  This
transformation preserves the hyperboloid. One can then choose the conformally flat patch of de
Sitter with origin $\vec x'$, and use the above construction. When $x,x'$ are not causally related,
the Green functions vanish in both cases.  Thus the Green function for global de Sitter is retarded
if the conformally flat Green function is, and both satisfy the same equations, because there is no
“wrap-around” in de Sitter.

The causal set Green function  we propose is well defined on a sprinkling into global de
Sitter.  Moreover, as we have shown, its continuum limit matches that of the Green
function into the conformally flat patch and thence from the above discussion, also the  Green function
of global  de Sitter spacetime.

In anti de Sitter (adS) spacetime there exist pairs of events $x'\prec x$ such that $\tau(x,x') $ is
finite, but $V(x,x')$ is infinite. While it is possible to obtain a Poisson sprinkling into such a
spacetime, the resulting poset is not locally finite and hence not strictly a causal set. Such an
interval is moreover not globally hyperbolic and hence falls outside the scope of our analysis.
However, the interior of a conformally flat patch of adS (the so-called half-space) is globally
hyperbolic and moreover, $V(x,x')$ is finite for every $x'\prec x$ in this region. Hence this patch
of adS has a causal set description.

In the conformally flat patch the adS metric takes the form 
\be 
ds^2= \frac{1}{(1+H x_3)^2} \biggl( -d x_0^2 + \sum_{i=1}^3 d x_i^2  \biggr), 
\ee 
where we have off set the coordinates $x_3 \rightarrow x_3 + \frac{1}{H}$ in order to connect  with the
de Sitter calculation. Again choosing $x'=(0, \vec 0)$, we can write the massless Green function as 
\be
 G_{0,\xi_c}(x,x')=\frac{1}{2\pi} \theta(x_0)\delta(\tau^2(x,x')) (1+Hx_3). 
\ee 
In the boosted coordinates, upto order $\alpha^2$, the conformal factor
\be 
\Omega^2(y)=\frac{1}{(1+Hy_3)^2} \approx \frac{1}{(1+A(\ty_0+\ty_3))^2}
\ee 
and is identical to that of de Sitter in the  calculation above. Moreover, to this order,
$(1+Hx_3)=(1+H(L-\epsilon)) \approx (1+HL)$, so that $\sqrt{V(x,x')}$
is given by Eqn (\ref{voldS}). The same argument can then be carried through to show that the massive causal set
Green function for arbitrary $\xi$ in the conformally flat patch of de Sitter is given by Eqn (\ref{deSgreen4d}). 

We have thus proved \textit{exact}  results in de Sitter spacetime and in a conformally flat patch of anti de Sitter spacetime, namely that
the expectation value of the causal set retarded Green function 
\be 
\bK_0(x,x')=\frac{1}{2\pi}\sqrt{\frac{\rho}{6}}
 \av{\bL_0(x,x')}
\ee 
is equal to the continuum massless conformally coupled Green function in the limit $\rho\rightarrow
\infty$. In addition, we make the proposal that the limit of the expectation value of 
${\cal{K}}^{(4)}(a,b)(x,x') $ with the appropriate $a$ and $b$ is the continuum massive Green function for arbitrary conformal coupling $\xi$. 
 
\section{Proposal for a Green function in $\mink^3$}
\label{3dgreen}
 
As a final illustration, we make a proposal for the causal set Green  function
in $d=3$ Minkowski spacetime. In continuum flat spacetime in 3 dimensions the massless scalar Green function is
\be 
\label{3dgf}
G^{(3)}_0(x,x')=\theta(t-t')\theta(\tau^2)\dfrac{1}{2\pi\tau(x,x')}, 
\ee 
where for now we ignore the singular behaviour at $\tau(x,x')=0$.  The causal set counterpart of the
proper time $\tau(x,x')$ in $\mink^d$ was given by Brightwell and Gregory \cite{Brightwell:1990ha} to be
proportional to the \textit{length}  $l(x,x')$ of the longest chain (LLC) from $x'$ to $x$. Explicitly
\be
\label{bg1}  
\lim_{\rho \rightarrow \infty} \av{\bl(x,x')} (\rho V(x,x'))^{-1/d} = m_d
\ee 
where $m_d$ is a dimension dependent constant bounded by
\be 
\label{bounds}
1.77\leq\frac{2^{1-\frac{1}{d}}}{\Gamma(1+\frac{1}{d})}\leq
m_d\leq\frac{2^{1-\frac{1}{d}}e\,(\Gamma(1+d))^{\frac{1}{d}}}{d}\leq 2.62.
\ee 
In $\mink^d$, $\rho V(x,x')= \zeta_d \tau^d(x,x')$ with $\zeta_d$ a dimension dependent constant, so that 
\be
\label{bg}
\lim_{\rho \rightarrow \infty} \rho^{-\frac{1}{d}} \av{\bl(x,x')}=\kappa_d\,\tau(x,x')
\ee
where $\kappa_d\equiv m_d(\zeta_d)^{1/d}$.  This \textit{suggests} that  the $d=3$ massless Green  function on $\mathcal{C}$ is 
\be 
\label{3dcsprop}
K_0(x,x') \equiv a H_0(x,x'). 
\ee 
where 
\be
\label{def}
H_0(x,x')\equiv
\left\{
\begin{array}{ll}
	\dfrac{1}{l(x,x')}  & \mbox{if } x' \prec x\\
	0 & \mbox{} \mathrm{otherwise}.
\end{array}
\right.
\ee
This will give us the desired $d=3$ Green function if it were also true that    
\be
\label{limit3d}
\lim_{\rho \rightarrow \infty} \bigg\langle\frac{1}{l(x,x')}\bigg\rangle \rho^{\frac{1}{3}}= \frac{1}{\kappa_3\tau(x,x')}
\ee
then comparison with Eqn (\ref{3dgf}) gives $a=\rho^{1/3} \dfrac{\kappa_3}{2\pi}=(\dfrac{\rho \pi}{12})^{1/3} \dfrac{m_3}{2\pi}$.

While we do not have an analytical proof of Eqn (\ref{limit3d}), we present simulations here to show that for large $\rho$, $\av{\frac{1}{l(x,x')}}\rightarrow\frac{1}{\av{l(x,x')}}$ and therefore it is indeed a good approximation. 

Starting with Eqn (\ref{limit3d}) we see that 
\be 
\lim_{N \rightarrow \infty} \bigg\langle\frac{1}{l(x,x')}\bigg\rangle
\bigg(\frac{N}{V(x,x')}\bigg)^{\frac{1}{3}}=
\frac{1}{m_3\,\zeta_3^{1/3}\tau(x,x')}
\ee 
where we have used $\rho=\frac{N}{V}=\frac{N}{\zeta_3\,\tau^3}$ and $\zeta_3=\frac{\pi}{12}$. Since the volume $V(x,x')$ is fixed, the limit $\rho \rightarrow \infty$ is the same as $N \rightarrow \infty$ and hence this simplifies to 
\be 
\lim_{N \rightarrow \infty} \bigg\langle\frac{1}{l(x,x')}\bigg\rangle=\frac{1}{m_3\,N^{1/3}}
\ee
Using Eqn (\ref{bounds}) we see that 
\be
b_l := \frac{1}{1.77\,N^{1/3}}\leq\lim_{N \rightarrow \infty}
\bigg\langle\frac{1}{l(x,x')}\bigg\rangle\leq\frac{1}{2.62\,N^{1/3}}
=: b_u
\ee
where we have defined $b_l$ and $b_u$ as the lower and upper bounds respectively.  

We calculate $\av{\frac{1}{\l(x,x')}}$ and  $\frac{1}{\av{\l(x,x')}}$ for sprinklings into a causal diamond in $\mink^3$, for $N$ values ranging from  $100$ to $50000$ in steps of $100$. For each $N$ value we perform  over 50 trials from which the averages are calculated. Our results are shown in Figs (\ref{plot_1})-(\ref{plot_5}). 

In Fig (\ref{plot_1}) we see that $\av{\frac{1}{l(x,x')}}$ is well within the  bounds $b_l$ and $b_u$.  In Fig (\ref{plot_2})  we show the percentage errors defined by 
\bea
\delta_l:=\frac{1}{b_l}\bigg(\bigg\langle\frac{1}{l(x,x')}\bigg\rangle-b_l\bigg)\times 100\quad\text{and}\quad\delta_u:=\frac{1}{b_u}\bigg(\bigg\langle\frac{1}{l(x,x')}\bigg\rangle-b_u\bigg)\times 100\nonumber
\eea
with respect to the lower and upper bounds. While there is a convergence for large $N$ the error does not go to zero for either of the bounds.  

It is also  useful to compare $\av{\frac{1}{l(x,x')}}$ to $\frac{1}{\av{l(x,x')}}$ since it is the theoretical bound on the latter which we are using. As shown in (Fig (\ref{plot_3})) we find an almost perfect matching of $\av{\frac{1}{l(x,x')}}$ with $\frac{1}{\av{l(x,x')}}$ even at relatively small $N$ values. We plot the percentage error in Fig (\ref{plot_4}) where 
\be
\Delta:=\bigg(\bigg\langle\frac{1}{l(x,x')}\bigg\rangle\bigg)^{-1}\bigg(\bigg\langle\frac{1}{l(x,x')}\bigg\rangle-\frac{1}{\av{l(x,x')}}\bigg)\times 100\nonumber
\ee
which is already very small for $N \sim 200$ and dies down further as $N$ grows. 

Using the ``FindFit'' function in Mathematica we find that the best fit value for  $m_3$  is in fact 1.854 for the range of $N$ that we have considered. As can be seen in Figure (\ref{plot_5}) the errors for this fit  are  very small. 

\bfig[h]   
\bsfig[b]{\textwidth} 
\caption{$\protect\av{\frac{1}{l(x,x')}}\,\text{vs}\,N$} 
\includegraphics[width=\textwidth]{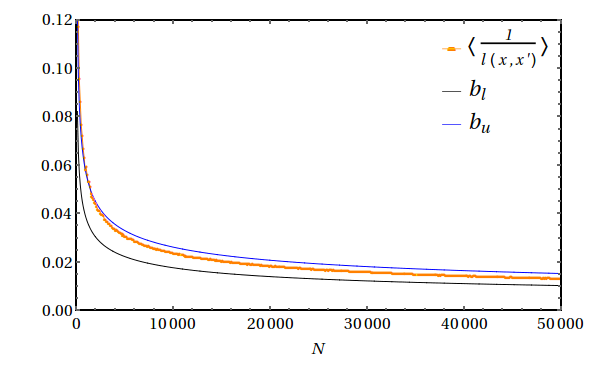}  
\label{plot_1}
\esfig
\caption{Fig (a) shows a comparison of $\av{\frac{1}{l(x,x')}}$ as a
	function of $N$, with the conjectured upper and lower bounds. }
\efig

\bfig
\ContinuedFloat
\bsfig[b]{0.5\textwidth} 
\caption{Errors in $\protect\av{\frac{1}{l(x,x')}}$ with respect to $\protect b_u\text{, }b_l$}   
\includegraphics[width=\textwidth]{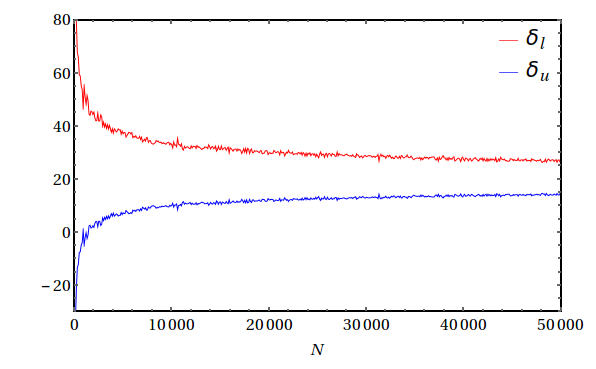} 
\label{plot_2}
\esfig
\bsfig[b]{0.5\textwidth}
\caption{Comparison of $\protect\av{\frac{1}{l(x,x')}}$ and $\protect\frac{1}{\av{l(x,x')}}$} 
\includegraphics[width=\textwidth]{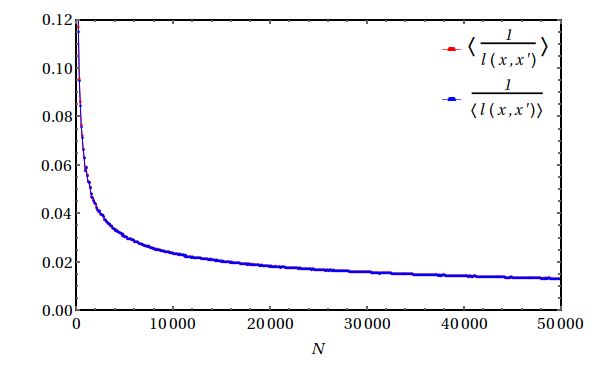}  
\label{plot_3}
\esfig
\caption{Fig (b) gives the  percentage error estimation with respect to these
	bounds. Fig (c) shows $\protect\av{\frac{1}{l(x,x')}}$ and
	$\protect\frac{1}{\av{l(x,x')}}$ vs $N$}
\efig

\bfig
\ContinuedFloat
\bsfig[b]{0.5\textwidth}
\caption{Error in $\protect\frac{1}{\av{l(x,x')}}$ with respect to $\protect\av{\frac{1}{l(x,x')}}$}
\includegraphics[width=\textwidth]{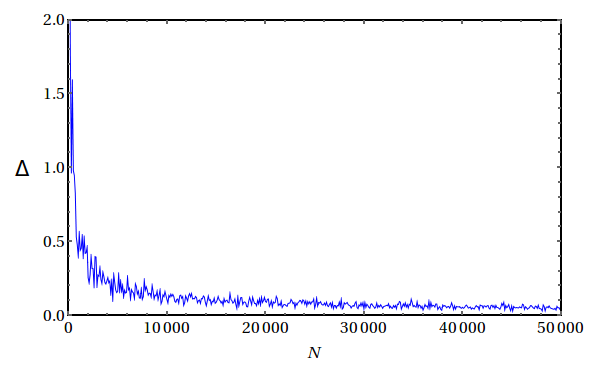} 
\label{plot_4} 
\esfig 
\bsfig[b]{0.5\textwidth}
\caption{Error in $\protect\av{\frac{1}{l(x,x')}}$ with respect to the Best Fit}
\includegraphics[width=\textwidth]{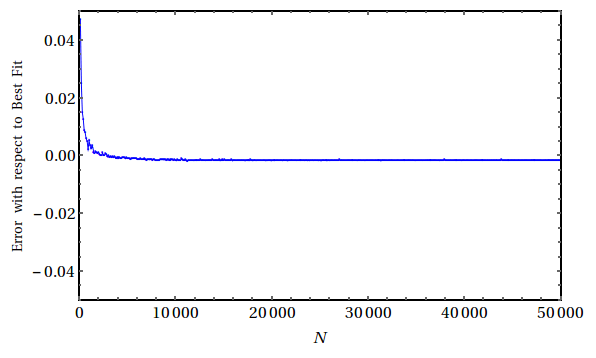} 
\label{plot_5} 
\esfig
\caption{Fig (d) shows the
	percentage error between $\protect\frac{1}{\av{l(x,x')}}$ and $\protect\av{\frac{1}{l(x,x')}}$. This rapidly goes to zero as $N$
	increases. Fig (e) shows the difference between $\protect\av{\frac{1}{l(x,x')}}$ and the best fit with $\protect m_3=1.854$, this too goes to zero rapidly.}
\efig

In order to extend this to the massive case, we need to ask what the analogue of the convolution in (\ref{numberchains}) is. Because of the non-trivial weight $\frac{1}{l(x,x')}$, we cannot simply count chains to get $C_k$. The convolution 
\be 
\av{\bHH_0}*\av{\bHH_0}=\rho \int d^3x_1 \av{\bHH_0(x,x_1)}{\av{\bHH_0(x_1,x')}}=\av{\bHH_1(x,x')}
\ee 
where 
\be 
H_1(x,x')=\sum_{x_1}H_0(x,x_1)H_0(x_1,x')=\sum_{x_1}\dfrac{1}{l(x,x_1)}C_0(x,x_1)\dfrac{1}{l(x_1,x')}C_0(x_1,x'). 
\ee 
counts instead the number of $1$-chains weighted by the inverse of the length of the longest possible chain in $\mathcal{C}$ between each successive pair of joints in the given chain. As in $d=2$ the trajectories are chains, but the $H_k(x,x')$ are not obtained by merely counting chains; each $k$-chain is weighted by the inverse of the length of the longest possible chain $C$ between each pair of joints in the given chain. 

Again defining 
\be
\label{3dgeneralab}
{\cal{K}}^{(3)}(a,b)(x,x') \equiv \sum\limits_{k=0}^\infty  a^{k+1} b^k H_k(x,x')\,,
\ee
we propose that this is the appropriate  causal set Green function for  the massive field for  $a=(\frac{\rho \pi}{12})^{1/3} \frac{m_3}{2\pi}$
and $b = -\frac{m^2}{\rho}$. 

In \cite{Johnston:2010su} a proposal for the d=3 Green function was made using the relationship between $\tau(x,x')$ and the volume $V(x,x')$, $\tau(x,x')\propto V(x,x')^{\frac{1}{3}}$. Our proposal uses instead the causal set analogue of $\tau(x,x')$ directly. In the large $\rho$ limit, one would expect both proposals to give the same result. 
 
\section{Discussion} 
\label{discussionch2} 

We showed that Johnston's hop-stop model for a Green function on a causal set can be generalized to an RNN in arbitrary curved spacetimes in 2 and 4 dimensions. In the $4d$ case we also constructed the massless causal set Green functions for global de Sitter spacetime and for a conformally flat patch of anti de Sitter spacetime and shown that they have the right continuum limits. The corresponding massive cases can, in principle, be obtained by evaluating a power series expansion. Finally, we proposed a potential Green function for the $3d$ Minkowski case and give numerical evidence in support of it.  

The following are a few ideas that need to be explored - 
\begin{itemize}
\item The $\mink^3$ case can be extended to an RNN if we know the analytic form for the LLC as we do for the chains. Alternatively, we could use Johnston's proposal for the Green function \cite{Johnston:2010su}, which only uses $C_1$ in the calculation.  
\item The causal set Green functions in $d>4$ can also be obtained. Massless Green functions in the continuum are derivatives of either $\frac{1}{\tau}$ or $\delta(\tau^2)$ depending on whether $d$ is odd or even. Since derivatives of $\delta(\tau^2)$ of any order can always be written as products of $\delta(\tau^2)$ and $\frac{1}{\tau}$, the knowledge of the causal set analogues of these two quantities with appropriate weights should suffice to write down the causal set Green function. 
\item We have ignored the causal set induced corrections to the retarded Green functions. Given that causal set theory posits a fundamental discreteness, the $\rho \rightarrow \infty$ limit is only a mathematical convenience. Indeed it is the large $\rho$ corrections to the continuum Green function which are phenomenologically interesting. This has been explored in \cite{Johnston2015} for cases of $d=2$ and $4$ Minkowski spacetime. Also, while our analysis has focused on the expectation value of the causal set Green function, we have not analyzed the fluctuations.
\item The retarded Green function is the starting point for quantum field theory in causal sets in its current form. It will be important to identify correct Green functions for astrophysically important spacetimes like FLRW and black hole spacetimes.
\item In this work we have used the continuum limit to identify which object in the causal set will behave as the correct Green function. As mentioned in the $\mink^3$ case, this object may not be unique. If we want to think of the causal set as fundamental, we must have a way to do this identification without reference to the continuum as well as a way to reconcile multiple Green function candidates. Recent work based on the idea of a \textit{preferred past} shows the construction of discretized wave operators \cite{Dable-Heath:2019sej}. This may help resolve the ambiguity.   
\end{itemize}

We now have the appropriate causal set Green functions for an RNN in general curved spacetimes as well as in dS and conformally flat patches of adS spacetimes. This makes it possible to construct the Sorkin-Johnston vacuum. As mentioned in the introduction, this is the next step in the construction of QFT on a causal set and has potential consequences for phenomenology in the early universe. We will see that unlike the present chapter, an analytic calculation will not be possible and we will rely largely on numerical studies.  

\chapter{The SJ Vacuum in deSitter Causal Sets}

As it is usually defined, the vacuum for QFT on a generic curved spacetime relies on a choice of observer or equivalently a choice of mode functions, and is hence non-unique. In  free scalar quantum field theory (FSQFT), the Sorkin-Johnston or SJ vacuum \cite{Johnston:2009fr, Sorkin:2011pn} is a proposal for an observer independent vacuum which {\it is} unique.  The  idea is to begin with the covariantly defined spacetime commutator or Peierls bracket 
\begin{equation}
    [\hP(x),\hP(x') ]=i\Delta(x,x'), 
\end{equation}
where the Pauli-Jordan (PJ) function $i\Delta(x,x')\equiv i \, ( \, G_R(x,x')-G_A(x,x')\,)$ and   $G_{R,A}(x,x')$  are the retarded and advanced Green functions. The  PJ function can be viewed as the integral kernel of a self-adjoint  operator $i\widehat\Delta$ on a bounded region $\mathcal{V}$ of spacetime.  Its non-zero eigenvalues thus come in positive and negative pairs, providing  a natural and covariantly defined mode decomposition into  ``SJ modes".  The positive part of the spectral  decomposition of $i\widehat\Delta$ is then defined to be the SJ Wightman or two-point function $\wsj(x,x')$. 

It is therefore of interest to ask what new role, if any, the SJ vacuum plays in FSQFT in 
cosmologically interesting spacetimes such as  de Sitter. Using a particular limiting procedure, it was argued in \cite{Aslanbeigi:2013fga} that the SJ vacuum for global de Sitter spacetime can be  identified with  one of the known Mottola-Allen $\alpha$-vacua \cite{Mottola, Allen:1985ux} for each value of $m^2 =\phm^2+\xi R >0$\footnote{Here $\phm$ is the physical mass. For a discussion on the meaning of mass in dS spacetime see \cite{Garidi2003}.} for spacetime dimensions $d\geq 2$, {\it except} for the conformally coupled massless case $m^2=m_c^2=\frac{(d - 2)}{4(d-1)} R\equiv\xi_cR$, where the SJ vacuum was argued to be ill-defined. Since there is no known de Sitter invariant Fock vacuum for the minimally coupled massless case $m=0$ \cite{Allen:1985ux}, they also suggest that the $m=0$ SJ vacuum is ill-defined. While  general infrared considerations might be consistent with the absence of an $\tm=0$ SJ vacuum, the situation for $\tm=\tm_c$ is puzzling. 

An important subtlety in the construction of the SJ vacuum is the use of a bounded region $\mathcal{V}$ of spacetime in defining $i\widehat{\Delta}$.  This  operator is Hermitian on the space of $\mathcal{L}^2$ spacetime functions, where 
\begin{equation}
    \braket{f,g}=\int_{\mathcal{V}}dV f^*(x)g(x)
\end{equation}
defines the $\mathcal{L}^2$ inner product and $\mathcal{V}$ is a finite volume region of the full spacetime $(\mathcal{M},g)$. Thus  the SJ vacuum of $(\mathcal{M},g)$ can be obtained only in the limit $\mathcal{V}\rightarrow \mathcal{M}$.  A pertinent  question is whether the SJ  construction is sensitive to exactly {\it when} this limit is taken.  

In the literature there have been two approaches to constructing the  SJ
vacuum  arising from the choice of when to take this ``IR limit". The first and more fundamental approach is what we dub the   ``ab initio'' calculation  where the eigenfunctions and eigenvalues of $i\hD$ are obtained in the  bounded region  $\mathcal{V}$. The  SJ vacuum $\wsj(x,x')$ is obtained as the positive part of $i\hD$.   If $\wsj(x,x')$ remains well-behaved when $\mathcal{V}\rightarrow \mathcal{M}$  then this gives the SJ vacuum in $(\mathcal{M},g)$.  This is the approach followed by \cite{Afshordi:2012ez} for the massless FSQFT  in the 2d causal diamond in Minkowski spacetime. The SJ two-point function was moreover  shown to be Minkowski-like near the center of the causal diamond, with the expected 2d logarithmic behaviour. The ab initio calculation is however computationally challenging since it is non-trivial to calculate the spectral (or eigen) decomposition of $i\hD$ explicitly. Indeed, the spectral decomposition  of $i\hD$ is known in very few examples other than the 2d causal diamond \cite{Fewster2012,Brum:2013bia,Buck:2016ehk, Mathur2019}. 

The second, more computationally accessible approach, which we dub the ``mode comparison" calculation,  was adopted extensively in \cite{Aslanbeigi:2013fga,Afshordi:2012jf}. The idea is to start with a  set  of  Klein Gordon (KG) modes $\{ \uqkg\}$ in  the full spacetime and  restrict them to  $\mathcal{V}$. The  SJ modes $\{ \uksj\}$ in $\mathcal{V}$ are obtained from $\{ \uqkg\}$  via a Bogoliubov transformation. 
The SJ modes are then assumed to  extend to the full spacetime only if the  coefficients of
this  transformation are well behaved in the IR limit.  
Furthermore, when the $\{ \uksj\}$ can themselves be identified with a known set of KG modes, the SJ vacuum is identified with the corresponding known KG vacuum in the full spacetime, rather than via an explicit calculation.  

In these two calculations, the IR limit is taken differently. In the former, it is taken after the finite SJ vacuum is constructed from the eigen decomposition in $\mathcal{V}$, while in the latter,  the limit is taken after the mode comparison in the full spacetime restricted to $\mathcal{V}$. In the 2d causal diamond both calculations give the same result away from the boundaries \cite{Afshordi:2012ez,Afshordi:2012jf}. However,  this is in general not guaranteed and needs to be checked case by case. The subtlety of when to take the limit was brought out in  \cite{Fewster2012} for the case of ultrastatic spacetimes. There, the finite $\mathcal{V}$ SJ vacuum was shown not to be equivalent to that constructed from a Hadamard state, and in some cases, to be in an inequivalent representation altogether. However, in taking the IR limit, both  yield  the same Hadamard vacuum.   
It is the aim here is to re-examine the de Sitter SJ vacuum from the perspective that the nature of  the SJ vacuum is sensitive to the manner in which the IR limit enters its construction. This study is significant for the definition of the SJ vacuum, since it is only if the ab initio calculation fails to survive the IR limit that we can  definitively say that there is no SJ vacuum.   

We begin with the two known $m=0$ vacua in de Sitter\footnote{There is also a de Sitter invariant and shift invariant vacuum defined in \cite{Page2012}. In this paper, we do not impose shift invariance.}: the $O(4)$ invariant Fock vacuum of \cite{Allen:1987tz} and the de Sitter invariant non-Fock vacuum of \cite{Kirsten:1993ug}. In the spirit of the mode comparison calculation, we show that the SJ modes cannot be obtained via a Bogoliubov transformation from the modes that define these two vacua. The calculation is done in a symmetric $[-T,T]$ slab of global de Sitter spacetime and the coefficients of the transformation are seen to diverge as $T\rightarrow \pi/2$ (the infinite volume limit). At present we do not have an analytic ab initio calculation of the SJ modes in de Sitter spacetime. Instead we use a causal set discretisation of a slab of de Sitter spacetime and obtain the causal set SJ vacuum via the ab initio calculation. In the massive theory in 2d, our results are in keeping with the findings of \cite{Aslanbeigi:2013fga} and agree very well with the continuum Mottola-Allen $\alpha$-vacua. On the other hand, while the $m=0$ SJ vacuum is well-defined, it appears to violate  de Sitter invariance. In the massive theory in 4d, our results show a substantial difference with the continuum expressions of \cite{Aslanbeigi:2013fga} and suggest that the causal set SJ vacuum, while de Sitter invariant, differs from the Mottola-Allen $\alpha$-vacua. For $m=0$ and $m_c$, interestingly, the SJ vacuum is well-behaved, and also does not violate de Sitter invariance. In particular, at and around $m=m_c$, the SJ vacuum behaves as a continuous function of $m$, suggesting no singular behaviour. While our numerical calculations are of course for a finite volume, by varying the IR cutoff we find  a convergence of the SJ vacuum,  which  supports our conclusions. 

In Section \ref{sjvac} we review the SJ construction, emphasising the role of the IR cutoff. In Section \ref{sec:dS} we show that the $m=0$ SJ modes in a slab of  de Sitter spacetime can neither be obtained 
from the $O(4)$-invariant Fock vacuum of \cite{Allen:1987tz} nor from the de Sitter invariant non-Fock vacuum of \cite{Kirsten:1993ug} via a Bogoliubov transformation.   
In Section \ref{sec:numerics} we present our results from numerical simulations using a causal set discretisation of a slab of de Sitter spacetime. Our analysis begins with the massless FSQFT in 2d and 4d causal diamonds in Minkowski spacetime. We show that the SJ vacuum looks like the Minkowski vacuum in a smaller causal diamond within the larger one, both in 2d and 4d. The former is consistent with the calculations of \cite{Afshordi:2012ez}. Next we calculate the SJ vacuum in slabs of 2d and 4d global de Sitter spacetime in the time interval $[-T,T]$ for different values of $m$. We vary $T$ as well as the density $\rho$ to look for convergence. We compare our results with the Mottola-Allen $\alpha$-vacua and show that while they agree well with the SJ vacuum (for $m>0$) in 2d, they differ significantly in 4d. We also examine the eigenvalues of the PJ operator in 2d and 4d de Sitter as a function of $m$ and find no significant changes around  $m=0$ and $m=m_c$. In section \ref{eom} we discuss the possibility of equations of motion on the causal set arising as a byproduct of the SJ construction.
In Section \ref{sec:discussion} we discuss the implications of our results.

Here we have used causal sets as a covariant discretisation of the continuum. In CST however, this discrete substratum is considered more fundamental than the continuum. From the CST perspective therefore the SJ de Sitter vacuum that we have obtained {is} physically more relevant to QFT in the early universe than any continuum vacuum. Our result that the causal set SJ vacuum differs significantly from the continuum vacua therefore suggests exciting new possibilities for CST phenomenology. 

\section{The  SJ vacuum} 
\label{sjvac}
We begin with a short introduction to the SJ vacuum construction for
FSQFT in a general globally hyperbolic, finite volume $\mathcal{V}$ region of spacetime $(\mathcal{M},g)$
\cite{Sorkin:2011pn, Aslanbeigi:2013fga, Afshordi:2012ez, Afshordi:2012jf, Sorkin:2017fcp}. 

The Klein Gordon (KG) equation in $(\mathcal{M},g)$ is  
\begin{equation} 
\biggl(\hB- m^2\biggr) \phi=0,
\end{equation} 
where $\hB \equiv g^{ab}\nabla_a \nabla_b$, and the effective mass $m^2=\phm^2 +\xi R$, where $\phm$ is the physical
mass, $R$ is the scalar curvature of $(\mathcal{M},g)$ and $\xi$ is the coupling.   
Let $\{\uqkg\}$ be a complete set of modes satisfying the KG equation in $(\mathcal{M},g)$ and orthonormal with respect to the KG symplectic form (or KG ``norm") 
\begin{equation}
\kgnorm{f,g}=\int_\Sigma (f^*\nabla_a g-g^*\nabla_a f) dS^a, 
\label{kgnorm}  
\end{equation}  
where $\Sigma$ is a Cauchy hypersurface in $(\mathcal{M},g)$. The field operator can be expressed as a mode expansion with respect to  the set $\{ \uqkg\}$    
\begin{equation}
\hP(x) \equiv \sum_{\mathbf{q}}\haq \uqkg(x) +\haqd\uqkg^*(x),
\label{field} 
\end{equation}
with $\haq,\,\haqd$
satisfying the  commutation relations
\begin{equation}
  [\haq,\haqpd]=\delta_{\mathbf{q}\mathbf{q}'},
  \quad [\haq,\haqp]=0\text{,
}\quad [\haqd,\haqpd]=0. \end{equation} 
The covariant commutation relations for the scalar field operator are
given by the Peierls bracket
\begin{equation} 
[\hP(x),\hP(x')] = i\Delta(x,x'), 
\end{equation} 
where the PJ
function is 
\begin{equation} 
i \Delta(x,x')  \equiv i( G_R(x,x')-G_A(x,x') ),
\end{equation} 
with $G_{R,A}(x,x')$ being the retarded and advanced Green
functions, respectively. In terms of the modes $\{\uqkg\}$  
\begin{equation} 
i \Delta(x,x')=\sum_{\mathbf{q}}\uqkg(x)\uqkg^*(x')-\uqkg^*(x)\uqkg(x'), 
\label{modeexp} 
\end{equation} 
and the two-point function associated with them is
\begin{equation}
    W(x,x') \equiv \sum_{\mathbf{q}}\uqkg(x)\uqkg^*(x'). 
\end{equation}
On the other hand the SJ state or equivalently the SJ two-point function $\wsj(x,x')$ for FSQFT, which as we will see below is constructed from the positive eigenspace of $i \hD$, is defined most generally by the following three conditions \cite{Sorkin:2017fcp} 
\begin{eqnarray} 
&&i\Delta(x,x') = \wsj(x,x') - \wsj(x',x),  \nonumber\\
&&\int_{\mathcal{V}} dV' \int_{\mathcal{V}} dV f^*(x') \wsj(x',x) f(x)  \geq 0,\quad(\text{Positive Semidefinite}) \nonumber \\ 
&&\int_\mathcal{V} dV' \wsj(x,x') \wsj^*(x',x'') = 0,\quad(\text{Ground state or Purity}) 
\label{eq:sjdefn}
\end{eqnarray} 
where the integrals are defined over a finite spacetime volume region $\mathcal{V}$ in the full spacetime $(\mathcal{M},g)$. 
In order to construct the SJ vacuum explicitly, the PJ
function is elevated to an integral operator in $\mathcal{V}$ 
\begin{equation}
i \hD \circ f \equiv i \int_{\mathcal{V}} \Delta(x,x')  f(x') dV_{x'}
\label{pjdef}
\end{equation} 
which acts on $\cL^2$ functions in $\mathcal{V}$ and where  
\begin{equation}
\braket{f,g}=\int_\mathcal{V} dV_x\,f^*(x)\,g(x) 
\end{equation}  
is the $\cL^2$ inner product.  
Since $\Delta(x,x')$ is  antisymmetric in its arguments, $i\hD$ is Hermitian on the space of $\cL^2$ functions in $\mathcal{V}$. Its non-zero eigenvalues, given by  
\begin{equation}
i \hD \circ \uksjn(x) =\int_{\mathcal{V}} dV_{x'}\,i\Delta(x,x')\uksjn(x')=\lk \uksjn(x)
\label{eq:eigen}
\end{equation} 
therefore come in pairs $(\lk, -\lk)$, corresponding to the
eigenfunctions $(\vkpjp, \vkpjm)$ where $\vkpjm=(\vkpjp)^*$.\footnote{We adopt the notation that the $\tilde{s}_k$ are the un-normalised (with respect to the $\mathcal{L}^2$ norm) SJ eigenfunctions,  whereas the $s_k$ without the tilde are the normalised SJ eigenfunctions.} This is
the central eigenvalue problem in the ab initio calculation of the SJ vacuum.  

It was shown in \cite{Sorkin:2017fcp} that 
\begin{equation} 
\kr (\hB -\phm^2) = \overline{\im (\hD)}, 
\label{eq:kerim} 
\end{equation} 
where the operators are defined in $\mathcal{V}$\footnote{In a spacetime of constant scalar curvature, $m$ defined above is constant, and hence this result continues to hold when $\phm$ is replaced by $m$.}. This means that the 
eigenvectors in the image of $i \hD$ (i.e., excluding those in $\kr (i\hD)$) span the full solution space of the
KG operator. One therefore has an {\sl intrinsic} and
coordinate independent separation of the space of solutions into the 
positive and negative eigenmodes of $i\hD$.\footnote{This is not unlike the polarisation in geometric quantisation.} 
The field operator thus has a 
{\it coordinate invariant} or observer independent decomposition 
\begin{equation}
\hP(x)=\sum_{\mathbf{k}}\hbk
\uksj(x)+\hbkd \uksj^*(x)
\label{field} ,
\end{equation}
where the SJ vacuum state is defined as 
\begin{equation} 
\hbk\ket{0_{SJ}}=0 \quad \forall\,\mathbf{k},
\end{equation} and  
\begin{equation} 
\uksj = \sqrt{\lk} \vkpjp   
\end{equation}  
are the normalised {\sl SJ modes} which form an orthonormal set in $\overline{\im(i\hD)}$ with respect to the
$\cL^2$ norm
\begin{eqnarray}
\braket{\uksj,\ukpsj}&=&{\lk}\delta_{\mathbf{k}\mathbf{k}'} \nonumber \\ 
\braket{\uksj^*,\ukpsj}&=&0.
\end{eqnarray}
Using the spectral decomposition 
\begin{equation} 
i \Delta(x,x') = \sum_{\mathbf k}  \uksj(x) \uksj^*(x') -
\uksj^*(x) \uksj(x') , 
\label{sjmodeexp}
\end{equation}
the SJ two-point function in $\mathcal{V}$ is the positive part of $i\hD$ 
\begin{equation} 
\wsj(x,x') \equiv  \sum_{\mathbf k} \uksj(x) \uksj^*(x'). 
\label{sjvacuum} 
\end{equation} 
If $\wsj(x,x')$ remains well-defined as the IR cutoff is taken to infinity, this defines the SJ vacuum in the full spacetime $(\cM,g)$. The SJ construction from the eigenvalue problem \eqref{eq:eigen} through to \eqref{sjvacuum} is the ab intio calculation referred to in the introduction. 

Alternatively, one can also obtain the SJ modes via a mode comparison calculation. Given the equality in \eqref{eq:kerim} between $\overline{\im(\hD)}$ and the
KG solution space, there must  exist a transformation between the KG modes $\{ \uqkg\}$ in $\mathcal{V}$ and
the SJ modes $\{\uksj\}$, even though the former need not be orthonormal with respect to the $\cL^2$ inner product.  Let 
\begin{equation}
\uksj(x)=\sum_{\mathbf{q}}\uqkg(x)\Aqk+\uqkg^*(x)\Bqk,
\label{uksjuqkg}
\end{equation}
where $\Aqk=\kgnorm{\uqkg,\uksj}$ and $\Bqk =\kgnorm{\uqkg^*,\uksj}$. Further, if we act with $i\Delta$ on \eqref{uksjuqkg} and use \eqref{modeexp}, we can also write $\Aqk= \dfrac{1}{\lambda_{\mathbf{k}}}\av{\uqkg,\uksj}$ and $\Bqk =-\dfrac{1}{\lambda_{\mathbf{k}}}\av{\uqkg^*,\uksj}$.
Using the fact that \eqref{sjmodeexp} and \eqref{modeexp} must be equal, we get the algebraic relations 
\begin{eqnarray}
\sum_{\mathbf{q}}\mathbf{A_{qk'}}\Aqk^*-\mathbf{B_{qk'}}\Bqk^*&=&\delta_{\mathbf{k}\mathbf{k}'}\nonumber\\
\sum_{\mathbf{q}}\mathbf{B_{qk'}} \Aqk -\mathbf{A_{qk'}} \Bqk &=&0.
\label{ABrelations} 
\end{eqnarray}
Additionally, if the KG modes themselves satisfy the $\cL^2$ orthonormality condition 
\begin{equation}
    \langle \uqkg,\uqpkg\rangle=\delta_{qq'},\indent \langle \uqkg^*,\uqpkg \rangle=0,  
\end{equation}
then the above equations simplify considerably as shown in \cite{Afshordi:2012jf}.\footnote{In assuming a  discrete index $\mathbf q$ we are already working in a bounded region of spacetime.} 
It is important to note that since the $\cL^2$ norm is defined for finite $\mathcal{V}$, the above calculations are limited to finite $\mathcal{V}$. Moreover, there are potential subtleties in identifying $\kr (\hB -m^2)$ in $\mathcal{V}$, starting from the solutions in the full spacetime.  

The question of course is whether the limits involved in the first and second approaches (that is, whether finding the SJ modes before or after taking the infrared limit) commute. 
A case in point is the 2d causal diamond in Minkowski spacetime where the SJ modes for the
massless scalar field are not simply linear combinations of plane
waves, but also include an important $\mathbf k$ dependent constant
\cite{Afshordi:2012ez, Johnston:2010su}, which {\it is} a solution for finite
$\mathcal{V}$. The two sets of eigenfunctions of $i\Delta$ are  
\begin{eqnarray} 
f_k(u, v) &= & e^{iku} - e^{ikv} \label{fmode}\\ 
g_k(u, v) &=& e^{iku} + e^{ikv} - 2\cos kL\label{gmode}, 
\end{eqnarray} 
where $u$ and $v$ are lightcone coordinates, and $2 L$ is the side length of the diamond. The eigenvalues are $\lk=L/k$ for both sets. For the $f$-modes,
$k$ is $k=n\pi/L$ with $n=\pm 1, \pm 2,...$ while for the $g$-modes $k$ satisfies the condition $\tan(kL)=2 kL$.       
In order to make contact with the IR limit, $W(x,x')$ was studied in a small region in the interior of the larger diamond, which to leading order was found to have the form of the (IR-regulated) 2d  Minkowski vacuum \cite{Afshordi:2012ez}. A similar conclusion was reached in \cite{Afshordi:2012jf} using the Bogoliubov prescription, and hence in this simple example, the results seem to be independent of the limiting procedure.  

\section{The massless de Sitter SJ vacuum } 
\label{sec:dS} 
In \cite{Aslanbeigi:2013fga} the mode comparison calculation was used to find the SJ modes in de Sitter spacetime. A restriction of the Euclidean modes \cite{Chernikov_1968} (which themselves are one of the $\alpha$-modes) in global de Sitter to a finite slab $\mathcal{V}$ was used as the starting point. Assuming that these modes are complete in  $\kr (\hB -m^2)$ when restricted to $\mathcal{V}$, they  solve \eqref{ABrelations} to get the SJ modes $\{ \uksj\}$, \eqref{uksjuqkg}. These can in turn be  identified with one of the
other (restricted to $\mathcal{V}$) $\alpha$-modes depending on the value of  $m$,  and thence the SJ vacuum is identified with the corresponding $ \alpha$-vacuum in the IR
limit for each $m$. Surprisingly, however, this identification fails in the conformally coupled massless case, $m_c=\frac{(d-2)}{4(d-1)} R$, since the Bogoliubov transformation breaks down. For this
and the minimally coupled massless case, $m=0$ (for which there is no $ \alpha$-vacuum), it is suggested that the SJ prescription itself
breaks down and that there is no de Sitter SJ vacuum.  
In both these cases however,  the
SJ modes must be well-defined when there is a finite $T$ IR cutoff. Strictly, it is only if an ab initio calculation
 of the SJ two-point functions fails to survive the IR limit that we can state that there is no SJ vacuum. 
 
The KG modes for the massive scalar field in global de Sitter are the Mottola-Allen $\alpha$-modes
which include the Euclidean modes as a special case. The mimimally coupled massless scalar field is
known not to admit a de Sitter invariant Fock vacuum (Allen's theorem)
\cite{Allen:1985ux}. Starting with a Fock vacuum defined with respect to an orthonormal basis $\{\phi_n(x)\}$ of the solution space of the KG equation, Allen shows that for the $m=0$ case the symmetric two-point function defined by
\begin{equation}
    G^{(1)}_\lambda(x,x')=\langle\lambda|\Phi(x)\Phi(x')|\lambda\rangle=\sum_n\,\phi_n(x)\phi^*_n(x')+\phi^*_n(x)\phi_n(x')
    \label{g1sum}
\end{equation}
    must satisfy 
    \begin{equation}
        G^{(1)}(x,x')+G^{(1)}(x,\Bar{x}')\neq C \quad\text{everywhere}
        \label{allencondition}
    \end{equation}
for some $C\in \mathbb{R}$, where $\Bar{x}'$ represents the antipodal point of $x'$. In \cite{Allen:1985ux} the de Sitter invariant $G^{(1)}(x,x')$ fails to satisfy the required condition \eqref{allencondition}, leading to the conclusion that the assumption that it is a Fock vacuum is false. Importantly the proof of condition \eqref{allencondition} relies on the use of the KG inner product.\footnote{The use of the $\cL^2$ inner product for the SJ modes is not a violation of Allen's theorem because the SJ modes are also KG orthogonal.}
     
     It is also worth mentioning at this point that because the $\cL^2$ inner product is only defined in a finite region of spacetime\footnote{Allen's theorem continues to hold in a finite region of spacetime as long as we choose this region to be symmetric about $\tau=0$, where $\tau$ is the time in hyperbolic coordinates \eqref{globmetric}.}, the entire prescription inherently breaks de Sitter invariance. In the case of global de Sitter with an IR cutoff at $[-T, T]$, this is certainly the case. Since the spatial part is compact we manage to preserve $O(4)$ invariance. However, the idea is, as in \cite{Aslanbeigi:2013fga}, to take the temporal cutoffs to infinity\footnote{In the causal set case we cannot take these temporal cutoffs to infinity, but we try to reach an asymptotic regime.} and make statements that have full de Sitter invariance. 
     
     On the other hand, as in the 2d diamond,
one might imagine that away from the boundaries, there is an
approximate isometry that is retained. However, 
even if the PJ operator is itself approximately invariant,  this does
not imply that the two-point
function is, since the latter is simply the positive part of the PJ operator. It is only if the isometries preserve the positive and
negative eigenspaces separately that this can be the
case. 

Let us address this question by asking if the known de Sitter violating
vacuum, the so-called $O(4)$ vacuum \cite{Allen:1987tz} is related to the SJ vacuum via a Bogoliubov transformation as in \cite{Aslanbeigi:2013fga}.  We work in the conformal coordinates \eqref{confmetric2}
\begin{equation}
ds^2=\frac{1}{H^2\sin^2\eta}[-d\eta^2+d\Omega^2(\chi,\theta,\phi)\,]\label{confmetric1},
\end{equation}
where we have shifted $\tilde{T}\rightarrow\eta=\tilde{T}+\pi/2$ so that $\eta\in[0,\pi]$ and $(\chi,\theta,\phi)$ are coordinates on $S^3$.
The  $O(4)$ modes are 
\begin{equation} 
u_{klm}(x)=HX_k(\eta)Y_{klm}(\chi,\theta,\phi) ,
\end{equation} 
where $k=0,1,...;\,l=0,1...k;\,m=-l,-l+1,...l-1,l$. For $k=0$,  
\begin{equation} 
X_0(\eta)=A_0\bigg(\eta-\frac{1}{2}\sin2\eta-\frac{\pi}{2}\bigg)+B_0, 
\end{equation} 
and for $k\neq 0$ 
\begin{equation} 
X_k(\eta)=\sin^{3/2}(\eta)(A_kP^{3/2}_{k+1/2}(-\cos\eta)+B_kQ^{3/2}_{k+1/2}(-\cos\eta)),   
\label{xk}
\end{equation} 
where $P^\mu_\nu(x),\,Q^\mu_\nu(x)$ are independent, associated Legendre
functions defined for real $x\in[-1,1]$ as in \cite{Gradshteyn}:  
\begin{eqnarray}
P^\mu_\nu(x)=\bigg(\frac{1+x}{1-x}\bigg)^{\mu/2}\frac{_2F_1(-\nu,\nu+1,1-\mu;(1-x)/2)}{\Gamma(1-\mu)},\\
Q^\mu_\nu(x)=\frac{\pi}{2\sin\mu\pi}\bigg(P^\mu_\nu(x)\cos\mu\pi-\frac{\Gamma(\nu+\mu+1)}{\Gamma(\nu-\mu+1)}P^{-\mu}_\nu(x)\bigg).
\end{eqnarray}    
Note that the $k\neq0$ modes are the same as the  Euclidean modes. The $Y_{klm}$ are spherical harmonics that satisfy 
\begin{equation}
\int d\Omega(\chi,\theta,\phi)Y_{klm}Y^*_{k'l'm'}=\delta_{kk'}\delta_{ll'}\delta_{mm'}.
\end{equation}
The coefficients for $k=0$ are $A_0=-i\alpha,\,B_0=(1/4+i\beta)/\alpha$, where
$\alpha,\beta\in\mathbb{R}$. The coefficients for $ k\neq 0$ are
\begin{equation} 
 A_k=\bigg(\dfrac{-1+i}{\sqrt{2}}\bigg)\sqrt{\dfrac{\pi}{4k(k+1)(k+2)}},
 \quad B_k=\dfrac{-2i}{\pi}A_k.
 \end{equation}

These $O(4)$ modes are orthonormal with respect to the KG inner product but as mentioned in the last section, the Bogoliubov coefficients are defined by their $\cL^2$ inner products so we must evaluate these. We also need a choice of the finite spacetime region $\mathcal{V}$ for the $\cL^2$ inner product, we consider a slab of dS spacetime such that $\eta\in(a,b)$, the infinite volume limit corresponds to $a\rightarrow 0,\,b\rightarrow\pi$. We have  
\begin{eqnarray}
\braket{u_{klm},u_{k'l'm'}}&=&H^2\int dV_x\,X^*_k(\eta)X_{k'}(\eta)Y^*_{klm}Y_{k'l'm'}\nonumber\\
&=&\frac{1}{H^2}\delta_{kk'}\delta_{ll'}\delta_{mm'}\int_a^b\frac{d\eta}{\sin^4\eta}X^*_k(\eta)X_{k}(\eta)\nonumber\\
&=&\delta_{kk'}\delta_{ll'}\delta_{mm'}T_k,
\label{tko4}
\end{eqnarray}
\begin{eqnarray}
\braket{u^*_{klm},u_{k'l'm'}}&=&\frac{(-1)^k}{H^2}\delta_{kk'}\delta_{ll'}\delta_{mm'}\int_a^b\frac{d\eta}{\sin^4\eta}(X_k(\eta))^2\nonumber\\
&=&\delta_{kk'}\delta_{ll'}\delta_{mm'}D_k.
\label{dko4}
\end{eqnarray}
The factor $(-1)^k$ in the second expression  is due to the choice of
spherical harmonics with the special property
$Y^*_{klm}=(-1)^kY_{klm}$ \cite{Aslanbeigi:2013fga}. These equations define $T_k$ and $D_k$ ($T_k$ is real by definition). Also note that $T_k$ and $D_k$ will necessarily blow up in the infinite volume limit.   

The Bogoliubov coefficients to obtain the SJ modes \eqref{uksjuqkg} from these $O(4)$ modes simplify to 
\begin{eqnarray}
A_{qk}&=&\dfrac{1}{\lambda_{k}}\sum_{n}\left(\delta_{qn}T_qA_{nk}+\delta_{qn}D_q^*B_{nk}\right)
          = \dfrac{1}{\lambda_{k}}(T_qA_{qk}+D_q^*B_{qk})\nonumber \\
B_{qk}&=&-\dfrac{1}{\lambda_{k}}\sum_{n}\left(\delta_{qn}D_qA_{nk}+\delta_{qn}T_qB_{nk}\right)=-\dfrac{1}{\lambda_{k}}(D_qA_{qk}+T_qB_{qk}),
\label{redAB}
\end{eqnarray}  
where the index $q$  implicitly contains the $l$ and $m$ indices and $\delta_{ll'}, \delta_{mm'}$ are omitted from the expressions. Inserting these expressions into \eqref{ABrelations} we find that 
\begin{eqnarray}
\sum_q\{(T_q^2-|D_q|^2)(A_{qk'}A_{qk}^*-B_{qk'}B_{qk}^*)\}&=&\lambda_k^2\delta_{kk'}
\nonumber  \\ 
\sum_q\{(T_q^2-|D_q|^2)(A_{qk}B_{qk'}-A_{qk'}B_{qk})\}&=&0. 
\label{simpleAB}
\end{eqnarray}
A convenient parameterisation is 
\begin{equation}
A_{qk}=\delta_{qk}\cosh\alpha_k,\quad\quad
B_{qk}=\delta_{qk}\sinh\alpha_k\,e^{i\beta_k}. 
\end{equation} 
From \eqref{simpleAB} this gives 
\begin{equation}
\lambda_k=\sqrt{T_k^2-|D_k|^2},
\label{lamds}
\end{equation}
which along with  \eqref{redAB} implies that 
\begin{eqnarray}
\lambda_k\cosh\alpha_k&=&T_k\cosh\alpha_k+D_k^*\sinh\alpha_k\,e^{i\beta_k}\nonumber\\
\text{or}\quad\tanh\alpha_k\,e^{i\beta_k}&=&\frac{\lambda_k-T_k}{D^*_k}=\frac{T_k-\lambda_k}{|D_k|}\,e^{i(\arg\,D_k+\pi)}. 
\end{eqnarray}
Defining $r_k\equiv \dfrac{D_k}{T_k}$, we see after some algebra and use of the double angle formula for $\tanh$ that  $\beta_k=\arg
r_k+\pi$ and  $\alpha_k=\frac{1}{2}\tanh^{-1}|r_k|$. Thus the Bogoliubov coefficients depend (via $\alpha_k$ and $\beta_k$) only on $r_k$, which can be finite in the infinite volume limit \textit{even if} $T_k$ and $D_k$ diverge. Note that if $|r_k|=1$, $\alpha_k$ and therefore the Bogoliubov coefficients diverge. When this happens the SJ vacuum cannot be obtained through a Bogoliubov transformation.

From \eqref{tko4} and \eqref{dko4} one can see that the Bogoliubov transformation does not mix different $k$'s. In particular, it does not mix $k\neq0$ modes with the $k=0$ mode. We already know from \cite{Aslanbeigi:2013fga} that the Euclidean modes (which are the same as the $O(4)$ modes for $k\neq0$) do not admit a well-defined Bogoliubov transformation to the SJ modes ($|r_k|=1$ for these modes) in the infinite volume limit. It immediately follows that the transformation from the $O(4)$ modes to the corresponding SJ state is ill-defined, and an SJ state with $O(4)$ symmetry cannot be derived in this way. Next, we calculate these transformations explicitly. We also find the $k=0$ transformation which turns out to be the only well-defined one.
\subsubsection*{Evaluation of $r_0$}
We put in the values of $A_0,B_0$ and substitute $\eta-\pi/2=x$, then
\bea
T_0&=&\frac{2\alpha^2}{H^2}\int_0^{b'}\frac{dx}{\cos^4x}\bigg\{\bigg(x+\frac{\sin2x}{2}\bigg)^2+t\bigg\},\\
D_0&=&\frac{-2\alpha^2}{H^2}\int_0^{b'}\frac{dx}{\cos^4x}\bigg\{\bigg(x+\frac{\sin2x}{2}\bigg)^2+d\bigg\},
\eea
where $t=\dfrac{1}{\alpha^4}\bigg(\dfrac{1}{16}+\beta^2\bigg)$ and $d=\dfrac{-1}{\alpha^4}\bigg(\dfrac{1}{4}+i\beta\bigg)^2$. So we have 
\bea
r_0=\frac{D_0}{T_0}=-\frac{\epsilon+d}{\epsilon+t}\quad\text{where}\quad\epsilon=\frac{\displaystyle\int_0^{b'}\dfrac{dx}{\cos^4x}\bigg(x+\dfrac{\sin2x}{2}\bigg)^2 }{\displaystyle\int_0^{b'}\dfrac{dx}{\cos^4x}}\,.
\eea
These integrals are well-behaved at the lower limit and diverge as $b'\rightarrow\pi/2$, so we can approximate them by their values near the upper limit. We get
$$\lim_{b'\rightarrow\pi/2}\epsilon=\frac{\pi^2}{4},$$
\be
r_0=-\frac{\pi^2+4d}{\pi^2+4t}\,.
\ee
which gives well defined Bogoliubov coefficients.
\subsubsection*{Evaluation of $r_k\,\,(k\neq0)$}
\be
T_k=\frac{1}{H^2}\int_0^\pi\dfrac{d\eta}{\sin^4\eta}\sin^3\eta\,(A_k^*P+B_k^*Q)(A_kP+B_kQ)
\ee
Here we have suppressed the indices and arguments on the Legendre functions $P$ and $Q$. We substitute $-\cos\eta=x\Rightarrow\sin\eta\,d\eta=dx$ and $\dfrac{d\eta}{\sin\eta}=\dfrac{dx}{1-x^2}$. We then get 
$$T_k=\frac{1}{H^2}(|A_k|^2T_k^{(1)}+(A_k^*B_k+B_k^*A_k)T_k^{(2)}+|B_k|^2T_k^{(3)}),$$ 
where
\bea
T_k^{(1)}&=&\int_{-1}^1\frac{dx}{1-x^2}(P^{3/2}_{k+1/2}(x))^2\\
T_k^{(2)}&=&\int_{-1}^1\frac{dx}{1-x^2}P^{3/2}_{k+1/2}(x)\,Q^{3/2}_{k+1/2}(x)\\
T_k^{(3)}&=&\int_{-1}^1\frac{dx}{1-x^2}(Q^{3/2}_{k+1/2}(x))^2.
\eea
Similarly $D_k=\dfrac{(-1)^k}{H^2}(A_k^2D_k^{(1)}+2A_kB_kD_k^{(2)}+B_k^2D_k^{(3)})$ with $D_k^{(i)}=T_k^{(i)}$.\\\\
From the definitions of the associated Legendre functions we have\footnote{We will write $F$ instead of $_2F_1$.}:
\bea
P_{k+1/2}^{3/2}(x)&=&\bigg(\frac{1+x}{1-x}\bigg)^{3/4}\frac{F(-k-1/2,k+3/2,-1/2;(1-x)/2)}{\Gamma(-1/2)}\\
Q_{k+1/2}^{3/2}(x)&=&\frac{\pi}{2}k(k+1)(k+2)\bigg(\frac{1-x}{1+x}\bigg)^{3/4}\frac{F(-k-1/2,k+3/2,5/2;(1-x)/2)}{\Gamma(5/2)}.\nonumber\\
\eea
The above integrals become 
\bea
T_k^{(1)}&=&\frac{1}{(\Gamma(-1/2))^2}\int_{-1}^1dx\,\frac{(1+x)^{1/2}}{(1-x)^{5/2}}\,F^2(-k-1/2,k+3/2,-1/2;(1-x)/2)\nonumber\\
T_k^{(2)}&=&\frac{\pi k(k+1)(k+2)}{2\Gamma(-1/2)\Gamma(5/2)}\int_{-1}^1dx\,F(-k-1/2,k+3/2,-1/2;(1-x)/2)\nonumber\\
&\times&\,F(-k-1/2,k+3/2,5/2;(1-x)/2)\nonumber\\
T_k^{(3)}&=&\frac{(\pi k(k+1)(k+2))^2}{(2\Gamma(5/2))^2}\int_{-1}^1dx\,\frac{(1-x)^{1/2}}{(1+x)^{5/2}}\,F^2(-k-1/2,k+3/2,5/2;(1-x)/2).\nonumber
\eea
All of the above integrals are divergent. However it turns out that the ratios $T_k^{(2)}/T_k^{(1)},\,T_k^{(3)}/T_k^{(1)}\rightarrow0$, therefore we have 
\bea
r_k=(-1)^k\frac{A_k^2}{|A_k|^2}=e^{i(\arg A_k+k\pi)} ,
\eea
whence we find that $|r_k|=1$ which implies that the Bogoliubov coefficients diverge.

In a similar manner, we also find that the modes that define the non-Fock but de Sitter invariant vacuum of Kirsten and Garriga \cite{Kirsten:1993ug} are unable to produce an SJ vacuum via the mode comparison method. The Kirsten and Garriga modes are closely related to the $O(4)$ modes, and in fact are identical to them for $k\neq 0$. For $k=0$, we have
\begin{equation}
    X_0=\frac{H}{\sqrt{2}}\left[Q+\left(\eta-\frac{1}{2}\sin 2\eta-\frac{\pi}{2}\right)P\right].
\end{equation}
We use the same notation as in \cite{Kirsten:1993ug}. The coefficients of $Q$ and $P$ are solutions to the field equation that satisfy the following commutation relations
\begin{equation}
\left[Q,P\right]=i,\indent \left[\hak,Q\right]=\left[\hak,P\right]=0,
\end{equation}
where $\hak$ are the annihilation operators associated to the $k\neq 0$ modes. Now we derive the transformation between the Kirsten and Garriga modes and the SJ modes. Again, we find that the $k=0$ transformation is the only well-defined one.

The PJ function in terms of the Kirsten and Garriga modes is
\be
i\Delta (x,x')=i\frac{H^2}{2} \left(f(x)-f(x')\right)+\sum_q u_q(x)u_q^*(x')-u_q^*(x)u_q(x),
\ee
where $f(x)=\eta_x-\frac{1}{2}\sin 2\eta_x-\frac{\pi}{2}$, and for simplicity $q$ refers to the principle index and we will omit the angular indices. The SJ modes then are
\be
s_k(x)=\frac{1}{\lambda_k}\langle i\Delta(x,x') ,s_k(x')\rangle=\sum_q\left(u_q(x) A_{qk}+u_q^*(x)B_{qk}\right)+i\frac{H^2}{2}C_k+i\frac{H^2}{2} f(x) D_k,
\label{skkg}
\ee
where $u_q$ are the $O(4)$ modes and $A_{qk}=\frac{1}{\lambda_k}\langle u_q,s_k\rangle$, $B_{qk}=-\frac{1}{\lambda_k}\langle u_q^*,s_k\rangle$, $C_k=\frac{1}{\lambda_k}\langle f,s_k\rangle$, and $D_k=-\frac{1}{\lambda_k}\langle 1,s_k\rangle$.
Using \eqref{skkg} we have the inner products
\be
\frac{1}{\lambda_{k'}}\langle s_k,s_{k'}\rangle=\sum_q \left(A^*_{qk}A_{qk'}-B^*_{qk}B_{qk'}\right)+i\frac{H^2}{2}\left(C^*_kD_{k'}-D^*_kC_{k'}\right)=\delta_{kk'}
\label{inkg2}
\ee
\be
\frac{1}{\lambda_{k'}}\langle s_k^*,s_{k'}\rangle=\sum_q \left(A_{qk'}B_{qk}-A_{qk}B_{qk'}\right)+i\frac{H^2}{2}\left(D_kC_{k'}-C_kD_{k'}\right)=0.
\label{inkg3}
\ee
Again using \eqref{skkg} and the definition of the coefficients, we have
\be
A_{qk}=\frac{1}{\lambda_k}\sum_{n\neq 0}\left(\langle u_q, u_n\rangle A_{nk}+\langle u_q^*, u_n\rangle^* B_{nk}\right)+i\frac{H^2}{2 \lambda_k}\cancelto{0}{\langle u_q,1\rangle} C_k+i\frac{H^2}{2 \lambda_k}\cancelto{0}{\langle u_q,f\rangle} D_k,
\ee
where the last two inner products vanish because $q\neq 0$ and $\langle Y_q,Y_0\rangle=0$, where the $Y$'s are  spherical harmonics. Similarly,
\be
B_{qk}=-\frac{1}{\lambda_k}\sum_{n\neq 0}\left(\langle u^*_q, u_n\rangle A_{nk}+\langle u^*_q, u_n\rangle^* B_{nk}\right).
\ee
The definitions of $A_{qk}$ and $B_{qk}$ for $q\neq 0$ and $k\neq 0$ are the same as in the $O(4)$ case, and they are therefore ill-defined.
\be
C_k=\frac{1}{\lambda_k}\sum_q\left(\cancelto{0}{\langle f, u_q\rangle}A_{qk}+\cancelto{0}{\langle f, u_q^*\rangle}B_{qk}\right)+i\frac{H^2}{2 \lambda_k}\langle f, 1 \rangle+i\frac{H^2}{2 \lambda_k}\langle f, f \rangle
\label{ccoef}
\ee
\be
D_k=-i\frac{H^2}{2 \lambda_k}\langle 1, 1 \rangle-i\frac{H^2}{2 \lambda_k}\langle 1, f \rangle.
\label{dcoef}
\ee
Let $C_k=D_k=0$ for $k\neq 0$\footnote{Justification for $C_k=D_k=0$ when $k\neq 0$: If $C_k=\frac{i}{H \alpha_k}e^{i\theta_k}$, $D_k=\frac{\alpha_k}{H}e^{i\theta_k}$, then from \eqref{inkg2} we need that $-i\frac{H^2}{2}\left(D^*_kC_{k'}-C^*_kD_{k'}\right)=e^{i(\theta_{k'}-\theta_k)}\propto\delta_{kk'}$. Therefore we must choose only one special value of $k$ for which $C_k$ and $D_k$ are not 0. From the equation in the second sentence of the next footnote, we see that this special value of $k$ is $k=0$.}, and $A_{qk}=B_{qk}=0$ for $k=0$\footnote{Justification for $A_{qk}=B_{qk}=0$ when $k=0$: Let $A_{qk}, B_{qk}\neq 0$ for some $q$. Then \eqref{inkg2} becomes $A^*_{q0}A_{q0}-B^*_{q0}B_{q0}-i\frac{H^2}{2}\left(C_0D^*_0-C^*_0D_0\right)=1$. But then $\langle s_0,s_q\rangle=\lambda_q\left(A^*_{q0}A_{qq}-B^*_{q0}B_{qq}\right)=0$. This is solved by either a) $A_{q0}=-\sinh\alpha_q e^{-i\beta_q}, B_{q0}=-\cosh\alpha_q$, or b) $A_{q0}=1/\cosh\alpha_q, B_{a0}=e^{i\beta_q}/\sinh \alpha_q$. But neither of these solutions yield vanishing $\langle s_0,s_q^*\rangle$. Therefore we must have $A_{qk}=B_{qk}=0$.}. We can then write
$A_{qk}=\delta_{qk}\cosh{\alpha_k}$, $B_{qk}=\delta_{qk}\sinh{\alpha_k e^{i\beta_k}}$, and \eqref{inkg2}-\eqref{inkg3} become
\be
-i\frac{H^2}{2}\left(D^*_0C_0-C^*_0D_0\right)=1,
\label{const1}
\ee
\be
C_0D_0-C_0D_0=0.
\label{const2}
\ee
The constraint \eqref{const2} is trivially satisfied, and \eqref{const1} is satisfied if we choose 
\be
C_0=\frac{i}{H\alpha}e^{i\theta}, \indent D_0=\frac{\alpha}{H}e^{i\theta}\indent(\alpha,\theta\in{\rm I\!R}).
\ee
Plugging these into \eqref{ccoef} we get 
\be
\frac{2\lambda_0}{i H^2} \frac{i}{\alpha H}=\langle  f, 1\rangle\frac{i}{\alpha H} e^{i\theta}+\langle f,f\rangle \frac{\alpha}{H} e^{i\theta}.
\ee
$\langle f, 1\rangle$ vanishes, leaving
\be
\alpha^2=\frac{2\lambda_0}{H^2\langle f, f\rangle}.
\label{alph1}
\ee
Similarly, from \eqref{dcoef} we get 
\be
\alpha^2=\frac{\langle 1, 1\rangle H^2}{2\lambda_0}.
\label{alph2}
\ee
Together \eqref{alph1} and \eqref{alph2} yield
\be
\alpha^2=\sqrt{\frac{\langle 1, 1\rangle}{\langle f, f\rangle}}=|\text{const}|,
\ee
where $\text{const}$ is a non-zero and finite constant.
Hence $C_0$ and $D_0$ are finite and well-defined.

\section{Causal Set SJ Vacuum from Simulations} 
\label{sec:numerics} 
While there is progress on finding the SJ modes via an ab initio calculation in some 2d as well as higher dimensional examples \cite{Buck:2016ehk, Mathur2019}, the calculation in  global de Sitter is considerably more difficult. In the absence of this, we can still carry out numerical calculations\footnote{The bulk of the simulations for this work were done using \texttt{Mathematica} \cite{WolframResearch}.} using causal sets to study the two-point
function. Causal sets are not only a natural covariant discretisation of the continuum, but also may contain important signatures of quantum spacetime. This makes the ab initio results in the causal set even more interesting than the ab initio results in the continuum.

Before we present the results, we carry out dimensional analysis that tells us the right quantities to compare. 
\subsubsection*{Dimensional analysis in the continuum}
The retarded Green function satisfies the KG equation so we have\footnote{$[\,]$ refers to length dimension.} $[G]=2-d=[\Delta]$. The eigenvalue equation for the PJ operator is 
\be
(i\Delta\,f_k)(x)=\int dV_y\,i\Delta(x,y)f_k(y)=\lambda_k\,f_k(x).
\ee
Therefore $[\lambda_k]=2$. The SJ two-point function is the positive part of the PJ operator and is given by 
\be
W(x,y)=\sum_k\lambda_k\,\tilde f_k(x)\tilde f^*_k(y)\quad\quad(\lambda_k>0),
\ee
where $\tilde f_k$ are the normalised (in the $L^2$ norm) eigenfunctions. So we have, $[\tilde f_k]=-d/2$, $[W]=2-d$.

\textbf{Note:} If we define the SJ modes as $f_k^{SJ}=\sqrt{\lambda_k}\,\tilde f_k$ then we get $[f_k^{SJ}]=1-d/2$.
\subsubsection*{Dimensional analysis in the causal set}
We can get the dimension of the massless causal set Green function $K_0$ by requiring that $[K_0m^2/\rho]=0$, where $[m^2/\rho]=d-2$. This gives $[K_0]=2-d=[G]=[i\Delta]$.
	
We use the following correspondence to define the analogs of integral operators in the causal set 
\be
\int dV_y\rightarrow\frac{1}{\rho}\sum_y\,.
\ee
The eigenvalue equation is given by a matrix equation
\be
\frac{1}{\rho}\,i\Delta f_k=\lambda_k\,f_k\,.
\label{csee}
\ee   
Here $[\lambda_k]=2$. These eigenvalues can be compared with the continuum eigenvalues.\footnote{Typically the $\frac{1}{\rho}$ factor in \eqref{csee} is omitted, which is why in the figures above showing the eigenvalues, the causal set spectra are divided by $\rho$.} 

As in the continuum, we have\footnote{This normalisation is obtained by taking the dot product of the vector with itself, divided by the density.} $[\tilde f_k]=-d/2$ and $[f_k^{SJ}]=1-d/2$. For the two-point function we have $[W]=2-d$. Therefore, $W$ can also be compared directly with its counterpart in the continuum.

We now present our numerical simulations for the causal set SJ vacuum in the causal diamonds in 2d and 4d Minkowski spacetime and slabs of 2d and 4d global de Sitter spacetime. 
Where visible, error bars in the binned data reflect the SEM. 
 
\subsection{ Causal Diamond in 2d Minkowski Spacetime}
We begin by revisiting the analysis of $\wsj$ for the massless FSQFT in a causal diamond in 2d Minkowski spacetime  \cite{Afshordi:2012ez}. 
The IR-regulated Minkowski two-point function is
\begin{equation}
  \text{Re}[\Wmink]=-\frac{1}{2 \pi} \ln(x) + c_1,\quad \quad x=\tau \,\text{or}\, |d|, 
  \label{eq:2dmink}
\end{equation}

where $c_1$ depends on the IR cutoff. In \cite{Afshordi:2012ez} it was shown that in a small subregion in the center of the causal diamond (i.e., away from the boundaries)

\begin{equation}
c_1\approx -\frac{1}{2 \pi} \ln(\lambda e^\gamma),
\end{equation}
where $\gamma$ is the Euler-Mascheroni constant and $\lambda\sim0.46/L$, and where $2L$ is the side length of the diamond.   

In our simulations, we work in units where the volume (in 2d this is an area) of the diamond is  unity, $L=1/2,\,V=4L^2=1$. Therefore, when we compare to the continuum function \eqref{eq:2dmink}, we set $c_1\approx -0.0786$.  

Our results are shown in figures \ref{2dcdeigens}-\ref{2dcdsub} and agree with the ab initio construction of \cite{Afshordi:2012ez}. Figure \ref{2dcdeigens} is a log-log plot of the positive causal set SJ eigenvalues, along with the positive continuum eigenvalues (discussed at the end of Section \ref{sjvac}). The two sets of eigenvalues are in agreement up to a characteristic ``knee" at which the causal set spectrum dips and ceases to obey a power-law with exponent $-1$. There is a clear convergence of the spectrum with causal set size $N$ except that the knee is pushed to smaller eigenvalues as $N$ increases. 

\bfig[H]
\begin{center} 
\includegraphics[width = .7\textwidth]{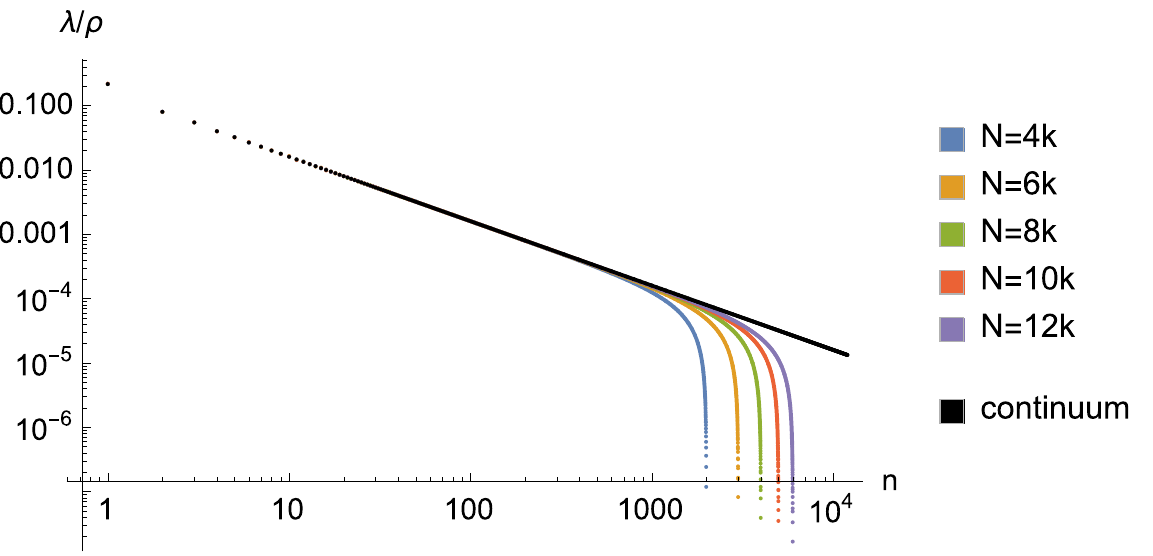}
\caption{Log-log plot of the eigenvalues of $i\Delta$ divided by density $\rho$ (except for the continuum), in the 2d causal diamond; $m=0$.}
\label{2dcdeigens}
\end{center} 
\efig
Figure \ref{2dcd} shows scatter plots of $\text{Re}[\wsj]$ for pairs of events that are causally and spacelike related; it also shows the binned and averaged plots where the convergence becomes clear. The convergence with $N$ is very good and tells us that we are in the asymptotic regime. This is the kind of convergence  we will look for when either a comparison with the continuum is not possible or when there is a marked discrepancy with the continuum. In order to compare with the continuum, $\wsj$ was calculated in \cite{Afshordi:2012ez} for pairs of points in a small causal diamond in the center of the larger causal diamond and it was shown that $\wsj$ agreed with the Minkowski vacuum in \eqref{eq:2dmink}. We carry out a similar comparison and the results are shown in figure \ref{2dcdsub}. This figure shows the scatter plots and the binned and averaged plots for $\wsj$ within a smaller  diamond of side length $1/4$ compared to that of the original diamond it is concentric to. The continuum IR-regulated Minkowski curve is also plotted. These plots confirm that away from the boundaries of the diamond $\tre[\wsj]$ indeed resembles the Minkowski vacuum, as was shown analytically and numerically in \cite{Afshordi:2012ez}.  
\bfig[H]
\bsfig[b]{0.4\textwidth}
	\includegraphics[width=\textwidth]{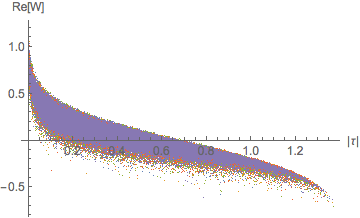}
	\caption{Causal}
\esfig
\bsfig[b]{0.4\textwidth}
	\includegraphics[width=\textwidth]{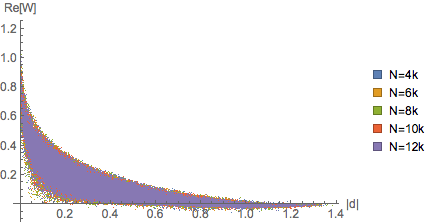}
	\caption{Spacelike}
\esfig\\
\bsfig[b]{0.4\textwidth}
	\includegraphics[width=\textwidth]{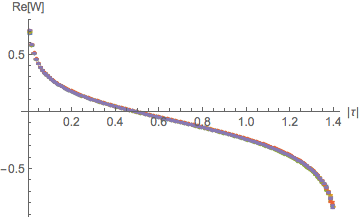}
	\caption{Causal}
\esfig
\bsfig[b]{0.4\textwidth}
	\includegraphics[width=\textwidth]{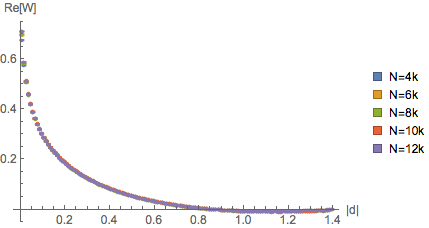}
	\caption{Spacelike}
\esfig
\caption{(a)-(b) represent $\tre[\wsj]$ vs. geodesic distance for a sample of $100000$ randomly selected pairs, in the 2d causal diamond; $m=0$. (c)-(d) are plots of the binned and averaged data with the SEM.}
\label{2dcd} 
\efig

\bfig[H]
	\bsfig[b]{0.4\textwidth}
		\includegraphics[width=\textwidth]{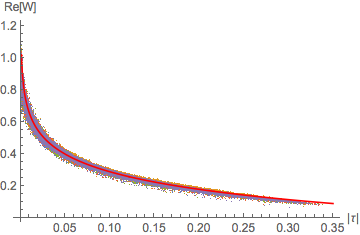}
		\caption{Causal}
	\esfig
		\bsfig[b]{0.4\textwidth}
			\includegraphics[width=\textwidth]{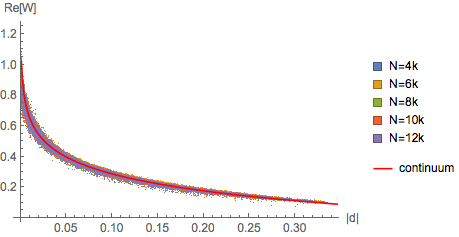}
			\caption{Spacelike}
		\esfig\\
			\bsfig[b]{0.4\textwidth}
				\includegraphics[width=\textwidth]{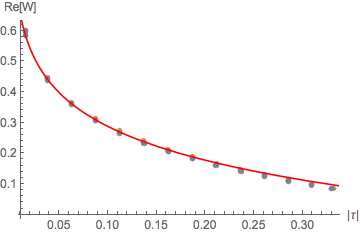}
				\caption{Causal}
			\esfig
				\bsfig[b]{0.4\textwidth}
					\includegraphics[width=\textwidth]{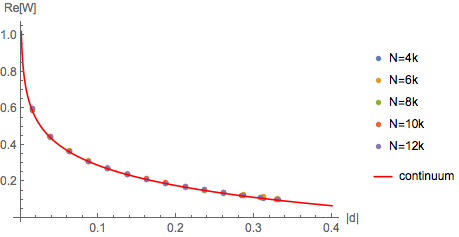}
					\caption{Spacelike}
				\esfig
\caption{(a)-(b) represent $\tre[\wsj]$ vs. geodesic distance for all pairs within a sub-diamond with side length $1/4$ of that the full diamond, in the 2d causal diamond; $m=0$. (c)-(d) are plots of the binned and averaged data with the SEM. In both cases, the continuum IR-regulated Minkowski Wightman function  \eqref{eq:2dmink} has also been shown.}
\label{2dcdsub}
\efig

\subsection{ Causal Diamond in 4d Minkowski Spacetime}
Next we examine the massless FSQFT in a causal diamond in 4d Minkowski spacetime. Unlike in 2d, we do not have an analytic ab initio calculation to compare with or refer to. We will instead rely on convergence properties and comparisons with the  continuum in a small causal diamond within the larger one. Another difference with the 2d case is that the causal set retarded Green function only agrees with the continuum one in the infinite density limit. This was discussed above in chapter 2.
 
The 4d Minkowski two-point function is 
\begin{equation}
    \text{Re}[\Wmink]=\frac{1}{4 \pi^2 x^2},\quad x=i\tau \,\text{or}\, |d| .
    \label{eq:4dmink}
\end{equation}

We work in units where the (top to bottom corner) height of the diamond is unity.
In figure \ref{4dcdgreen} we plot binned and averaged values for the causal set retarded Green function \eqref{massless4d} along with its  expectation value at finite density \eqref{linkexp}. The corresponding continuum Green function \eqref{cont4d} has a delta function on the lightcone and is therefore infinitely sharply peaked there. While this is not the case in the causal set, the discrepancy grows smaller as the density is increased.

\bfig[H]
\begin{center} 
\includegraphics[width = .56 \textwidth]{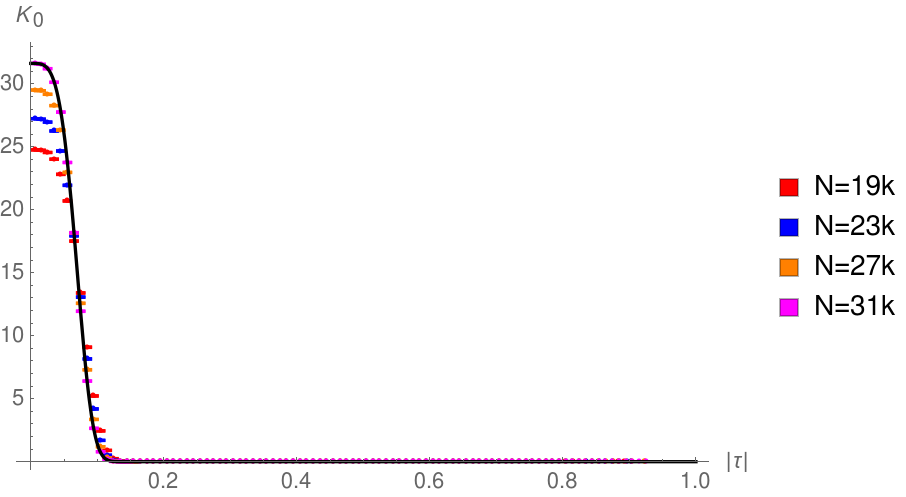}
\caption{The binned and averaged plot for $K_0$ vs. $|\tau|$ as $N$ is varied, in the 4d causal diamond. The black curve represents the expectation value  \eqref{linkexp} for $N=31k$. We see an excellent match.}
\label{4dcdgreen}
\end{center} 
\efig
In figure \ref{4dcdeigens} we show the log-log plot of the SJ spectrum. This spectrum is qualitatively similar to the spectrum in the 2d diamond, in that it obeys a power-law in the large eigenvalue regime,  while exhibiting a knee in the UV (smaller eigenvalue regime) where it dips. It moreover converges well as $N$ is increased, except near the knee which, as in the 2d diamond, shifts to the UV as $N$ increases. This suggests that we are in the asymptotic regime. 

\bfig[H]
\begin{center} 
\includegraphics[width = .6 \textwidth]{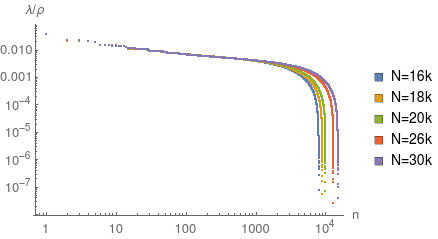}
\caption{Log-log plot of the eigenvalues of $i\Delta$ divided by density $\rho$, in the 4d causal diamond; $m=0$.}
\label{4dcdeigens}
\end{center} 
\efig
In figure \ref{4dcd} we show the scatter and binned plots for $\text{Re}[\wsj]$ as $N$ is varied. The convergence with increasing density suggests that the larger $N$ values are approaching the asymptotic regime. The Minkowski two-point function \eqref{eq:4dmink} is also included in this plot and it clearly does not agree with $\wsj$ in the full diamond. The small distance behaviour shows an interesting departure from the continuum, softening the divergences. 

\bfig[H] 
		\bsfig[b]{0.46\textwidth}
		\includegraphics[width=\textwidth]{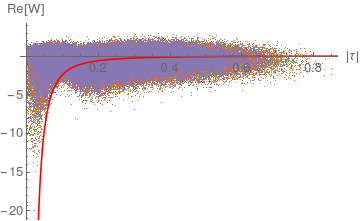}
		\caption{Causal}
		\esfig
		\bsfig[b]{0.53\textwidth}
		\includegraphics[width=\textwidth]{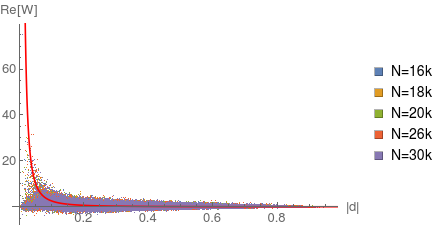}
		\caption{Spacelike}
		\esfig\\
		\bsfig[b]{0.46\textwidth}
		\includegraphics[width=\textwidth]{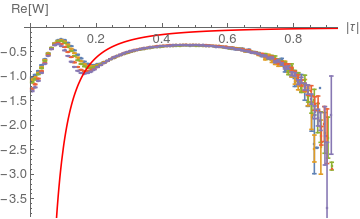}
		\caption{Causal}
		\esfig
		\bsfig[b]{0.53\textwidth}
			\includegraphics[width=\textwidth]{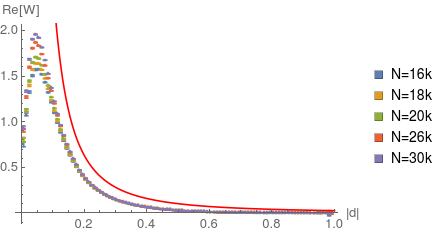}
			\caption{Spacelike}
			\esfig
\caption{(a)-(b) represent $\tre[\wsj]$ vs. geodesic distance for a sample of $100000$ randomly selected pairs, in the 4d causal diamond; $m=0$. (c)-(d) are plots of the binned and averaged data with the SEM. In both cases, the continuum Minkowski Wightman function \eqref{eq:4dmink} has also been shown in red.}
\label{4dcd} 
\efig
Figure \ref{4dcdsub} shows the scatter and binned plots for a smaller causal diamond of side length $1/2$ compared to the larger diamond it is in the center of. Although the agreement of $\wsj$ with $\Wmink$ is not as good as in 2d, we see that as $N$ increases, there is a convergence of $\wsj$ to $\Wmink$. This suggests that as in 2d, the 4d diamond also shows an agreement with the Minkowski vacuum far away from the boundary. 

Figure \ref{4dcdpairs} shows the distribution of pairs of points in the diamond as a function of the proper time and  distance. From this plot one can see that there are many fewer pairs of points at small and large proper distance and times than in the intermediate regimes. Nevertheless, the scatter plots and the error bars on the binned plots do not show significant deviation in these regimes. 
\bfig[H]
	\bsfig[b]{0.45\textwidth}
	\includegraphics[width=\textwidth]{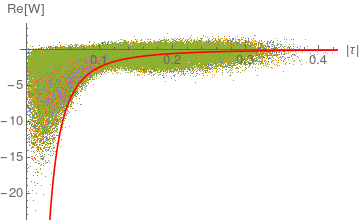}
	\caption{Causal}
	\esfig
	\bsfig[b]{0.54\textwidth}
	\includegraphics[width=\textwidth]{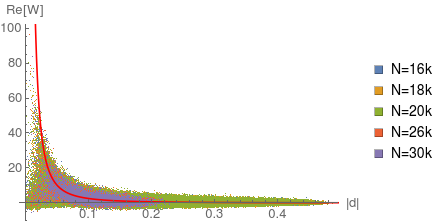}
	\caption{Spacelike}
	\esfig\\
		\bsfig[b]{0.45\textwidth}
		\includegraphics[width=\textwidth]{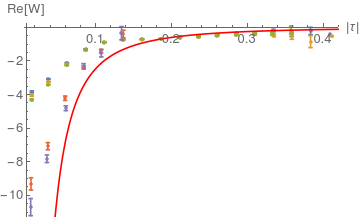}
		\caption{Causal}
		\esfig
		\bsfig[b]{0.54\textwidth}
		\includegraphics[width=\textwidth]{comparison_wspace_with_N_m0_4dhalfraw}
		\caption{Spacelike}
		\esfig
\caption{(a)-(b) represent $\tre[\wsj]$ vs. geodesic distance for all pairs within a sub-diamond with height $1/2$ of the full diamond, in the 4d causal diamond; $m=0$. (c)-(d) are plots of the binned and averaged data with the SEM. In both cases, the continuum Minkowski Wightman function  \eqref{eq:4dmink} has also been shown.}
\label{4dcdsub}
\efig

\bfig
	\bsfig[b]{0.5\textwidth}
	\includegraphics[width=\textwidth]{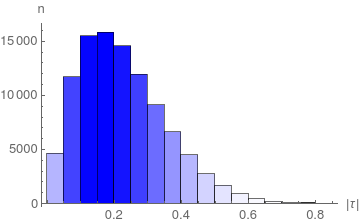}
	\caption{Causal pairs}
	\esfig
	\bsfig[b]{0.5\textwidth}
	\includegraphics[width=\textwidth]{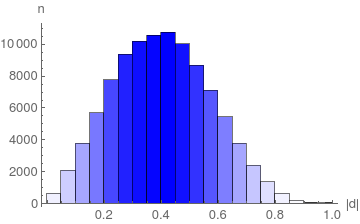}
	\caption{Spacelike pairs}
	\esfig
\caption{Distribution of the number of causal and spacelike pairs $n$ with magnitude of the geodesic distance for $N=30k$, in the 4d causal diamond.}
\label{4dcdpairs} 
\efig

\subsection{Slab of 2d de Sitter Spacetime}
The simulations in the 2d and 4d causal diamond help set the stage for the simulations in  slabs of 2d and 4d de Sitter spacetime, which we turn to in this and the next subsection. As in the causal diamond examples, we will look for convergence of the causal set calculation with $N$ to establish that we are in the asymptotic regime. The slab in de Sitter spacetime lies within the region $[-T,T]$\footnote{$T$ is the cutoff in the conformal time defined in  \eqref{confmetric2}.} and we will probe our results' sensitivity to  $T$. We will also look for convergence with $T$ at fixed $\rho$, to show that the results are independent of the cutoff. 

The Wightman function for the Euclidean vacuum in $d$ spacetime dimensions is given by\footnote{The expression for $W_E$ in equation $B.36$ of \cite{Aslanbeigi:2013fga} has a minor typographical error: the factor of $4\pi$ should be raised to the power of $d/2$. See for example \cite{Bousso2002}.}
\be
W_E(x,y)=\frac{\Gamma[h_+] \Gamma[h_-]}{(4 \pi)^{d/2}\ell^2 \Gamma[\frac{d}{2}]}\,  {}_2F_1\left(h_+,h_-, \frac{d}{2};\frac{1+Z(x,y)+i\epsilon\, \text{sign}(x^0-y^0)}{2}\right),
\label{we}
\ee
where $Z(x, y)$ is defined by \eqref{zdef},  $h_\pm=\frac{d-1}{2}\pm\nu$, $\nu=\ell\sqrt{m_*^2-m^2}$, $m_*=\frac{d-1}{2\ell}$ and $_2F_1(a,b,c;z)$ is a hypergeometric function. The 
symmetric two-point function, or Hadamard function, for any other Allen-Mottola $\alpha$-vacuum is \cite{Aslanbeigi:2013fga}
\be
    H_{\alpha\beta}(x,x')=\cosh2\alpha\,H_E(x,x')+\sinh2\alpha\,[\cos\beta\,H_E(\bar{x},x')-\sin\beta\,\Delta(\bar{x},x')]
    \label{walpha},
\ee
where $\bar{x}$ is the antipodal point of $x$. The Wightman function is related to $H$ by $2W=H+i\Delta$. We will make comparisons with the $\alpha$-vacua found to correspond to the SJ vacuum in \cite{Aslanbeigi:2013fga}. Since we  work in even dimensions, these are
$\alpha=0$ for $m\geq m_*$ (yielding the Euclidean vacuum), and 
\be
    \alpha=\frac{1}{2}\tanh^{-1}|\sin\pi\nu|\quad\text{and}\quad\beta=\pi[\frac{d}{2}+\theta(-\sin\pi\nu)]
\ee
for $m<m_*$.

In this subsection we consider 2d de Sitter spacetime, and work in units in which the de Sitter radius $\ell=1$. In 2d,  $m_*= 0.5$, and the conformal mass $m_c=0$. Hence the minimally coupled and the conformally coupled massless cases coincide.
Our simulations span slabs of different heights given by $T$ values ranging from $1$ to $1.5$, while our $N$ values range from $8k$ to $36k$. We show the log-log plots of the PJ spectrum for the massless $m=0$ and for the massive $m=2.3$ cases in figure \ref{fig:2ddSspectrum}. As in the 2d diamond, the causal set spectrum exhibits a characteristic knee. The spectrum converges very well for both sets of masses, with the knee shifting to the UV as $N$ increases, as expected. We also compare the causal set spectrum with the finite $T$ continuum spectrum obtained via the mode comparison method in \cite{Aslanbeigi:2013fga}. As shown in figure \ref{fig:2ddSspectrum} this spectrum does not seem to agree with the causal set spectrum even though the latter convergences with $N$. 

\bfig[H]
	\bsfig[b]{0.43\textwidth}
	\includegraphics[width=\textwidth]{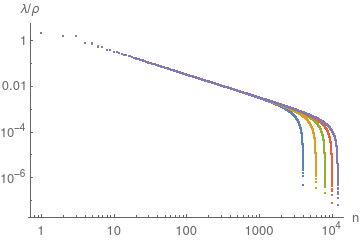}
	\caption{$m=0$}
	\esfig
		\bsfig[b]{0.57\textwidth}
		\includegraphics[width=\textwidth]{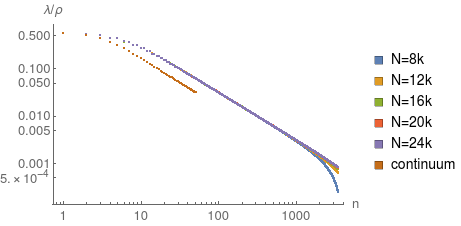}
		\caption{$m=2.3$}
		\esfig
\caption{Log-log plot of the positive eigenvalues of $i\Delta$ at $T=1$, in 2d de Sitter. In the massive case on the right we plot the largest 3500 positive eigenvalues and the corresponding continuum eigenvalues from the finite T mode comparison results of \cite{Aslanbeigi:2013fga}.} 
\label{fig:2ddSspectrum}
\efig

In the simulations whose results we present below, we examine two masses in detail: $m=0$ and $m=2.3$\footnote{This is an arbitrary choice of mass with no special physical significance. It  allows for comparisons with \cite{Aslanbeigi:2013fga} who use a similar mass in their 2d de Sitter causal set simulations.}, and vary over both the slab height $T$ as well as the density $\rho$. For $m=2.3$, as can be seen in the scatter plots of figures \ref{2ddSSJ1}, \ref{2ddSSJ15} and \ref{2ddST156}, $\wsj$ agrees very well with the SJ vacuum expected from the calculation in \cite{Aslanbeigi:2013fga} (the Euclidean vacuum). Furthermore, it appears that $\wsj$ for a given $T$ is simply the restriction of $\wsj$ for a larger $T$. This is also in agreement with the simulation results of \cite{Aslanbeigi:2013fga}. 

For the massless case, the scatter plots of $\wsj$ in figures \ref{2ddSSJ1zero}, \ref{2ddSSJ15zero} and \ref{2ddST156zero} do not show convergence, but instead fan out, as a function of the proper time and distance. As the density decreases, for $T=1.56, N=36k$, the scatter plot figure \ref{2ddST156zero} shows a clustering into two distinct sets. This shows that $\wsj$ may not just be a function of proper time and distance, and hence {\it may not be} de Sitter invariant.

In figure \ref{fig:2ddSsem} the binned and averaged plots for $\wsj$ show  very good convergence with $N$. While this is consistent with the narrowing of the $m=2.3$ scatter plots at higher densities, the convergence for $m=0$ is not (since the $m=0$ scatter plots do not narrow much). Hence both the scatter plots and the binned plots are important in determining convergence as well as understanding the nature of $W_{SJ}$. 

\bfig[H]  
\bsfig[b]{0.42\textwidth}
\includegraphics[width=\textwidth]{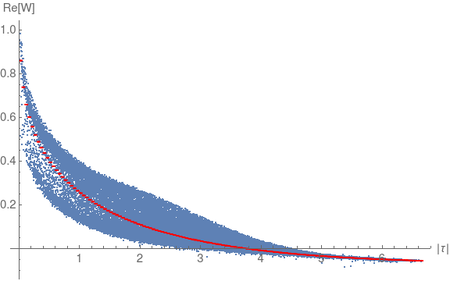}
\caption{Causal $m=0$}
\esfig
\bsfig[b]{0.42\textwidth}
\includegraphics[width=\textwidth]{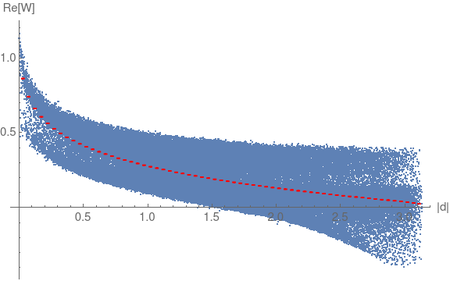}
\caption{Spacelike $m=0$}
\esfig\\
\bsfig[b]{0.4\textwidth}
\includegraphics[width=\textwidth]{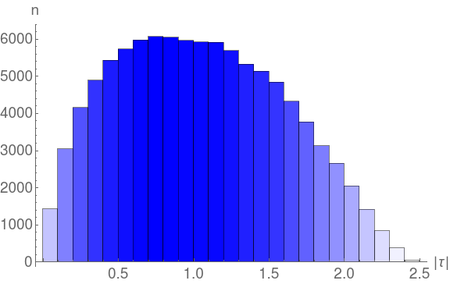}
\caption{Causal pairs}
\esfig
\bsfig[b]{0.4\textwidth}
\includegraphics[width=\textwidth]{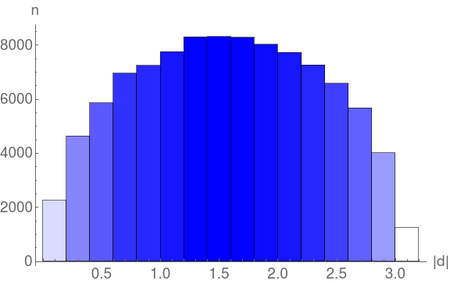}
\caption{Spacelike pairs}
\esfig
\caption{$N=32000, T=1, \rho=1635.08$, in 2d de Sitter. (a)-(b) represent $\tre[\wsj]$ vs. geodesic distance for a sample of 100000 randomly selected
pairs, and the red curve represents the mean values with the SEM. (c)-(d) are plots of the distribution of pairs.}
\label{2ddSSJ1zero}
\efig

\bfig[H]
\bsfig[b]{0.45\textwidth}
\includegraphics[width=\textwidth]{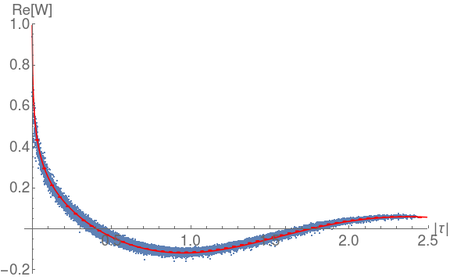}
\caption{Causal $m=2.3$}
\esfig
\bsfig[b]{0.45\textwidth}
\includegraphics[width=\textwidth]{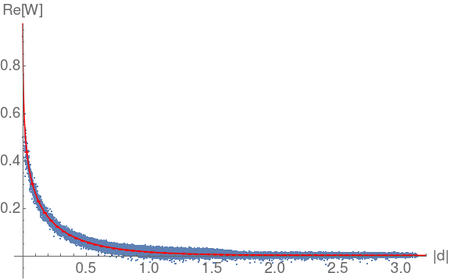}
\caption{Spacelike $m=2.3$}
\esfig
\caption{$N=24000,\, T=1,\, \rho=1226.31$, in 2d de Sitter. The scatter plot is $\tre[\wsj]$ vs. geodesic distance for a sample of 100000 randomly selected
pairs. The red curve represents the continuum $W_E$ from \eqref{we}.}
\label{2ddSSJ1} 
\efig
\bfig[H]
\bsfig[b]{0.45\textwidth}
\includegraphics[width=\textwidth]{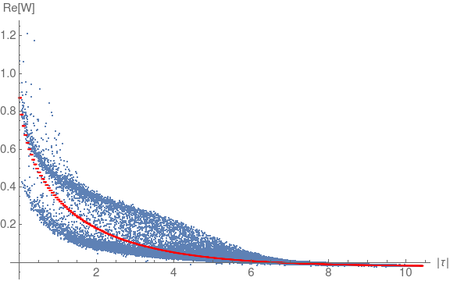}
\caption{Causal $m=0$}
\esfig
\bsfig[b]{0.45\textwidth}
\includegraphics[width=\textwidth]{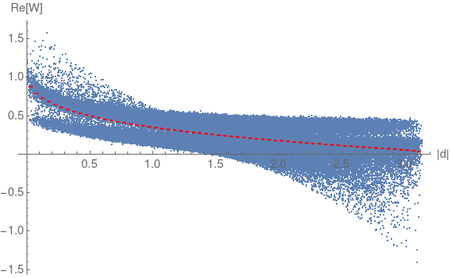}
\caption{Spacelike $m=0$}
\esfig\\
\bsfig[b]{0.45\textwidth}
\includegraphics[width=\textwidth]{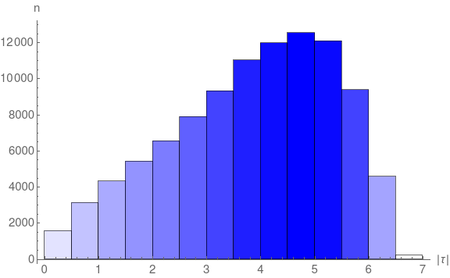}
\caption{Causal pairs}
\esfig
\bsfig[b]{0.45\textwidth}
\includegraphics[width=\textwidth]{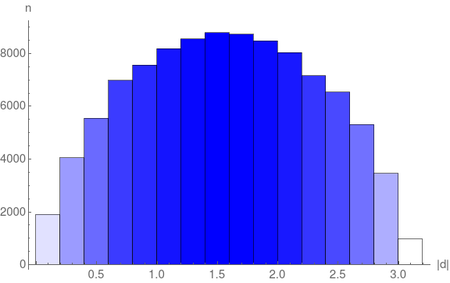}
\caption{Spacelike pairs}
\esfig
\caption{$N=36000, T=1.5, \rho=203.15$, in 2d de Sitter. (a)-(b) represent $\tre[\wsj]$ vs. geodesic distance for a sample of 100000 randomly selected pairs. The red curve represents the mean values with the SEM. (c)-(d) are plots of the distribution of pairs.}
\label{2ddSSJ15zero} 
\efig
\bfig[H]
\bsfig[b]{0.45\textwidth}
\includegraphics[width=\textwidth]{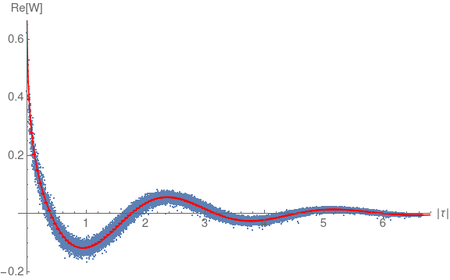}
\caption{Causal $m=2.3$}
\esfig
\bsfig[b]{0.45\textwidth}
\includegraphics[width=\textwidth]{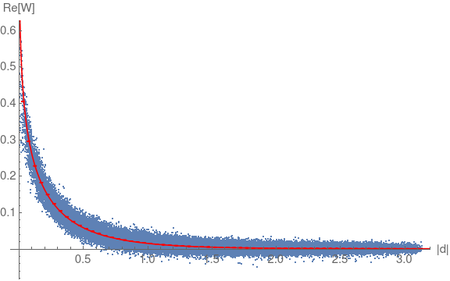}
\caption{Spacelike $m=2.3$}
\esfig
  \caption{$N=36000,\, T=1.5,\, \rho=203.15$, in 2d de Sitter.  $\tre[\wsj]$ vs. geodesic distance for 100000 randomly selected
pairs. The red curve represents the continuum $W_E$ from \eqref{we}.}
  \label{2ddSSJ15} 
 \efig
\bfig[H]
\bsfig[b]{0.45\textwidth}
\includegraphics[width=\textwidth]{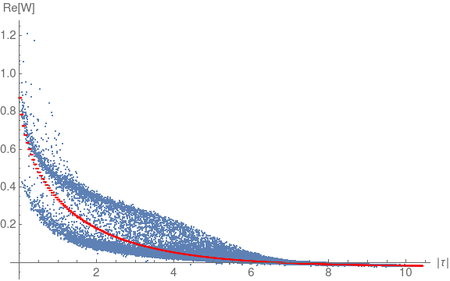}
\caption{Causal $m=0$}
\esfig
\bsfig[b]{0.45\textwidth}
\includegraphics[width=\textwidth]{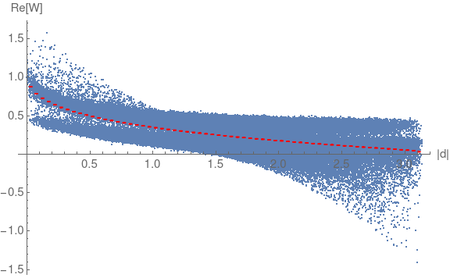}
\caption{Spacelike $m=0$}
\esfig\\
\bsfig[b]{0.45\textwidth}
\includegraphics[width=\textwidth]{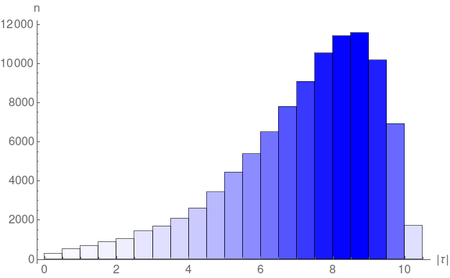}
\caption{Causal pairs}
\esfig
\bsfig[b]{0.45\textwidth}
\includegraphics[width=\textwidth]{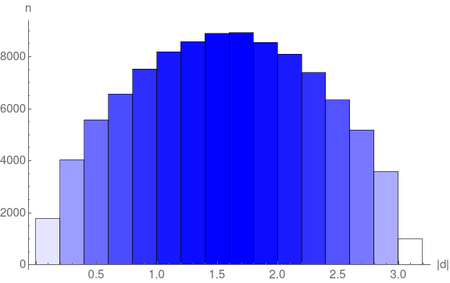}
\caption{Spacelike pairs}
\esfig 
\caption{$N=36000, -1.56<\tilde{T}<1.56, \rho=30.93$, in 2d de Sitter. (a)-(b) represent $\tre[\wsj]$ vs. geodesic distance for a sample of  100000 randomly selected pairs. The red curve represents the mean values (of the data) with the SEM. (c)-(d) are plots of the distribution of pairs.}
  \label{2ddST156zero} 
 \efig
\bfig[H]
\bsfig[b]{0.45\textwidth}
\includegraphics[width=\textwidth]{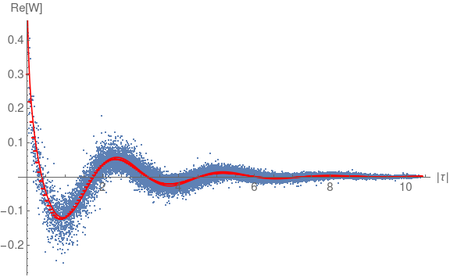}
\caption{Causal $m=2.3$}
\esfig
\bsfig[b]{0.45\textwidth}
\includegraphics[width=\textwidth]{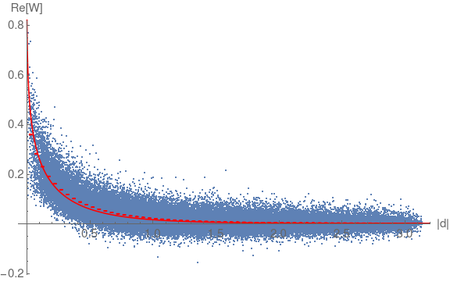}
\caption{Spacelike $m=2.3$}
\esfig
\caption{$N=36000, T=1.56, \rho=30.93$, in 2d de Sitter. $\tre[\wsj]$ vs. geodesic distance for a sample of 100000 randomly selected pairs. The red curve represents the continuum $W_E$ from \eqref{we}.}
\label{2ddST156} 
 \efig
 \bfig[H]
  \bsfig[b]{0.46\textwidth}
  \includegraphics[width=\textwidth]{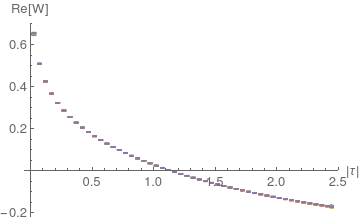}
  \caption{Causal $m=0$}
  \esfig
  \bsfig[b]{0.53\textwidth}
  \includegraphics[width=\textwidth]{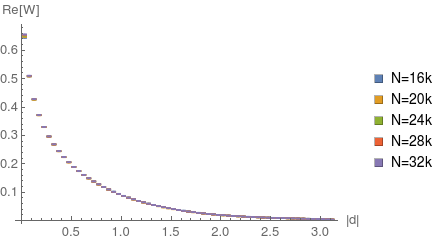}
  \caption{Spacelike $m=0$}
  \esfig\\
  \bsfig[b]{0.46\textwidth}
  \includegraphics[width=\textwidth]{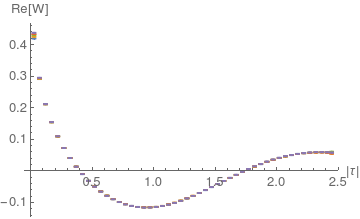}
  \caption{Causal $m=2.3$}
  \esfig
  \bsfig[b]{0.53\textwidth}
  \includegraphics[width=\textwidth]{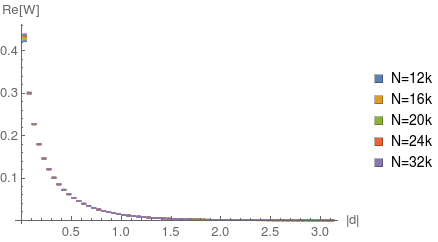}
  \caption{Spacelike $m=2.3$}
  \esfig 
      \caption{Variation of binned and averaged $\tre[\wsj]$ with density at $T=1$, in 2d de Sitter.}
      \label{fig:2ddSsem}
  \efig
 
\subsection{Slab of 4d de Sitter Spacetime}

Finally, we examine the 4d de Sitter SJ vacuum. Again, we work in units in which the de Sitter radius $\ell=1$. In 4d, $m_*=1.5$ and $m_c=\sqrt{2}\approx 1.41$. 

In figure \ref{fig:4ddSgreen} we show the scatter plot of the causal set retarded Green function \eqref{deSgreen4d}, taking the conformally coupled massless case as an example. While the small $\tau$ discrepancy with  the continuum expression is expected and attributed to the local finiteness of the causal set, the behaviour for large $\tau$ compares well with the continuum.
Figure \ref{fig:4ddSeigen} shows the log-log plot of the SJ spectrum for $m=0$ and $m=2.3$ for various $N$. We find that there is excellent convergence with $N$ in both cases, and again, as in the other cases we have seen, there is a knee which shifts to the UV as $N$ is increased. However, there is poor agreement with the continuum values of the finite $T$ spectrum calculated via the mode comparison method in \cite{Aslanbeigi:2013fga}, as in the 2d case. In figure \ref{fig:4ddSeigenwithm} we also show the spectrum for $m$ varied around $m=0$ and $m=m_c\approx1.41$. There is no unusual behaviour close to these masses.    
\bfig[H]
\begin{center}
\includegraphics[width=.52\linewidth]{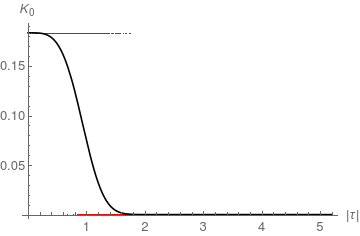}
\caption{$K_0$ vs. $|\tau|$ for $N=32k, T=1.42,\rho=7.978,m=m_c=\sqrt{2}$, in 4d de Sitter. The black curve represents the expectation value \eqref{linkexp}.}
\label{fig:4ddSgreen}
\end{center} 
\efig 

\bfig[H]
	\bsfig[b]{0.46\textwidth}
	\includegraphics[width=\textwidth]{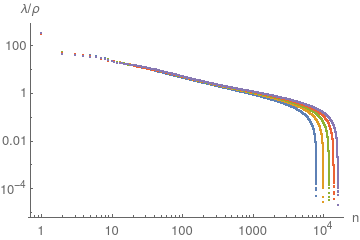}
	\caption{$m=0$}
	\esfig
	\bsfig[b]{0.53\textwidth}
	\includegraphics[width=\textwidth]{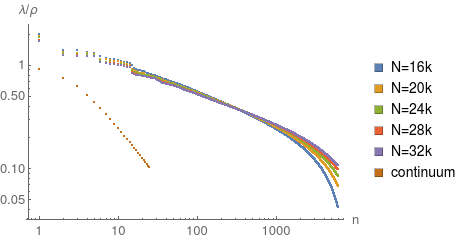}
	\caption{$m=2.3$}
	\esfig
    \caption{Log-log plot of the positive eigenvalues of $i\Delta$, in 4d de Sitter. In the massive case on the right we plot the largest 6000 positive eigenvalues and the corresponding continuum eigenvalues from the finite T mode comparison results of \cite{Aslanbeigi:2013fga}.}
    \label{fig:4ddSeigen}
\efig

Figures \ref{fig:4ddSzero} and \ref{fig:4ddS} are sample scatter plots of $\wsj$ for $m=0$ and $m=2.3$.
In figure \ref{fig:4ddSfixedT} we fix $T$ for $m=0$ and for $m=m_c\approx1.41$ and vary $N$ to check for convergence with density; for smaller proper times and distances, the convergence is not as good as it is for larger proper times and distances. For $m=1.41$ we also plot the  Wightman function associated with the  Euclidean vacuum $W_E$ in \eqref{we}. $W_E$ does not compare well with the causal set $\wsj$.
Next, in figure \ref{4ddSfixedrho} we fix the density $\rho=9$ and check the convergence with $T$, which we vary from $1.2$ to $1.42$. We find good convergence for various $m$ values. However, the Wightman function associated with the  $\alpha$-vacuum \eqref{walpha} as well as the Euclidean vacuum $W_E$ once again do not compare well with the causal set $\wsj$ for any of these masses. This is somewhat surprising, since the discrepancy occurs well away from the massless minimally and conformally coupled cases.  
\bfig[H]
\bsfig[b]{0.47\textwidth}
\includegraphics[width=\textwidth]{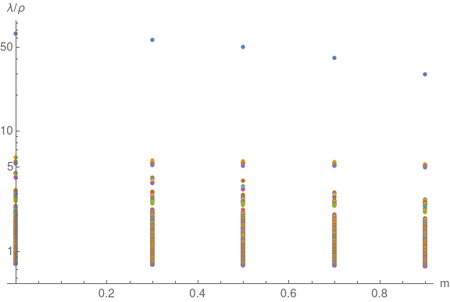}
\caption{$m=0$}
\esfig
\bsfig[b]{0.47\textwidth}
\includegraphics[width=\textwidth]{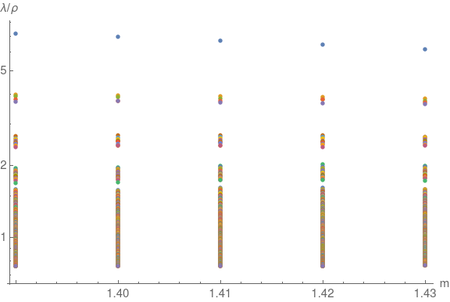}
\caption{$m=1.41$}
\esfig
    \caption{Log-linear plot of the first 500 positive eigenvalues of $i\Delta$ at $T=1.42,\, \rho=9$, in 4d de Sitter.}
    \label{fig:4ddSeigenwithm}
\efig
\bfig[H]
\bsfig[b]{0.48\textwidth}
\includegraphics[width=\textwidth]{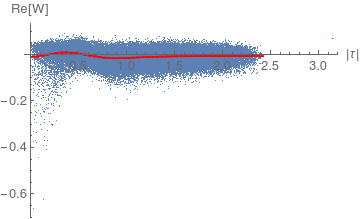}
\caption{Causal $T=1$}
\esfig
\bsfig[b]{0.48\textwidth}
\includegraphics[width=\textwidth]{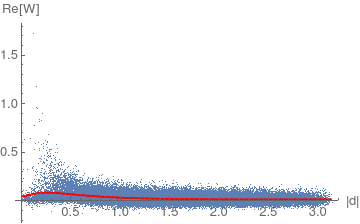}
\caption{Spacelike $T=1$}
\esfig\\
\bsfig[b]{0.48\textwidth}
\includegraphics[width=\textwidth]{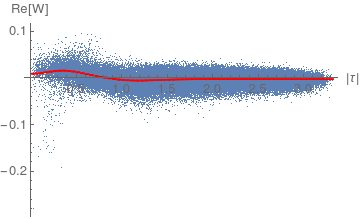}
\caption{Causal $T=1.2$}
\esfig
\bsfig[b]{0.48\textwidth}
\includegraphics[width=\textwidth]{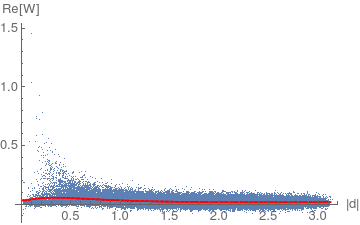}
\caption{Spacelike $T=1.2$}
\esfig
\caption{$m=0,\,N=32000$, in 4d de Sitter. $\tre[\wsj]$ vs. geodesic distance for   100000 randomly selected pairs,  and the red curve represents the mean
values with the SEM.}
\label{fig:4ddSzero}
\efig

\bfig[H]
\bsfig[b]{0.48\textwidth}
\includegraphics[width=\textwidth]{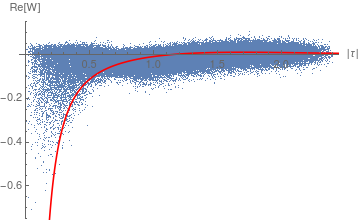}
\caption{Causal $T=1$}
\esfig
\bsfig[b]{0.48\textwidth}
\includegraphics[width=\textwidth]{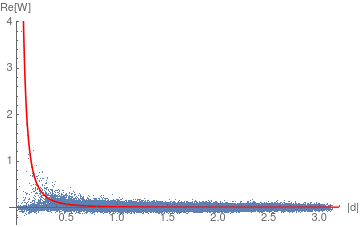}
\caption{Spacelike $T=1$}
\esfig\\
\bsfig[b]{0.48\textwidth}
\includegraphics[width=\textwidth]{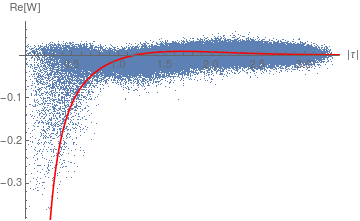}
\caption{Causal $T=1.2$}
\esfig
\bsfig[b]{0.48\textwidth}
\includegraphics[width=\textwidth]{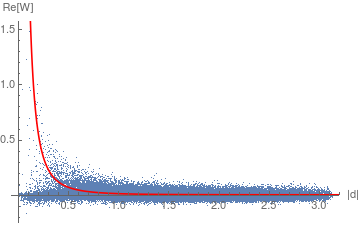}
\caption{Spacelike $T=1.2$}
\esfig
\caption{$m=2.3,\,N=32000$, in 4d de Sitter. $\tre[\wsj]$ vs. geodesic distance for a sample of  100000 randomly selected pairs. The red curve shows the Euclidean two-point function $W_E$ from \eqref{we}.} 
\label{fig:4ddS}
  \efig 
Further, in figure \ref{4ddSfixedrhom} we look at $\wsj$ for varying masses at fixed $T=1.42$ and $\rho=9$. We find that $\wsj$ looks like a continuous function of $m$ even as $m$ is varied around $m_c$. Indeed, the  large distance behaviour for all the masses is exactly the same. At smaller distances, there is an interesting bifurcation as $m$ changes: $\text{Re}[W]$ is positive for small masses and negative for large masses.
This figure also shows the number of pairs as a function of distances. The discrepancies in the small distance behavior could be attributed to the small number of pairs there.

Our simulations thus strongly suggest that the causal set 4d de Sitter $\wsj$  differs from the Mottola-Allen $\alpha$-vacua for all masses. 
\bfig[H]
\bsfig[b]{0.46\textwidth}
\includegraphics[width=\textwidth]{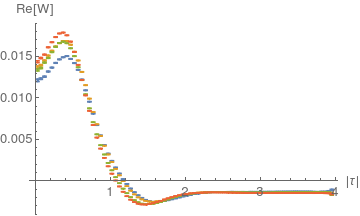}
\caption{Causal $m=0,T=1.3$}
\esfig
\bsfig[b]{0.53\textwidth}
\includegraphics[width=\textwidth]{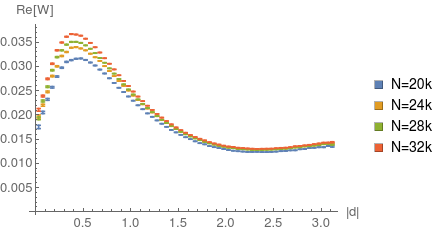}
\caption{Spacelike $m=0,T=1.3$}
\esfig\\
\bsfig[b]{0.46\textwidth}
\includegraphics[width=\textwidth]{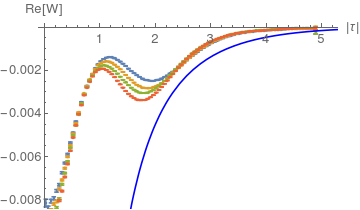}
\caption{Causal $m=1.41,T=1.4$}
\esfig
\bsfig[b]{0.53\textwidth}
\includegraphics[width=\textwidth]{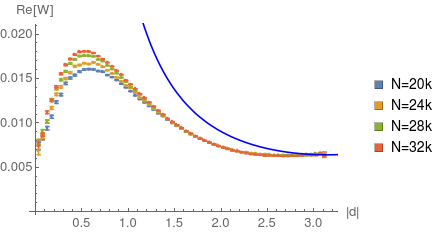}
\caption{Spacelike $m=1.41,T=1.4$}
\esfig
         \caption{$\tre[\wsj]$ vs. geodesic distance with varying density, in 4d de Sitter. The  blue curve shows the Euclidean two-point function as a reference.}
  \label{fig:4ddSfixedT}
\efig
\bfig[H] 
\bsfig[b]{0.46\textwidth}
\includegraphics[width=\textwidth]{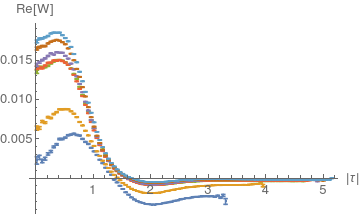}
\caption{Causal $m=0$}
\esfig
\bsfig[b]{0.53\textwidth}
\includegraphics[width=\textwidth]{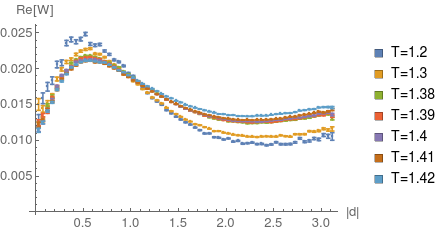}
\caption{Spacelike $m=0$}
\esfig 
\efig

\bfig[H] 
\bsfig[b]{0.46\textwidth}
\includegraphics[width=\textwidth]{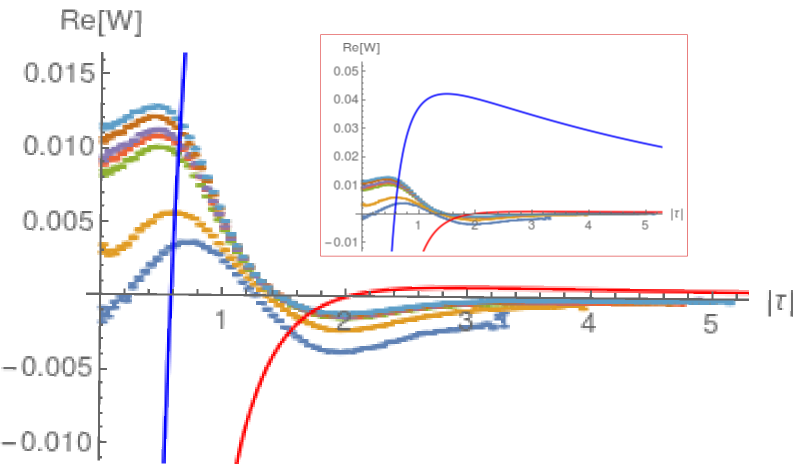}
\caption{Causal $m=0.7$}
\esfig
\bsfig[b]{0.53\textwidth}
\includegraphics[width=\textwidth]{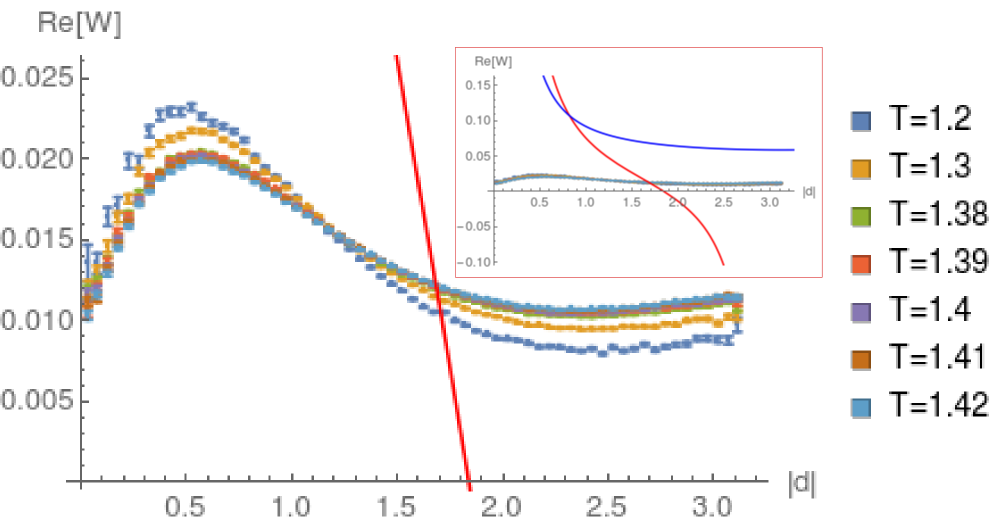}
\caption{Spacelike $m=0.7$}
\esfig\\
\bsfig[b]{0.46\textwidth}
\includegraphics[width=\textwidth]{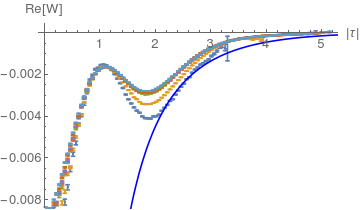}
\caption{Causal $m=1.41$}
\esfig
\bsfig[b]{0.53\textwidth}
\includegraphics[width=\textwidth]{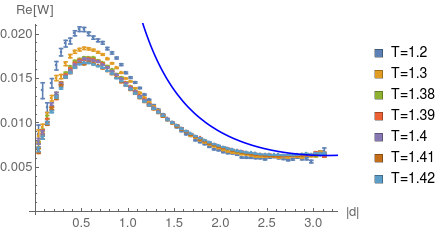}
\caption{Spacelike $m=1.41$}
\esfig\\
\bsfig[b]{0.46\textwidth}
\includegraphics[width=\textwidth]{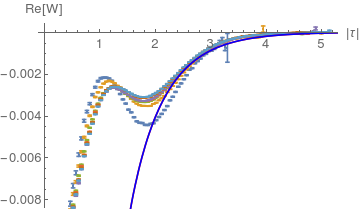}
\caption{Causal $m=1.5$}
\esfig
\bsfig[b]{0.53\textwidth}
\includegraphics[width=\textwidth]{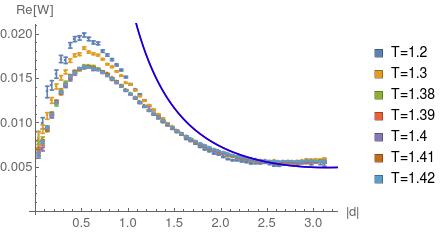}
\caption{Spacelike $m=1.5$}
\esfig
\caption{$\tre[\wsj]$ vs. geodesic distance with varying $T$ for various $m$ at $\rho=9$, in 4d de Sitter. The red and blue curves represent the corresponding continuum $\alpha$- and Euclidean two-point functions respectively. The inset figures represent the zoomed-out versions. In (e)-(f), for $m=\sqrt{2}$ there is no corresponding $\alpha$-vacuum, and in (g)-(h) the $\alpha$-vacuum and Euclidean vacuum coincide.}
\label{4ddSfixedrho} 
\efig
 
\bfig[!ht]
\setcounter{subfigure}{0}
\bsfig[b]{0.46\textwidth}
\includegraphics[width=\textwidth]{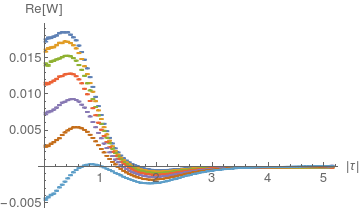}
\caption{Causal $T=1.42$}
\esfig
\bsfig[b]{0.53\textwidth}
\includegraphics[width=\textwidth]{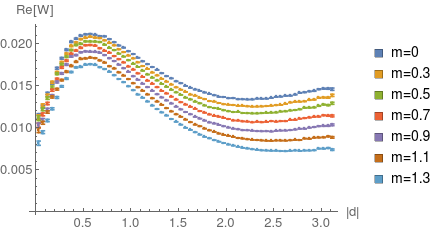}
\caption{Spacelike $T=1.42$}
\esfig\\
\bsfig[b]{0.46\textwidth}
\includegraphics[width=\textwidth]{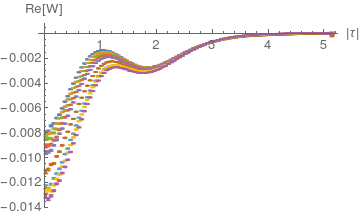}
\caption{Causal $T=1.42$}
\esfig
\bsfig[b]{0.53\textwidth}
\includegraphics[width=\textwidth]{comparison_wcausal_with_mass_concentrated_h1p42_rho9}
\caption{Spacelike $T=1.42$}
\esfig\\
\bsfig[b]{0.5\textwidth}
\includegraphics[width=\textwidth]{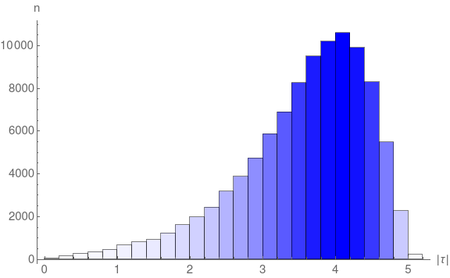}
\caption{Causal pairs}
\esfig
\bsfig[b]{0.5\textwidth}
\includegraphics[width=\textwidth]{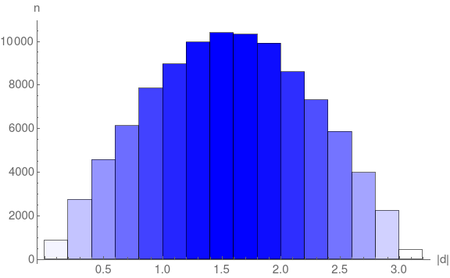}
\caption{Spacelike pairs}
\esfig
\caption{$\tre[\wsj]$ vs. geodesic distance with varying $m$ at $\rho=9$, in 4d de Sitter. (e)-(f) show the distribution of pairs.}
\label{4ddSfixedrhom}
\efig 

\section{Equations of Motion on the Causal Set}
\label{eom}

Before concluding this chapter we comment on another idea arising from the SJ construction - the causal set analogs of the equations of motion in the continuum\footnote{Approaches using the causal set analog of the d'Alembertian operator in the causal set have been studied elsewhere \cite{Benincasa:2010ac, Glaser:2013xha, Aslanbeigi2014, Dable-Heath:2019sej}.}. We saw that the relation 
$$
\kr (\hB -\phm^2) = \overline{\im (\hD)},
$$
with the operators defined in $\mathcal{V}$ can be used to give a coordinate independent decomposition of the field operator. We would like to explore if this relation can be used as an evolution equation in the causal set. The region of interest in this discussion will be the causal diamond in $\mink^d$ but in principle it is applicable to a globally hyperbolic region of any general spacetime.  

Since the causal set is a discrete structure, a solution of the field equation (like the KG equation) $b$ is a column matrix with indices representing spacetime points. Since $b\in\kr (\hB)$ it also lies\footnote{in the discrete case the closure is irrelevant} in $\im (\hD)$ therefore we must be able to write it as
\be
\label{expb}
b_x=\sum_{k=1}^r a_k s_{kx}
\ee   
where $s_{kx}$ span $\im (\hD)$ and $r=\text{dim}(\im (\hD))$. We have used $x$ as an index here instead of a coordinate to emphasize that we are dealing with matrices and not functions. If can think of this equation as an initial value problem by assigning known values to some ``initial'' points in the causal set i.e., some values in the solution vector $b$ can be assigned as initial values. 

In the continuum, we assign initial data on either a Cauchy surface or an initial null hypersurface. In order to define any meaningful initial value problem in the causal set we must first find analogs of these 2 concepts. An inextendible antichain comes closest to the idea of a Cauchy surface, however as shown in Fig \ref{ivp} there can be relations in the causal set that do not intersect an inextendible antichain, making it porus. We could however consider artificially thickening this antichain by adding in more elements until we get rid of this porosity. The analog for the initial null hypersurface can be the layer of nearest neighbours of the minimal element $\mathbb{L}_1$ defined in Eq.\eqref{layer}. 
\bfig[!t]
\bsfig[b]{0.5\textwidth}
\includegraphics[width=\textwidth]{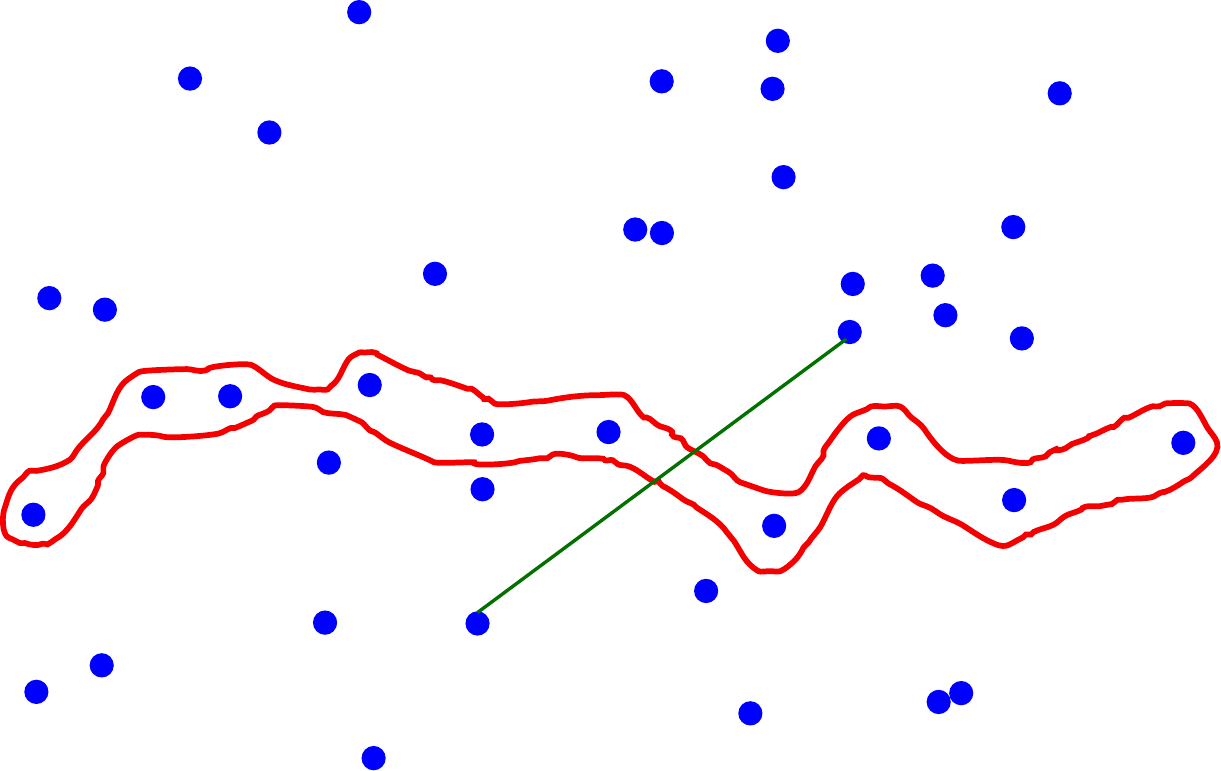}
\caption{}
\esfig
\bsfig[b]{0.5\textwidth}
\includegraphics[width=\textwidth]{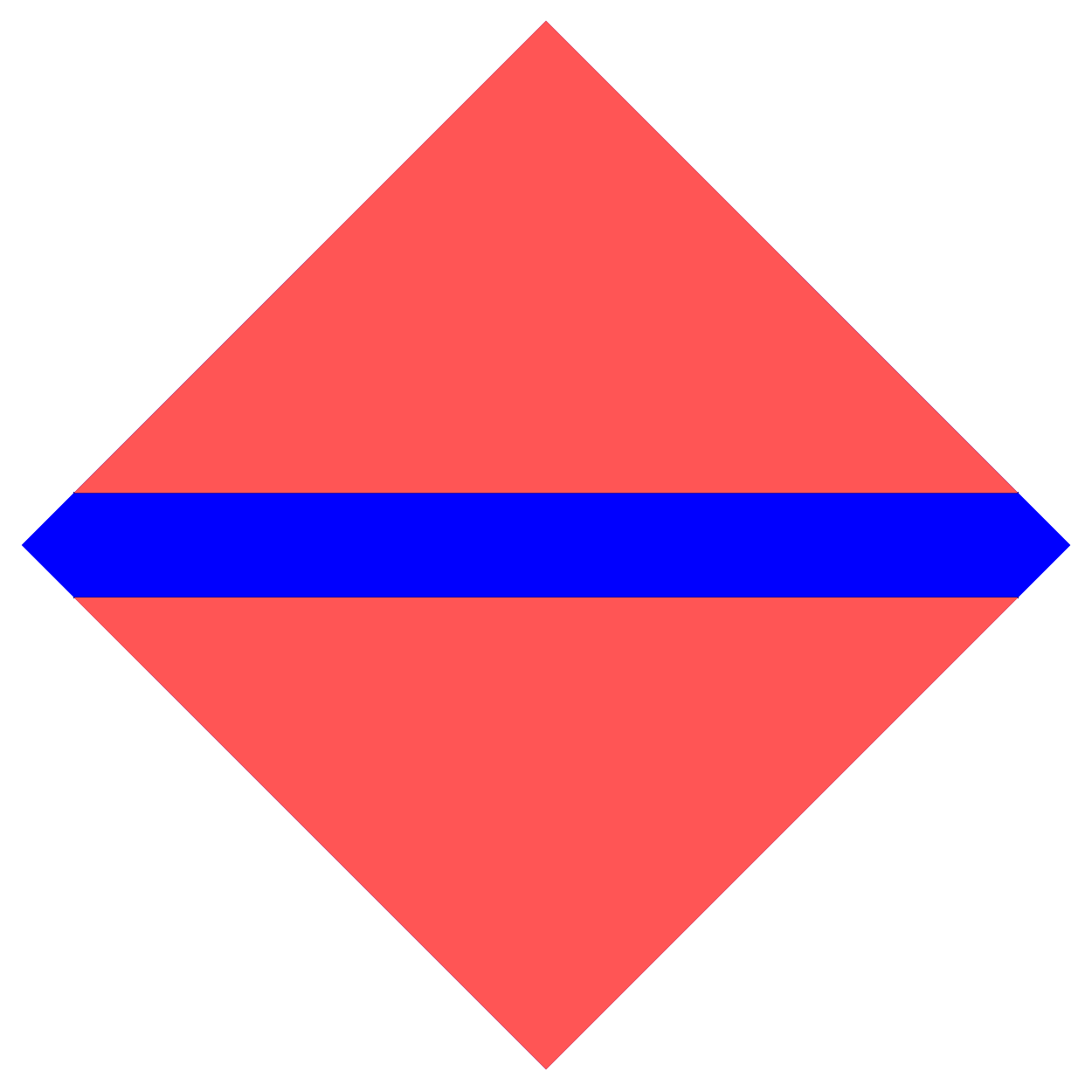}
\caption{}
\esfig
\caption{(a) An inextendible antichain is marked in red and an example of a relation that does not "intersect" the antichain is shown. (b) A thickened region around the center in $\diam_\ell^2$, all points in a sprinkling that fall in this region have to be initialized to solve the causal set initial value problem.}
\label{ivp} 
\efig
Consider an $N$-element causal set, then $b^\intercal=(b_1,b_2,....b_N)$. As already mentioned, which of these are to be given as initial data will depend on our choice. Here, let us assume that we initialize the first $i$ elements, then $b^\intercal=(b_1^*,b_2^*,...b_i^*,...b_N)$, where we used an asterix to denote known values. Now Eq.\eqref{expb} can be written as       
\be
\begin{pmatrix}
	b_1 \\
	\vdots \\
	b_N
\end{pmatrix}
=
\begin{pmatrix}
	s_{11} & s_{21} & \dots & s_{r1}\\
	\vdots & \vdots & \ddots & \vdots\\
	s_{1N} & s_{2N} & \dots & s_{rN}
\end{pmatrix}
\begin{pmatrix}
	a_1 \\
	\vdots \\
	a_r
\end{pmatrix}
\ee
This equation can be now split into 2 sets - one with the known initial data $(b_1^*,b_2^*,...b_i^*)$ and the other with the unknowns $(b_{i+1},...b_N)$. A natural way to proceed will be to use the first set of equations to determine the coefficients $a_k$ and then use these in the second set to determine the the remaining entries in $b$.

The first set contains $r$ unknowns and $i$ equations. The requirement that this set must give unique solutions for $(a_1,a_2,...a_r)$ already places strong constraints on $i$. We require that $i=r$ and that the rank of the now $r\times r$ square matrix 
$$
\begin{pmatrix}
s_{11} & s_{21} & \dots & s_{r1}\\
\vdots & \vdots & \ddots & \vdots\\
s_{1r} & s_{2r} & \dots & s_{rr}
\end{pmatrix}
$$  
be $r$. If either $i\neq r$ or the rank$\neq r$, we have either no solution or infinitely many solutions. 

The first condition is a physical requirement which says that the number of initial values that need to be given must be equal to the size of $\im (\hD)$. While it is hard to find this number exactly, we appeal to the ideas in chapter 4 for an estimate. In $\diam_\ell^2$ we expect the size of $\im (\hD)$ to be given in terms of the spatial volume and is\footnote{See chapter 4 for details.} $\sim\sqrt{N}$. Since we want to initialize $\sim\sqrt{N}$ points in the solution $b$, we will also estimate how thick the ``waist'' of diamond must be in order to accommodate these many points. Consider a region of height $h$ in the middle of $\diam_\ell^2$, the area of such a region is given by $h(1-h/2)$. Since the density $\rho=N/V=2N$ is fixed, we require that 
\be
h\bigg(1-\frac{h}{2}\bigg)\times\,2\,N=\sqrt{N}.
\ee 
The 2 solutions of this quadratic equation represent a horizontal strip along the $x$-axis and a vertical strip along the $t$-axis. The physically relevant solution is the horizontal one and it is given by
\be
h=\frac{1}{2\sqrt{N}}+\frac{1}{16N}+\mathcal{O}\bigg(\frac{1}{N^{3/2}}\bigg).
\ee
We see that in the continuum limit $N\rightarrow\infty\implies h\rightarrow0$ and we recover the spatial Cauchy surface. Therefore we see that in order to solve for $b$ as an initial value problem we will have to initialize all its components that lie in a region of thickness $h$ around the center of $\diam_\ell^2$.

A similar analysis can be carried out in the case where the initialization is to be done on the initial null surfaces. There we will need to estimate how many layers from the minimal element have to be initialized to get the required $\sqrt{N}$ points.

The second condition rank$\neq r$ may be violated when there are either \textit{singleton elements} or \textit{non-Hegelian pairs} in the causal set. A singleton element is one which is unrelated to any other element in the causal set. Since they are causally disconnected form the main causal set they can be assigned any value irrespective of the rest of the elements and should not contribute to the equations of motion in any way. Non-Hegelian pairs are a set of elements $\{e_1 , . . .  e_n\}$ that are mutually unrelated and have identical (strict) causal relations to the rest of the causal set. That is $(u \prec e_i \prec w)\implies (u \prec e_j \prec w),\,\forall  i, j = 1, . . . ,n$. As far as the causal set is concerned all such elements, having the same relation to the rest, can be treated as one element and hence assigning values to such elements more than once is redundant.

\section{Discussion}
\label{sec:discussion} 
Our simulations suggest that the CST 4d de Sitter SJ vacuum for all masses, while de Sitter invariant, is not equivalent to any of the Mottola-Allen $\alpha$-vacua. Moreover, contrary to the conclusions of \cite{Aslanbeigi:2013fga} which are based on a mode comparison calculation, we find that the CST SJ vacuum is well-defined both for $m=0$ and $m=m_{c}$ in 2d and 4d de Sitter. In 2d, where these two masses are equal, the CST SJ vacuum does not seem to be de Sitter invariant. In 4d on the other hand, as already mentioned above, the massless (as well as $m=m_{c}$) de Sitter CST SJ vacuum is de Sitter invariant. 

Our simulations are by necessity limited to a finite region of de Sitter, given by the IR cutoff $T$ and a finite density $\rho$. However, the convergence results we find are convincing and indicate that the CST SJ vacua will not change significantly as $T  \rightarrow \pi/2$ (the infinite volume limit). The convergence with density is especially good at larger proper times and distances. At smaller proper times and distances there is an approach to an asymptotic form, though not exact convergence. Put together these results suggest that the CST SJ vacuum converges to a continuum SJ vacuum with the two-point function approximately given by figures \ref{fig:4ddSfixedT} in Section \ref{sec:numerics}.    

Our results show a discrepancy with the results of \cite{Aslanbeigi:2013fga} in 4d de Sitter spacetime. One possibility, as with any numerical finding, is simply that our densities and $T$ values are not large enough to make the comparison. However, it seems an unlikely explanation given the apparent convergence we have found with density and $T$. We believe that it instead arises from the differences in how IR limits enter into the ab initio versus the mode comparison calculations\footnote{An analytic form has been found recently which gives results in agreement with the mode comparison method of~\cite{Aslanbeigi:2013fga}. This suggests that the discrepancy might only manifest in the CST SJ vacuum and not in the continuum.}. Thus, our work strongly suggests that the CST SJ state for 4d de Sitter is an altogether new de Sitter vacuum. 

The SJ vacuum in de Sitter spacetime clearly requires further study. The following ideas need to be explored further -
\begin{itemize} 
\item The symmetry properties of the SJ vacuum can be tested by carrying out numerical studies in distorted versions of the dS slab. One simple example would be to boost the spacelike boundaries of the slab that we consider (i.e., use slant boundaries instead of straight). Even though the SJ vacuum is covariant by construction, it could be the case that using finite regions of spacetime breaks symmetries in more subtle ways.
\item From the CST perspective, our results bring new light to questions of relevance to early universe phenomenology. Given that the continuum is an approximation to an underlying causal set, the natural vacuum for FSQFT on a 4d de Sitter-like causal set {\it is} the SJ vacuum we have obtained. Since this CST SJ vacuum differs markedly from the standard continuum 4d de Sitter vacua, it suggests that early universe phenomenology could be very different from what one expects from standard continuum calculations. 
\item A general understanding of the SJ vacuum from a representation theory perspective will help us understand the consequences of various choices made during the construction (the choice of inner product for example). Some work on this has been done by Fewster \cite{Fewster:2018ltq}. 
\item The proposal for the equations of motion made in section \ref{eom} needs to be studied in greater detail. First, it will improve our understanding of Eq. \ref{eq:kerim} as applied to causal sets - this is the starting point of the SJ construction. Second, it will shed some light on the relevance of the concept of an initial value problem, which is inherently non-covariant, to causal set theory. Third, it gives us an alternate, algebraic approach to the d'Alembertian approach. Fourth, it may help clarify the need for augmenting $\kr (i\hD)$ (or truncating $\im (i\hD)$ ) in the causal set.          
\end{itemize}
In the next chapter we study the entanglement properties of the CST SJ vacuum. In particular, we consider dS event horizons and restrict $\wsj$ to left and right Rindler wedges. This gives us an entangled vacuum state and allows us to study the behaviour of its entanglement entropy.    

\chapter{Spacetime Entanglement Entropy for deSitter Horizons}

Cosmological horizons {in de Sitter ($\deS$) spacetime} share several  key features with black hole horizons \cite{Bekenstein:1972tm, Bekenstein:1973ur,
  Bekenstein:1974ax}, as first suggested in \cite{gibbons}. Classically, both can be associated with a temperature, {as
well as} 
an entropy proportional to the horizon area based on a mathematical analogy {with} the laws of thermodynamics. Quantum
mechanically, observers outside both horizons can detect thermal radiation characterised by the horizon temperature.
However there are also key differences \cite{Davies:1988dma}. Most obvious is the fact that different observers in de
Sitter have different corresponding horizons. Moreover, the thermality of $\deS$  radiation is not reflected in the
stress energy tensor of the quantum state and is instead red-shifted by the expansion. {Despite this, the entropy-area
relationship is robust and can moreover be extended to all causal horizons \cite{Jacobson:2003wv}.} 

The interaction of matter fields with black hole horizons also exhibits thermodynamic features. As in the case of a
black body, incoming radiation is scattered into thermal radiation at around the black hole temperature
\cite{panangaden}.  In \cite{1983ee} Sorkin proposed that the dominant contribution to black hole entropy  can potentially come from the entanglement entropy (EE) of a non-gravitational
field. This EE was defined using the reduced density matrix of the exterior region. An
explicit calculation for a scalar field was carried out in \cite{Bombelli:1986rw} and seen to give rise to an
area law after imposing a UV cutoff. An area dependence arises naturally from complementarity and is an
important feature of EE. It has been shown to hold for a diverse range of quantum systems \cite{cardy}.

Numerous researchers have since studied the connection between {EE} and black hole entropy \cite{ted, solod, Emparan, Jacobson:2012yt}. In \cite{jacobson1994} Jacobson suggested that the ``species puzzle'' can be resolved
by showing that the renormalisation of the gravitational constant appearing in the Bekenstein-Hawking entropy is
similarly species dependent. In recent years, the idea of holographic EE has gained considerable ground starting with
the work of Ryu and Takayanagi \cite{Ryu2006}. The EE in $\deS$  was first calculated in \cite{Maldacena:2012xp} and shown to
exhibit the area law relation, both for a free massive field theory using the $\deS$  Euclidean vacuum, as well as for strongly coupled field theories with holographic duals (see also \cite{Dong:2018cuv}).

All these calculations of the EE use the density matrix specified on a partial Cauchy hypersurface ${\Sigma}$, with the entropy attributed to its spacetime domain of dependence $\mathcal{D}({\Sigma})$.
However, it is desirable to define the EE in a more covariant language, since horizons are intrinsically spacetime in character. In \cite{Sorkin:2012sn} Sorkin proposed a spacetime EE, which we term the Sorkin Spacetime Entanglement Entropy or SSEE for short, defined for a Gaussian free scalar field theory. The SSEE between a globally hyperbolic subregion $O$ in a globally hyperbolic compact\footnote{The compactness condition on $\mathcal{M}$ is important in defining the SSEE, since the domain of the integral operator $i\hat \Delta$ is the space of compactly supported functions, while its range includes functions that are not of compact support.} spacetime region $\mathcal{M}$ and its causal complement is given by  
\begin{equation}
\ssee=\sum_{\mu} \mu\ln |\mu|,  \quad W_O(x,x') v=i\mu\Delta_O(x,x') v,\indent \Delta_O v\neq 0,
\label{s4}
\end{equation}
where $W_O(x,x')$ and $ i\Delta_O(x,x')$ denote the restrictions to $O$ of the Wightman function
$W(x,x')=\langle0|\phi(x)\phi(x')|0\rangle$, and the Pauli-Jordan function $i\Delta(x,x')=[\phi(x),\phi(x')]$, respectively.
Recently this formula has been shown to be valid up to first order in perturbation theory for generic perturbations away from the free field Gaussian theory as well \cite{ngsee}; in this case the Gaussian free field correlation functions are replaced with their perturbation-corrected counterparts. 

\begin{figure}
 	\centering
 	\includegraphics[scale=0.85]{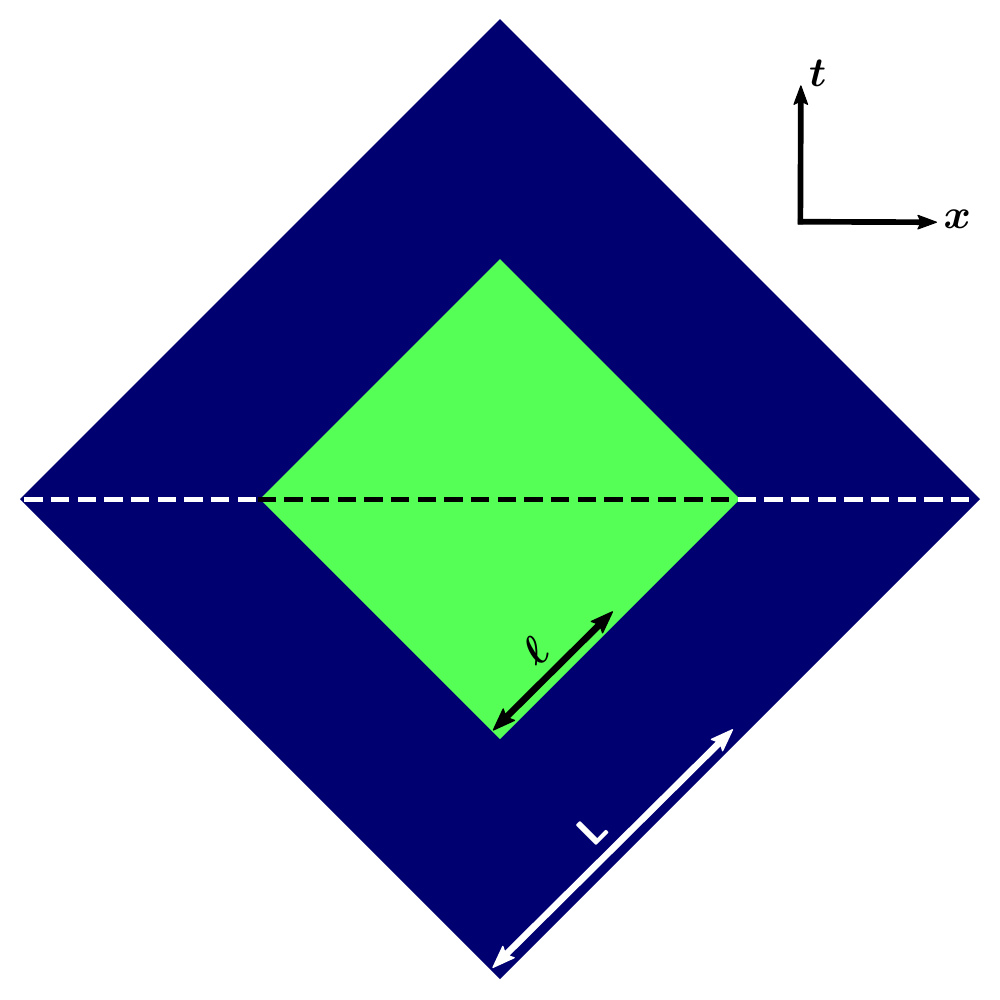}
 	\caption{The causal completion or domain of dependence $\mathcal{D}$ of two line segments, one contained within the other.}
 	\label{diamonds}
 \end{figure}

In \cite{Saravani:2013nwa}  $\ssee$  was calculated for nested  causal diamonds in $d=2$
continuum Minkowski spacetime $\mink^2$, $\diam_\ell^2 \subset \diam_L^2$, which are each the domain of dependence of nested  spatial intervals of lengths $2\sqrt{2}\ell$
and $2\sqrt{2}L$, respectively, as shown in  Figure~\ref{diamonds}. Rather than the Minkowski vacuum, the calculation
of \cite{Saravani:2013nwa} used the covariantly defined Sorkin-Johnston (SJ) vacuum for free scalar fields \cite{Johnston:2009fr,Sorkin:2017fcp}.
 As in other calculations of EE, $\ssee$ can be calculated in the continuum only after imposing a UV cutoff.  The SJ
 vacuum offers the choice of a covariant cutoff in the eigenspectrum of the Pauli-Jordan operator $i\hD$ (the SJ spectrum), which is
 at the heart of the SJ construction. Using this cutoff it was shown in  \cite{Saravani:2013nwa}  that the $\ssee$ satisfies the
 expected $d=2$ ``area'' law.

 Since a causal set which is approximated by a continuum spacetime comes with a built-in covariant spacetime cutoff,
 one might expect that the  SSEE for a causal set doesn't need further regularisation. While it is finite for a finite causal
 set, it was shown in \cite{Sorkin:2016pbz} that the SSEE in the causal set version of the  calculation in
 \cite{Saravani:2013nwa} obeys a spatial area law only after a suitable ``double truncation'' of the causal set SJ spectrum both in
 $\diam_L^2$ and in $\diam_\ell^2$.  Without this, the SSEE follows a spacetime volume law and thus violates
 complementarity.

\begin{figure}[!h]
	    \centering
	    \includegraphics[width=0.85\textwidth]{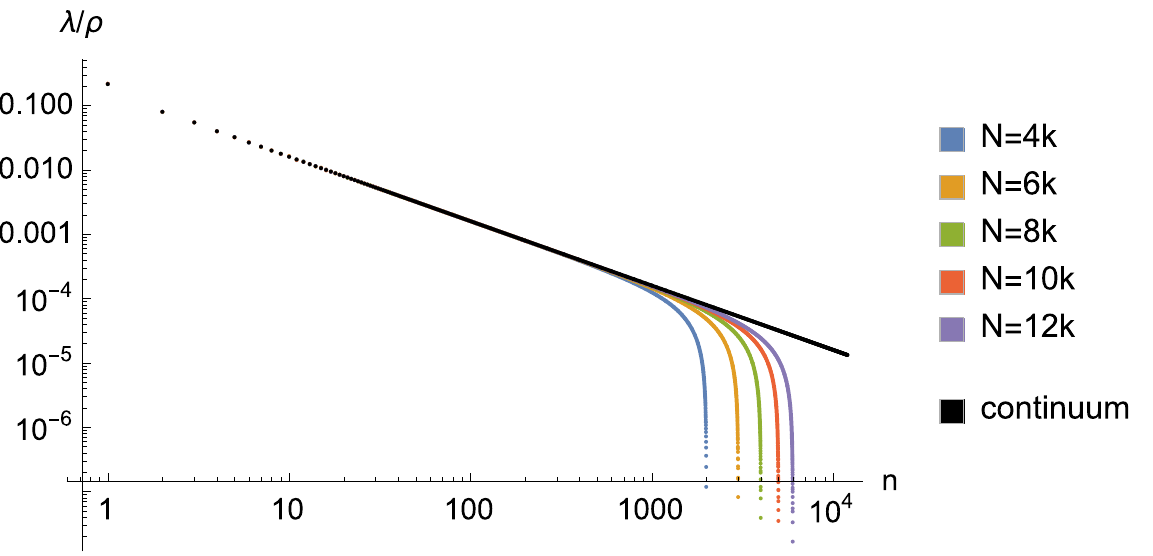}
	    \caption{Log-log plot of the normalised  SJ spectrum for the $2d$ causal diamond of side length $2L=1/\sqrt{2}$, for both the continuum as well as for causal sets of size  $N$.}
	    \label{fig:knee}
          \end{figure}

 The double truncation used in \cite{Sorkin:2016pbz} was motivated by comparing the SJ spectra of the 
 continuum with that of the causal set in $\diam_L^2$.  The latter possesses a characteristic ``knee" at which the
 eigenvalues dramatically drop to small but non-zero values (see Figure \ref{fig:knee}).  It is roughly around  this knee that the discrete and continuum spectra begin to disagree.  
Importantly, while the formula  for the SSEE  \eqref{s4} excludes {solutions with strictly zero eigenvalues}, it does
not exclude {those with} finite {\it near} zero eigenvalues, which characterise the post-knee {causal set} SJ spectrum. These
modes can be shown to contribute to large $\mu$ values in \eqref{s4} which then dominate the SSEE. If we include eigenfunctions $v$ that lie in the kernel of $i\Delta$ but not necessarily of $W$, this gives an infinite
contribution to $\ssee$, since the equation can only be satisfied for $\mu \rightarrow \infty$ (this is also discussed in 
\cite{Belenchia:2017cex}). Thus, in the causal set,
the contribution from a $v$ which is {\it almost} in the kernel, i.e., $||i \Delta v|| \approx 0$, lends itself to a very large (though finite) value of $\mu$, and hence to a much larger SSEE. Extending this work to gravitational horizons is of course very  important, not least because causal sets provide a covariant  UV cutoff, essential to the finiteness of EE.
 
We calculate the causal set SSEE for the $\deS$  horizon, for a conformally coupled, massless, free scalar field in dimensions $d=2,4$ and for $\mink^4$.  We find that, as for nested causal diamonds in $\mink^2$, the SSEE obeys an area law only
after a suitable double truncation, without which it follows a spacetime volume law.  The truncation scheme used in \cite{Sorkin:2016pbz} used the explicit analytic form of the
SJ spectrum in the $2$d flat spacetime causal diamond to motivate the truncation in the causal set SJ spectrum. We motivate the choice of truncation scheme for the $\deS$  causal set SJ spectrum by requiring the causal set SSEE to satisfy both an area law as well as complementarity. As we will see, satisfying both criteria is non-trivial.

Section \ref{prelims} provides a background for  our work. In Section \ref{areas}  {we begin with a discussion of
  area laws and complementarity. We define the
  two complementary  Rindler-like wedges in $\deS$  and the corresponding Bekenstein-Hawking  area law which we} might 
expect to recover from the SSEE. In Section \ref{ssee_in_cs} we set up the calculation of {the} SSEE in {a}  finite causal set. In Section
\ref{2dmink_review} we review the results of the calculation of SSEE  for nested causal diamonds in
$\mink^2$ \cite{Sorkin:2016pbz} {and} the {critical} role {played by the} double truncation {procedure} in obtaining the area law. In
Section \ref{gen_trunc} {we propose generalisations of the truncation scheme of  \cite{Sorkin:2016pbz} for general
  spacetimes}.

In Section \ref{results} we present the results of extensive numerical simulations for the causal set SSEE for $\deS_{2,4}$ horizons and for nested causal diamonds in $\mink^4$. Our investigations of different truncation schemes show that an area law compatible with the
Bekenstein-Hawking entropy of the horizon is not easy to satisfy. Complementarity is guaranteed in dS, up to Poisson fluctuations, by the fact that the Rindler-like wedges are identical in the continuum. In the Minkowski case it turns out to be non-trivial. We present a few truncation schemes and discuss their relative merits. In section \ref{acausality} we comment briefly on causality violation which is a consequence of the truncation procedure. We end with  open questions in Section \ref{discussion}. 

\section{Preliminaries} 
\label{prelims} 
\subsection{Complementary Regions in Global $\deS$  } 
\label{areas}

\begin{figure}[!h]
	    \centering
	    \includegraphics[width=0.6\textwidth]{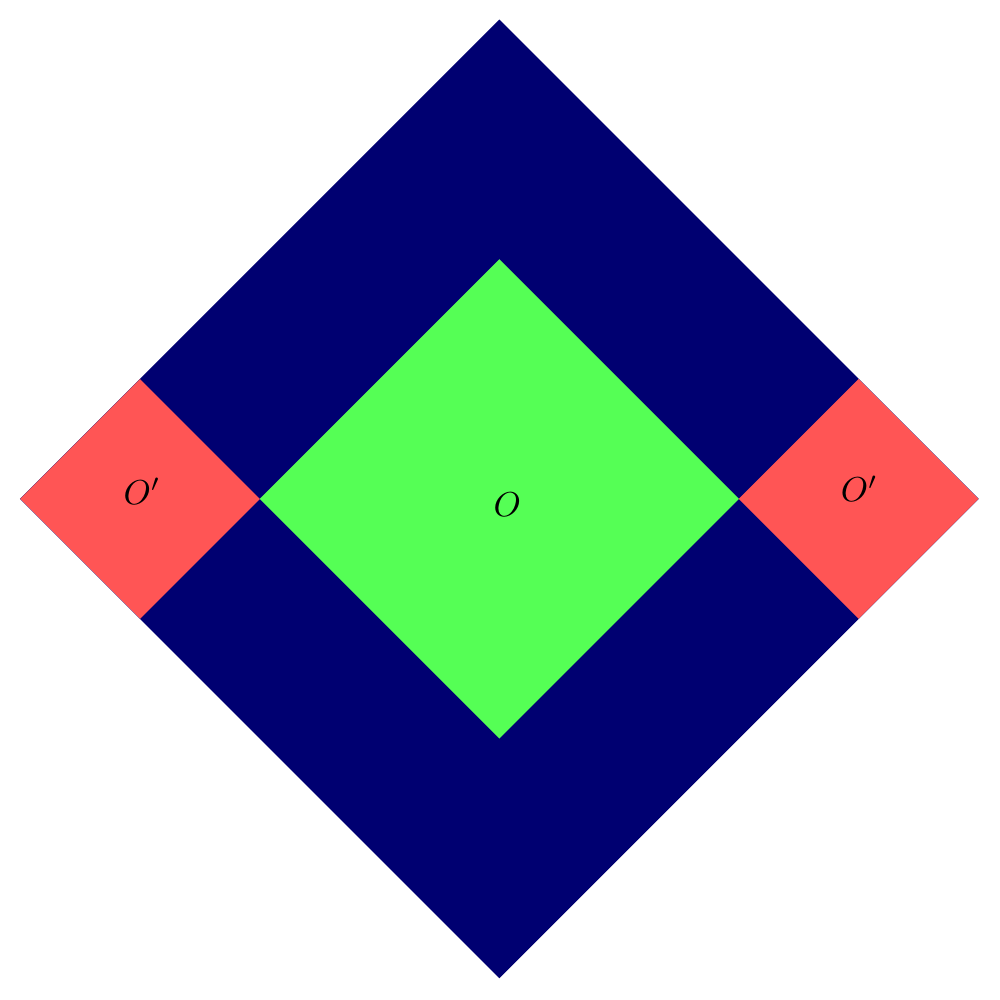}
	    \caption{A nested causal diamond $O$ and its complement $O'$ in $\mink^2$.}
	    \label{fig:2ddiamonds}
          \end{figure}

Let $(\mathcal{M},g)$ be a globally hyperbolic spacetime region and $O$ a globally hyperbolic subregion $O \subset \mathcal{M}$.
The  SSEE of
$O$ is defined with respect to its causal complement $O'$, where
$O' \subset O^c\subset \mathcal{M}$ such that $ \, \, x \in O'
\Leftrightarrow x$ is spacelike to $O$.  Since $(\mathcal{M},g) $ is globally hyperbolic, so is $O'$  and hence
the EE of $O'$ can also be defined with respect to $O$,   which is {\it its} causal complement.
$O$ and $O'$ are said to be {\sl complementary} to each
other, where  we now use the term ``complementarity'' to denote causal complementarity.  Figure \ref{fig:2ddiamonds}
shows an example of a smaller causal diamond $\diam^2_\ell$  nested inside a larger one $\diam_L^2$ in $\mink^2$. The complement $O'$  to $O \sim \diam_\ell^2$ is a union of two disconnected causal diamonds.  
In \cite{Saravani:2013nwa} and \cite{Sorkin:2016pbz} the SSEE of $O$ with respect to $O'$ was calculated in the
continuum and in the causal set, respectively.   Note that in the standard definition  the spatial 
complementary regions $\Sigma_O$ and $ \Sigma_{O'}$ are used to define EE, where $\Sigma_{O}$ denotes a partial Cauchy
hypersurface of the region $O$  in  $\Sigma$, a Cauchy hypersurface of $(\mathcal{M},g)$. However, because $O$ is globally
hyperbolic,  the ``information content'' of $\Sigma_{O}$ is the same as that of $O$.

A feature of bipartite EE is that it satisfies complementarity, i.e.,  that the EE of $O$ with respect to its complement
$O'$ is the same as that of $O'$ with respect to $O$. This in turn implies the area law since the two complementary
regions only share a spatial boundary separating them. The gross feature of this boundary is its  ``area'' or $d-2$
spatial volume, which means that the EE satisfies an area law.\footnote{The EE could also depend on the more detailed
  geometry of the boundary, but we will ignore this possibility in our work. See \cite{Sorkin:1999yj} for a discussion
  on this.} Conversely, a scaling of the EE with the spatial or spacetime volume of the region means that
complementarity is not satisfied, since in general the volumes of $O$ and $O'$  can be unequal.

\begin{figure}[!h]
	    \centering
	    \includegraphics[width=0.58\textwidth]{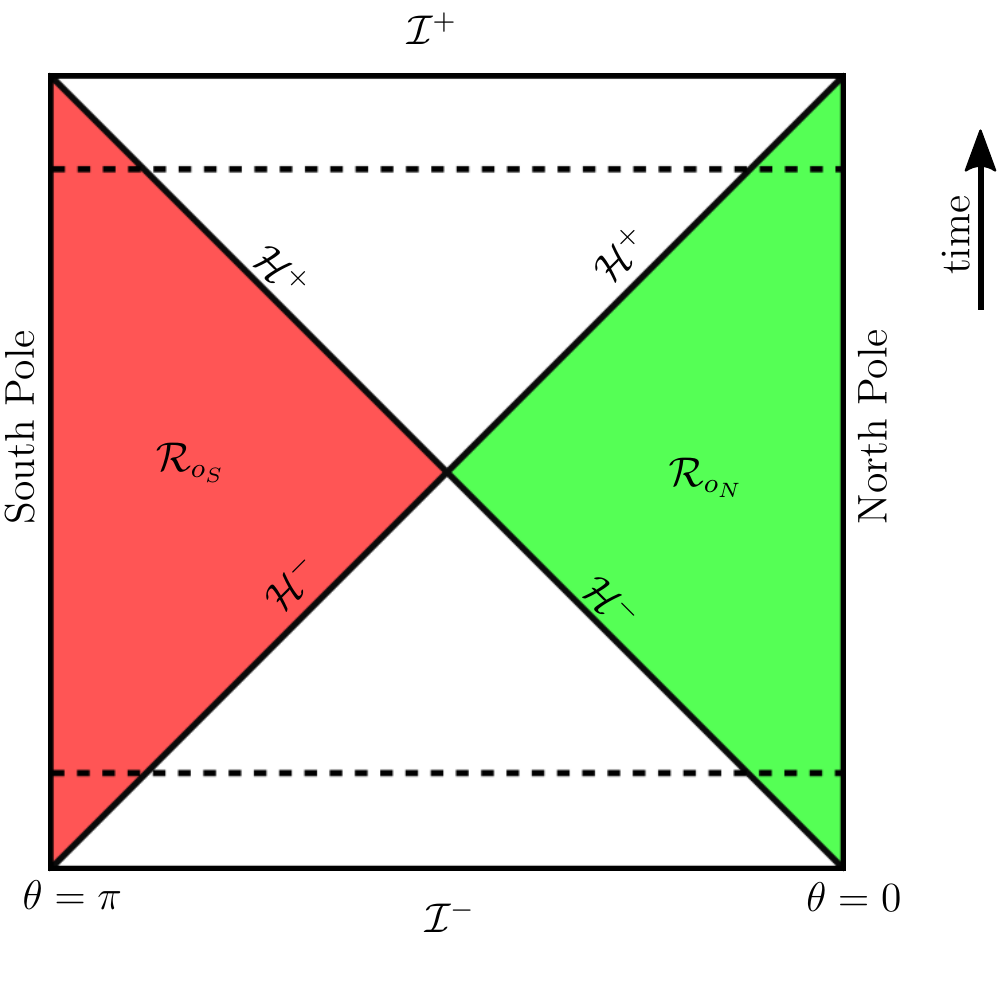}
	    \caption{The entangled Rindler-like wedges in $\deS$   corresponding to observers at the north and south pole. The dashed lines correspond to the boundaries of the slabs we consider.}
	    \label{fig:wedges}
          \end{figure}

In $\deS$, one wishes to calculate the SSEE between the two Rindler-like wedges which intersect at the
bifurcate horizon. As shown in the conformal diagram in Figure \ref{fig:wedges}, associated with any time-like observer $o$ is a future/past horizon $\cH_{\pm} = \partial (J^\pm(\gamma_o))$ where $\gamma_o$ is the world line of $o$. The Rindler-like wedge $\cR_o \equiv J^+ (\gamma_o)\cap  J^- (\gamma_o))$ has a boundary which intersects $\cH_+$ and $\cH_-$ at a bifurcate horizon, whose area is $A=4 \pi l^2$ in $4$d. 
Let us assume that the observer is at the south pole $o_S$. The Rindler-like wedge $\cR_{o_N}$ associated with its antipode at the north pole, $o_N$, is then the  complement of $\cR_{o_S}$. The SSEE we wish to calculate is from the entanglement between these two identical  Rindler-like wedges, which should therefore also satisfy complementarity. 

The EE for $\deS_2$ should have the same form as that for flat space. It contains a logarithmic UV cutoff dependence and in the case with two boundaries is given by \cite{solod, cardy}
\begin{equation}
  S= \frac{1}{3} \ln \biggl( \frac{l}{a} \biggr) + b,
  \label{2dentropy} 
\end{equation}
where $a$ is a  UV cutoff and $b$ a non-universal constant. For $\deS_4$, the entropy-area relation, which is the same as the Bekenstein-Hawking  entropy, is expected to be \cite{gibbons, Bousso2002}
\begin{equation} 
S = \frac {A}{4\ell_p^2}=\frac{ c^3}{4 G\hbar} A = \pi l^2\indent \,\,    (\textrm{for}\,\, G=\hbar=c=1).
    \label{sh}
\end{equation} 
We will compare our results with the causal set analog of this formula and ask if the causal set SSEE we find can account for the expected behaviour. It is understood that when \eqref{sh} is compared with the EE, the area of the entangled region in the EE is in units of the UV cutoff so that the entropy is dimensionless.

\subsection{Causal Set SSEE}
\label{ssee_in_cs}

Next, we set up the calculation of the horizon SSEE in a causal set approximated by  $\deS$. For a particular realisation $C$, we denote by $C_O$ the sub-causal set approximated by the subregion $O \subset \mathcal{M}$, and its cardinality by $\Ns$.

Since our calculations are numerical, we are limited by the size $N$ and hence to finite volumes $V$ of $\deS$ .  As in \cite{Surya:2018byh} we pick a symmetric ``slab'' of $\deS$  with  $\Tilde{T} \in [-T, T]$, so that in $4$d
\begin{equation}
  V_{slab}=\frac{4 \pi^2 l^4}{3} f(T), \quad f(T)=\tan T \biggl( \cos 2 T + 2 \biggr)\sec^2 T.
  \label{eq:slabvol}
\end{equation}
The causal sets we obtain from this sprinkling therefore have  finite $N$.

We can obtain the causal set analog for the Bekenstein-Hawking entropy \eqref{sh} in $d>2$ by translating the continuum area to the discrete one using a factor of the density $\rho$ as follows
\begin{equation}
 \langle \mathrm{S}^{(c)} \rangle= \rho^{\frac{2}{d}} \frac A4. 
  \label{eq: causalsetentropy}
\end{equation}
where we have set $\ell_p^2=1$.
Using the slab volume \eqref{eq:slabvol} in  $\deS_4$,  this translates into the discrete entropy 
\begin{equation}
\av{\mathrm{S}^{(c)}} = \frac{1}{2} \sqrt{\frac{3}{f(T)}} \sqrt{N}.  
    \label{4dcsentropy}
\end{equation}

In $d=2$, the discrete entropy is given by taking the  cutoff  $a$ in \eqref{2dentropy} to be  $\sqrt{\frac{1}{\rho}}=\sqrt{\frac VN}$. In $\deS_2$, therefore, the discrete entropy should take the universal $d=2$ form  
\begin{equation}
    \mathrm{S}^{(c)}  = \frac{1}{6} \ln N +b .
    \label{2dcsentropy}
  \end{equation}

We now review the definition of the SSEE  associated with a Gaussian scalar field on a causal set $C$.  We begin with the discrete 
Pauli-Jordan function $i\Delta_C(x,x')$ which is  the difference between the causal set retarded and advanced Green functions
$G_{R,A}(x,x')$, for $x,x'\in C$, defined using the order relations in $C$.  The SJ prescription then associates a
unique state, or  Wightman function $W_C(x,x')$ as the positive part of $i\Delta_C(x,x')$.
Next, consider any
causally convex subset $C_O \subset C$  and the  restrictions  $i\Delta_{C_O}(x,x'), W_{C_O}(x,x')$  of
$i\Delta_C(x,x')$ and $W_C(x,x')$ to $C_O$. Importantly, although $W_C(x,x')$ is a pure state, this is not true of
$W_{C_O}(x,x')$ which is not the positive part of $i\Delta_{C_O}(x,x')$. 
The simplest form\footnote{It is possible for example to have finite $N$ corrections to this formula, which vanish in
  the continuum limit.}  that the  causal set SSEE $\sseec$ takes is then 
\begin{equation}
  \sseec=\sum_{\mu} \mu \, \ln |\mu|,  \quad W_{C_O} \circ v=i\mu \Delta_{C_O}\circ v,\indent \Delta_{C_O} \circ v\neq 0,  
\label{s4c}
  \end{equation} 
where for $x \in C_O$,  $A_{C_O} \circ v (x) \equiv \sum_{x'\in C_0} A(x,x')v(x')$.   We will henceforth refer to the above equation and its continuum counterpart as the SSEE equation and $v$ and $\mu$ as {\sl generalised} eigenvectors and eigenvalues.  
Note  that in adapting the continuum formula to the causal set, we have retained the {strict} requirement that $v$ cannot lie in the kernel of $\Delta_{C_O}$.

An important aspect of the calculation of the EE is the introduction of a UV cutoff, which renders it finite. An unregulated quantum field in the continuum consists of infinitely many UV degrees of freedom which would yield an unbounded EE. When the regulated EE satisfies an area law, it is proportional to the spatial area of the entangled regions in units of the UV cutoff. This gives the scaling $S\propto a^{2-d}$ (for $d>2$), where $a$ is the UV cutoff in length dimensions and $d$ is the spacetime dimension.\footnote{For a heuristic argument for why the EE will in general be proportional to the spatial area of the entangling surface in units of the UV cutoff see \cite{Chandran:2015vuk}.} This is also true in the case of quantum theories with local interactions \cite{Eisert:2008ur}.

The causal set provides a natural cutoff length scale $a=\rho^{-1/d}\propto N^{-1/d}$. Based on this,  the
expected UV-dependence of the entanglement entropy $S$ of a scalar field in various dimensions  is as shown in the table
below, with \eqref{4dcsentropy} and \eqref{2dcsentropy} being special cases.  Since the leading area term in $d=2$ is a constant (as the spatial boundary of the entangling region is one or
two points), one also considers the subleading contribution $c_1\ln a$, where $c_1$ is a universal constant.\footnote{When
  the spatial boundary is a single point $c_1=-1/6$, and when it is two points  $c_1=-1/3$.} 
 \begin{table}[h]
\label{comtable}
\begin{centering}
\begin{tabular}{|c|c|c|}
\hline
 {\text{Spacetime Dimension}} & $S(a)$ &  $S(N)$\\ \hline
 		$d=2$ &  $c_1\ln a+\text{const}$ &  $-c_1\ln \sqrt{N}+\text{const}$\\ \hline
 		$d=3$ &  $1/a$ &  $N^{1/3}$ \\ \hline
 		$d=4$ &  $1/a^2$ &  $\sqrt{N}$ \\
 \hline
\end{tabular}
\caption{The dependence of the entropy $S$ on the UV length cutoff $a$ and causal set size $N$.}
\end{centering}
\end{table}

If instead  $S(N)\sim N$,  this means 
that $S$ satisfies a  spacetime {\it volume} rather than an area law. Interestingly, we will see that this is what commonly happens in the causal set when we compute the SSEE without any truncations. 

\subsection{Review of Causal Set SSEE for Nested Causal  Diamonds in $\mink^2$}
\label{2dmink_review}

In order to set the stage we review the results of  \cite{Sorkin:2016pbz} for causal sets approximated by the nested
causal diamonds $\diam_\ell^2\subset \diam_L^2\subset \mathbb{M}^2$, with side lengths $2L>2\ell$ (see Figures \ref{diamonds}
and \ref{fig:2ddiamonds}).   In this special case, one can make comparisons with the continuum results of
\cite{Saravani:2013nwa} which made use of the fact that the continuum SJ modes for $\diam_L^2$ are explicitly known \cite{Johnston:2010su}:
	\begin{eqnarray}
	f_k(u,v)=e^{iku} - e^{ikv} &|& k=\frac{n\pi}{L},\quad\, n\in\mathbb{Z^\pm}\nonumber\\
	g_k(u,v)=e^{iku} + e^{ikv}-2\cos kL &|& k\in \text{ker} (\tan(kL)-2kL)\nonumber\\ && \xrightarrow[]{m\rightarrow\infty}\bigg(m-\frac{1}{2}\bigg)\frac{\pi}{L}\approx\frac{m\pi}{L}\,,\,\,m\in\mathbb{Z^\pm}. \nonumber 
	\end{eqnarray}
In the UV limit, i.e.,  for large $k$,  the SJ spectrum takes the simple form $\lambda_k=\frac{L}{k}$ for both sets of
modes. {In this limit, these modes} moreover become linear combinations of the same plane waves, but are out of phase. Thus the UV part of the SJ spectrum for both modes can be characterised by an integer $n$, with $k=\frac{n \pi}{L}$.    

For a causal set approximated by $\diam_L^2$, the SJ spectrum was calculated using the $d=2$ causal set retarded Green function \cite{Sorkin:2016pbz}. Figure \ref{fig:knee} shows a comparison of the continuum and causal set SJ spectra for $\diam_L^2$ which match up to the characteristic ``knee'' mentioned in the Introduction. As the sprinkling density $\rho=N/V$ increases, the knee in the causal set SJ spectrum occurs at larger $k$ values.   

The continuum SSEE was calculated in \cite{Saravani:2013nwa} for $\diam^2_\ell \subset  \diam^2_L$ using a cutoff
$a=1/k_{\mx}$ and shown to satisfy the expected ``area'' law of \eqref{2dentropy}. 
However, the analogous  calculation in the causal set, yielded a volume law,  $\sseec \propto N$,  rather than an area law \cite{Sorkin:2016pbz}. 
        
This surprising feature, which is markedly different from the continuum result, can be traced to the shape of the causal
set SJ spectrum. As evident in Figure \ref{fig:knee}, beyond the knee the causal set SJ spectrum contains a large number
of near zero eigenvalues, which are absent in the continuum. In the causal set SSEE,  \eqref{s4c}, the generalised
eigenvector $v$ is required to lie outside the kernel of $\Delta_{C_O}$. However, because of the nature of the causal
set discretisation, fluctuations near the cutoff scale $\rho^{-1}$ can yield  eigenvectors that are ``almost'' but not
strictly in the kernel. This is true in general of the  discrete-continuum correspondence: as one  gets closer to the
spacetime discreteness scale $\rho^{-1}$, the relative fluctuations get larger. Thus, it is reasonable to expect that at
such scales, the causal set SJ spectrum will deviate significantly from the continuum. 

Indeed, as  shown in \cite{Sorkin:2016pbz}, { truncating}   the SJ spectrum of both $i\Delta_C$
as well as $i\Delta_{C_O}$ around {this}  knee has the effect of giving back the expected $d=2$ ``area law'' as in the continuum. Such a truncation can be motivated by appealing to the fact that the SJ modes $f_k,g_k$ are combinations of plane wave modes with wavenumber $k=\frac{2 \pi}{\nu}$, where $\nu$ is the wavelength. The causal set discreteness then gives a natural choice for the minimum wavelength $\nu_\mn \sim \rho^{-1/2}=2L/\sqrt{N}$. 
Since $k\sim \frac{ n \pi}{L}$ for large $k$, this suggests a truncation to retain as many modes as  $n_\mx \sim
\sqrt{N}$. The dimensionless causal set  {SJ eigenvalue} $\lambda^{cs}$ is related to the dimensionful continuum
{SJ eigenvalue}  $\lambda$ by $
\lambda^{cs}=\rho^{\frac{2}{d}}\lambda$. This means that $\lambda_\mn=\frac{L^2}{ \pi n_{max}}$ corresponds to $\lambda^{cs}_\mn \sim \frac{\sqrt{N}}{4 \pi}$ when we  
truncate the SJ spectrum with   $n_{\mx} \sim  \sqrt{N}$. 
     
The choice of $\sqrt{N}$ modes can also be justified by appealing to another aspect of the continuum picture. In the conventional  spatial way of understanding a quantum field and its EE, the field is quantised on a spatial Cauchy hypersurface and the contributions to the EE come from the field modes on that Cauchy hypersurface. In the continuum we do not expect to have {\it more} field modes contribute to the SSEE than in the spatial case, when we are working with domains of dependence. While the space of our solutions $\dim(\Delta)=N$ is larger, the space of independent solutions given by the $\mathrm{Im}(i \Delta)$ should remain the same as in the spatial picture. We expect the latter to be given in terms of the spatial volume (here the length) of the Cauchy hypersurface, so that the number of non-redundant solutions $\sim \sqrt{N}$ (where we have singled out the time-symmetric $t=0$ diameter of the causal diamond). Alternatively, since $\lambda$  has a dimension of $(\mathrm{length})^2$, we may assume that it is more generally the product of an IR scale and a UV scale,  $\lambda^{cs}_\mn\sim \rho^{\frac{1}{2}}L\sim \sqrt{N}$.  We  note  that since the number of the eigenvalues is $\sim N$, the reduction to $\sqrt{N}$ modes is a very non-trivial restriction.   

Thus, we have a {\sl number truncation} characterised by $n_{\mx}$ which gives the number of (largest in magnitude) eigenvalues that are retained, or alternatively, a {\sl magnitude truncation} $\lambda^{cs}_\mn$ which gives the minimum magnitude of the eigenvalues that are retained. These are related in $\diam_L^2$ by 
\begin{equation} 
  \lambda_\mn^{cs} = \frac{N}{4\pi n_{\mx}},
  \label{eq:magtrunc}
          \end{equation} 
but this relation may not hold more generally.

Once the truncation scheme is decided, the truncation needs to be implemented {\it twice}. This is the {\sl
  double truncation} followed in \cite{Sorkin:2016pbz}  which we describe in some detail below for the specific case
of $\diam_\ell^2 \subset \diam_L^2$. {Our notation is a little heavy for the sake of clarity, but we will shed it
  for simpler notation subsequently.} 

The first truncation $n_\mx\sim\sqrt{N}$ or $\lambda_\mn^{cs} \sim \frac{\sqrt{N}}{4\pi}$  is on the SJ  spectrum in
$\diam_L^2$, which therefore also truncates the {operator $i\Delta_L^t$ and therefore the} SJ  Wightman function
$W_{L}$ to $W_L^t$.   After the first truncation the region beyond the knee in the SJ spectrum of $i\Delta_L$ 
{is removed, leaving behind a residual} power law {behaviour}. {Next,} when $i\Delta_L^t$ is restricted to
$\diam_\ell^2$, i.e., $i\Delta_\ell^t(x,x')\equiv \Delta_L^t(x,x')|_\ell$ {the}  knee reappears once again in the
spectrum of the corresponding integral operator {$i\Delta_\ell^t$ in $\diam_\ell^2$}. Hence a second
truncation with $n^\ell_\mx\sim\sqrt{N_\ell}$ is necessary {in the spectrum of}  $i \Delta_\ell^t$, {which we
  denote by  $i \Delta_\ell^{tt}$}.  Finally, the
restriction $W_l^t\equiv W_L^t|_\ell$  of $W_L^t$ to $\diam_\ell^2$,  must then be further projected onto this smaller
(double) truncated subspace of the eigenbasis of $i \Delta_\ell^t${, to give us $W_l^{tt}$}. {Note that $i
  \Delta_\ell^{tt}$ is {\it not} the operator obtained after truncating the spectrum of the
  Pauli-Jordon operator $i
  \Delta_\ell$ in $\diam_\ell^2$.} 
    
The reappearance of the knee in the spectrum of $\mathrm{Im} (i \Delta_\ell^t)$ can be traced to the fact that the Pauli-Jordan integral {\it operators} $i\hat{\Delta}_L$ and $i\hat{\Delta}_\ell$ are defined over different integral domains and hence the  spectrum of  $i\hat{\Delta}_\ell$ cannot be obtained from a restriction of that of $i\hat{\Delta}_L$. This ``non-locality'' is an important feature of the SJ vacuum. 
{Most importantly}, without this second truncation, the full set of ``near zero'' elements in $\mathrm{Im} (i \Delta_\ell^t)$ is not removed and this gives rise to a too-large SSEE.  

This gives us a template for implementing the double truncation procedure more generally, for any $C_O \subset C$. Thus, the first truncation is performed on  the SJ spectrum of $i\hat{\Delta}_C$ to give the  truncated operator  $i\hat{\Delta}^t_C$, and its associated Wightman function $W_C^t(x,x')$. The restriction of the truncated Pauli-Jordan function $i\Delta^{t}_{C_O}(x,x')= i{\Delta}^t_C(x,x')|_{C_O}$ corresponds to an  operator $i\hat{\Delta}^{t}_{C_O}$ in $C_O$, i.e., for $x \in C_O$,  $i\hat{\Delta}^{t}_{C_O}\circ v(x) = i \sum_{x'\in C_O}\Delta^{t}_{C_O}(x,x')v(x')$. 
    
{The second truncation is then performed on the spectrum of $i\hat{\Delta}^{t}_{C_O}$, which yields the operator
$i\hat{\Delta}_{C_O}^{tt}$, as well as the  projection  $W_{C_O}^{tt}$ of the restriction $W_C^t(x,x')|_{C_O}$ to this
second truncated eigenbasis. Thus} the double truncated SSEE {version}  of \eqref{s4c} is 
\begin{equation} 
\sseec=\sum_{\mu} \mu \, \ln |\mu|,  \quad W_{C_O}^{tt} \circ v=i\mu \Delta_{C_O}^{tt}\circ v,\indent
\Delta_{C_O}^{tt} \circ v\neq 0,  \label{s4ct}
\end{equation}
where $tt$ denotes the double truncation procedure described above. {We now  drop the ``${tt}$''
  superscript for simplicity of notation, and refer to the spectrum as either truncated or untruncated.}  

\subsection{Generalised Truncation Schemes}
\label{gen_trunc}
	
In what follows, we discuss ways in which to generalise the truncation procedure in $\diam_L^2$ without explicit
knowledge of the SJ spectrum in the continuum. Out of the several possibilities, the ones that would closely mimic the
continuum would be those that satisfy  an area law relation for the SSEE compatible with the Bekenstein-Hawking entropy, as well as complementarity.

We consider causal sets obtained by sprinkling into the finite volume ``slab''  between $[-T, T]$ in $\deS$. As discussed in Section \ref{areas}, the south and north Rindler-like wedges $\cR_{o_S}$ and $\cR_{o_N}$ are complementary to each other, and intersect only at the equator of the $t=0$ $3$-sphere. In the $\deS$  slab, these regions have hyper-hexagonal boundaries (see Figure \ref{fig:wedges}). Both $\cR_{o_S}$ and $\cR_{o_N}$ are also time-symmetric and are the domains of dependence of time-symmetric $t=0$ Cauchy slices, which are the Southern and Northern hemispheres of the 3-sphere, respectively.

Since the SJ spectrum of the hyper-hexagon is not known, we cannot resort to comparisons with the continuum as in the
nested diamonds. {In the course of our investigations we tried a very large number of different truncation schemes. Of these
  we} focus on two particular schemes which we think are physically motivated and simple to generalise and
  moreover, give an SSEE which satisfies an area law.  

The first choice we make is an estimation of the number truncation $n_{max}$, inspired by the {nested $d=2$
  diamonds, where}   $n_{\mx}=\sqrt{N}$ for each of the two sets of modes.  { This was motivated by the fact that the 
number of modes should be proportional to the spatial volume of a Cauchy hypersurface.} A natural generalisation of this is
\begin{equation}
  n_{\mx} = \mathrm{\con} {N}^{\frac{d-1}{d}}.
  \label{eq:numtrunc} 
\end{equation}  
Note that the identification of the spatial volume is neither uniquely nor
covariantly defined, and hence  there is no unique choice of $\con$; in particular one can deform the Cauchy hypersurface to one that has arbitrarily small spatial volume.   
In our investigations of the nested causal diamonds in $\mink^4$ (Section \ref{mink.app}) we experimented with several values of $\con$, including that corresponding to the volume of the time symmetric slice. In the de Sitter case, this latter factor turns out to be too large, leading to too small a truncation. As a result, we focus here only on values of $\con$ which give the most reasonable results, i.e., $\con=1,2$.

Our second choice is a new truncation scheme, which we dub the linear  scheme.  Since the SJ spectrum in the causal
set is a power law and therefore linear in the log-log plot, up to a characteristic knee,  it is reasonable to truncate the spectrum at the
point where this linear regime ends.  This requires an estimation of  the end of the linear regime in the log-log
plot. One method is to use the change in the slope of the logarithms of the
data. We implement this in the following way: First, the logarithms of each $n^{th}$ eigenvalue are taken along with the logarithm of its label $n$. Then the
slope of the line between each nearest neighbor pair\footnote{An alternative method, which yields similar results, is
  to take the slopes of more than than a pair (say, every 50) of nearest eigenvalues.} of data points is computed. Due
to fluctuations in the causal set data, these slopes also fluctuate when going from one pair's slope to the next, even
in the (approximate) power law regime. In order to smooth out these fluctuations, the slopes are binned and
averaged. Then, a smooth interpolating function is fit to the averaged slopes. This interpolating function can then be
used to track the drop in the slope and set the truncation number or magnitude. The region of nearly constant
  (negative) slope $m$ is first  identified, and the estimation of the knee corresponds to a drop to a more negative
  $m'$. A choice is then made of the fractional drop $\delta=\frac{m-m'}{m}$ to obtain the knee.
We take the magnitude of the eigenvalue {at this estimated knee} as our magnitude
truncation, or the number $n_{\mx}$ (rounded to the nearest integer) at which this happens as our number
truncation. We have explored various choices of
  $\delta$, also allowing it to be different in the slab and in the Rindler-like wedge. 

An advantage of the linear truncation over the {generalised number}   truncation \eqref{eq:numtrunc} is
that it is covariantly defined, without appealing to any features of a Cauchy surface and {the associated}
ambiguity of {choosing} a proportionality constant. There is of course the fine tuning that comes with the
  choice of $\delta$ and the hope is to be able to find a suitable range of values, 
as much as the quality of data allows.

\begin{figure}[!h]
	\centering
	\includegraphics[width=0.65\textwidth]{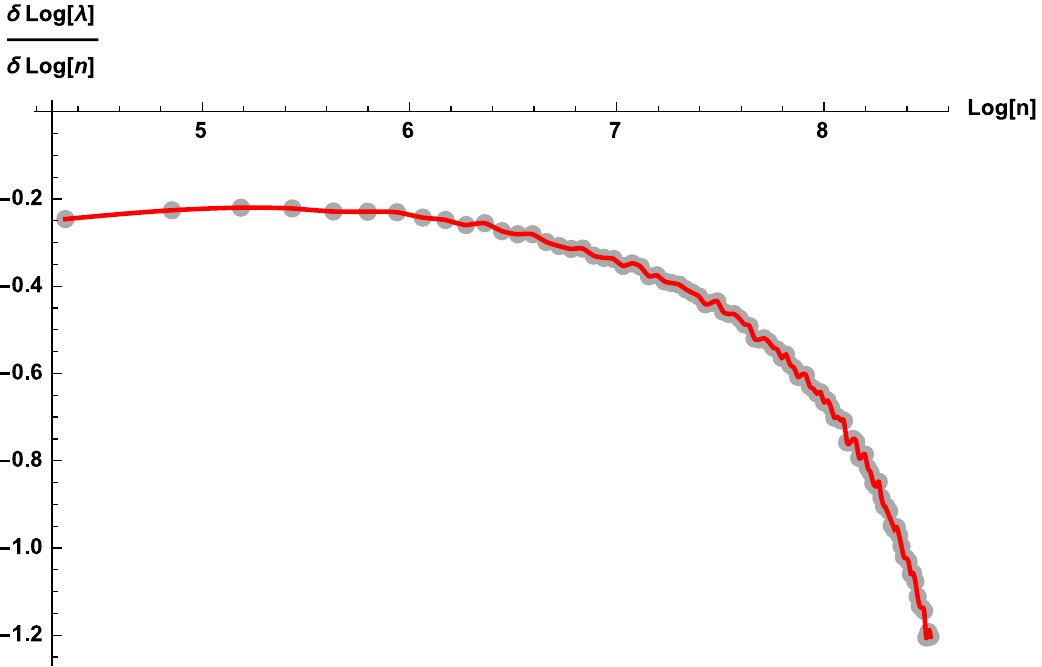}
	\caption{Slopes of the log-log SJ spectrum in $\deS_4$. Data points are binned averages and the curve is an interpolating function fit to the data.}
	\label{fig:slopes}
\end{figure}  

In all the cases we study, the numerically generated causal set SJ spectrum can additionally be used to estimate the power law behaviour of $\lambda^{cs}$ as a function of $n$. Rescaling the spectrum by $\rho^{-\frac{2}{d}}$ collapses the data in the linear regime, so that 
\begin{equation}
   \rho^{-2/d}\lambda^{cs}=\frac{b}{n^a},
  \label{eq:csmagtrunc}
\end{equation}
where the exponent $a$ and the constant $b$ can be determined empirically. For $\diam_L^2$, for example, $a=1$ and
$b=1/(4 \pi)$. For the $\deS_2, \deS_4 $ slabs and associated Rindler-like wedges 
these values are given in the following table{, where the slab height has been chosen to be $T=1.2$.}

\begin{table}[h]
\begin{centering}
\begin{tabular}{|c|c|c|}
\hline
 {Spacetime} & Slab &  Wedge\\ \hline
 		$\deS_2$ &  $a\sim 1,\,b\sim 1.68$ &  $a\sim 1,\,b\sim 0.26$\\ \hline
 		$\deS_4$ &  $a\sim 0.25,\,b\sim 2.16$ &  $a\sim 0.36,\,b\sim 0.78$ \\ \hline
\end{tabular}
\caption{The values of parameters $a$ and $b$ in \eqref{eq:csmagtrunc} determined from the spectrum of $i\Delta$ in the regions considered.}
\end{centering}
\end{table}
This also allows us to translate $n_\mx$ (picked either by the number or linear truncation method) into a magnitude truncation $\lambda^{cs}_\mn$ for this choice of $T$. We have not however studied the effect of varying $T$ on the parameters $a$ and $b$ and whether or not the spectrum can be collapsed to a universal form.
  
\section{Results}
\label{results}   

The simulations presented here were performed using Mathematica on an HP Z-8 workstation with 320GB pooled RAM. For larger $N$ values, {a significant fraction} of this pooled memory was used in the simulation, when all the trials for fixed $N$ are parallelised. The results presented here are the culmination of {extensive exploration of various truncation schemes, including certain magnitude truncations not described in Section \ref{gen_trunc}. Here we only present results from the two described in Section \ref{gen_trunc} and for  choices of $\con$ and $\delta$ which best satisfy the criterion of an area law compatible with the Bekenstein-Hawking entropy. As mentioned in Section \ref{gen_trunc}, the two Rindler-like wedges are indentical and hence complementarity should be automatically satisfied. In our investigations, we also calculated  the SSEE for a causal diamond in the slab spacetime, whose complement is not necessarily a causal diamond, but for this work we present results only from  the Rindler-like wedges, since these are of most interest for the $\deS$ horizons. For completeness we also present the results for the nested causal diamonds in $\mink^4$.

\subsection{$\deS_2$}
\label{ds2}

In $\deS_2$,  the two complementary regions $\cR_{o_S}$ and $\cR_{o_N}$  are each conformal to causal diamonds.  The
simulation results we present are for a slab of $\deS_2$ of height $T=1.2$ into which we sprinkle  causal sets with
sizes $\langle N \rangle$ ranging from $2000$ to $16000$.

Figure \ref{fig:ds2notrunc} shows the
dependence of the SSEE with $N$ without truncating the SJ spectrum.  The SSEE clearly scales
linearly with $N$ and therefore obeys a spacetime volume law, as in the case of the $d=2$ nested diamonds \cite{Sorkin:2016pbz}. 
 \begin{figure}[!htp]
 	\centering
   \includegraphics[width=0.6\textwidth]{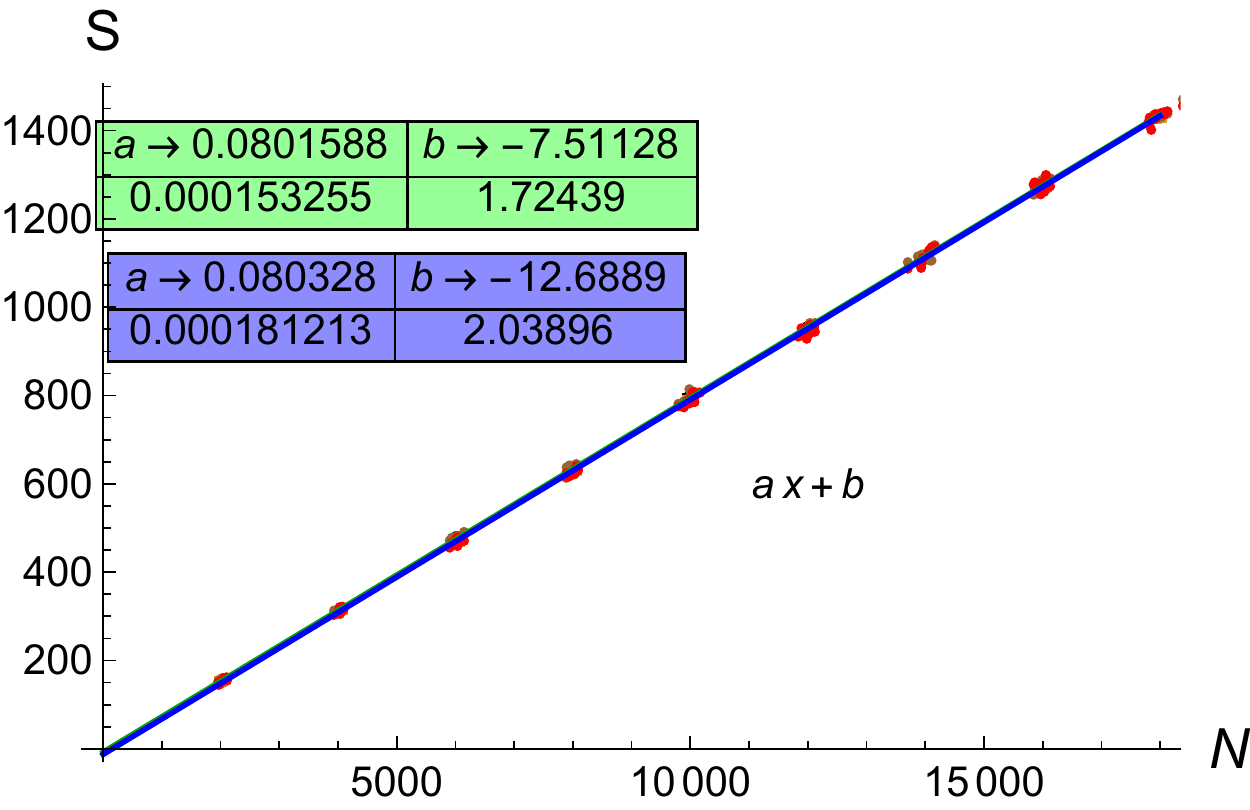}
  \caption{Untruncated SSEE vs. $N$ in the $\deS_2$ slab of height $T=1.2$ for the two
    Rindler-like wedges $\cR_{o_S}$ and $\cR_{o_N}$ (shown in green and blue).  The best fits are shown.}
   \label{fig:ds2notrunc}
  \end{figure}

 Next we implement the truncation schemes discussed in Section \ref{gen_trunc} for the SJ spectrum for a causal  region
 of cardinality $N_s$. For each  $\langle N \rangle$, we run  $10$ simulations for the number truncation while we run
$5$ simulations for the linear truncation. For the  latter,  the estimation of the linear regime is done for the SJ  spectrum
in the slab as well as for the SJ spectrum in the Rindler-like wedges.

For the number truncation \eqref{eq:numtrunc} we  work with {$\con$} values of $1$ and $2$, the latter being the analogue of the $2$d causal diamond truncation.\footnote{Note that while $n_\mx$ in our review of the $2$d causal diamond denoted the maximum number of modes of each family of $f$ and $g$ eigenfunctions, here we refer to it as the total number of eigenfunctions irrespective of  degeneracies. Hence the two-fold degeneracy of the $2$d diamond amounts to keeping a total of $2\sqrt{N}$ eigenvalues in the terminology henceforth.}
  
For the linear truncation scheme, different values of $\delta$ were explored. We found that the one most compatible with the area law is $\delta \sim 0.1$} in both the slab and the Rindler-like wedge.

\begin{figure}[!htp]
  \begin{subfigure}[b]{0.5\textwidth}
  \includegraphics[width=\textwidth]{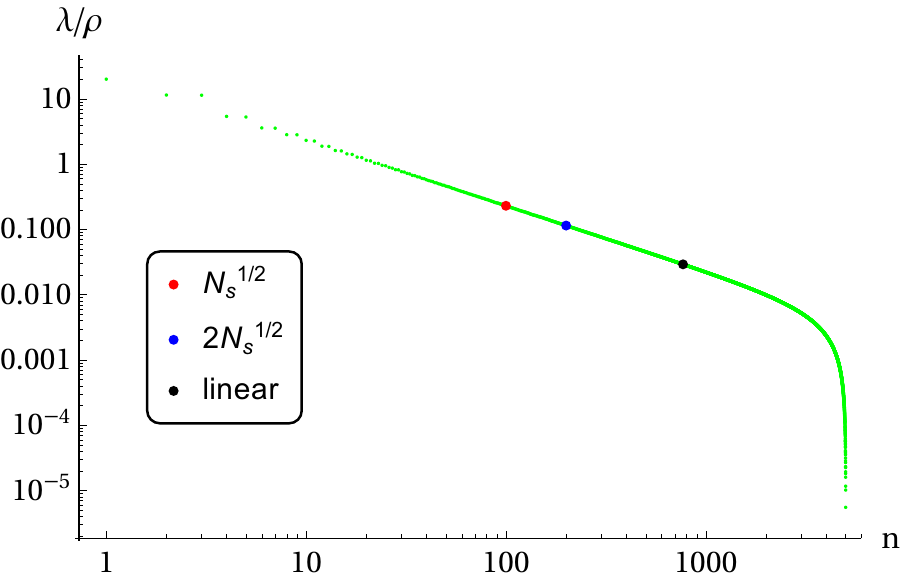}
  \caption{}
  \label{ds2spectrummarked}
  \end{subfigure}
  \begin{subfigure}[b]{0.5\textwidth}
  \includegraphics[width=\textwidth]{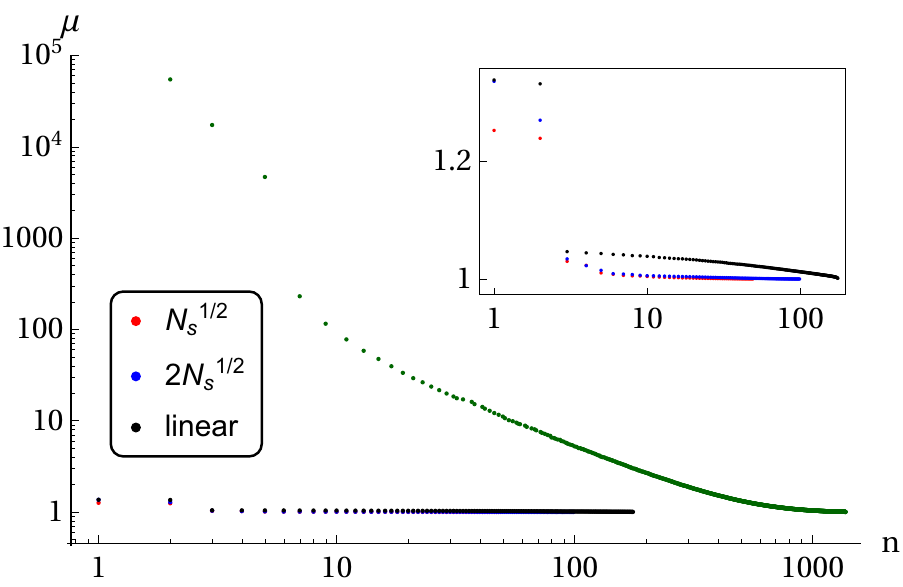}
  \caption{}
  \label{ds2 generalized spectrum}
  \end{subfigure}
  \caption{ (a) The SJ spectrum for an $N=10^4$ causal set sprinkled into the $\deS_2$ slab. Three
    different truncations choices are marked. 
    (b) The spectrum for the SSEE \eqref{s4c} with and without these truncations.}
  \end{figure}

In Figure \ref{ds2spectrummarked} we show the log-log plot of the untruncated causal
 set SJ spectrum of the $\deS_2$ slab,  with these three choices for truncation marked. All three clearly lie in the
 linear regime, with the linear truncation being the closest to the knee. In Figure \ref{ds2 generalized spectrum} we show the
log-log spectrum of the generalised eigenvalue equation before and after truncation. What is striking is the drastic
reduction in not only the number but also the magnitude of the eigenvalues.  It is this feature that seems to make it
possible to recover an area law after truncation. 

\begin{figure}[!h]
  \begin{subfigure}[b]{\textwidth}
  \includegraphics[width=0.5\textwidth]{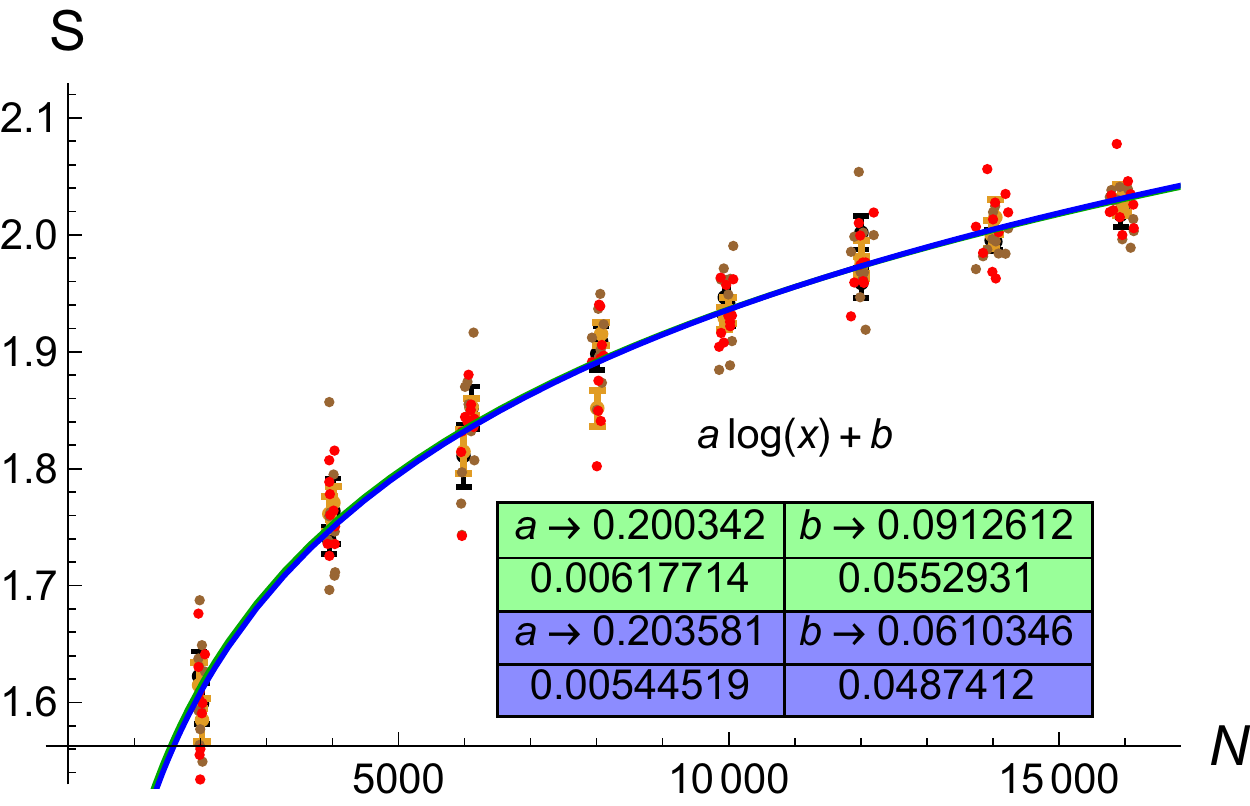}
  \includegraphics[width=0.5\textwidth]{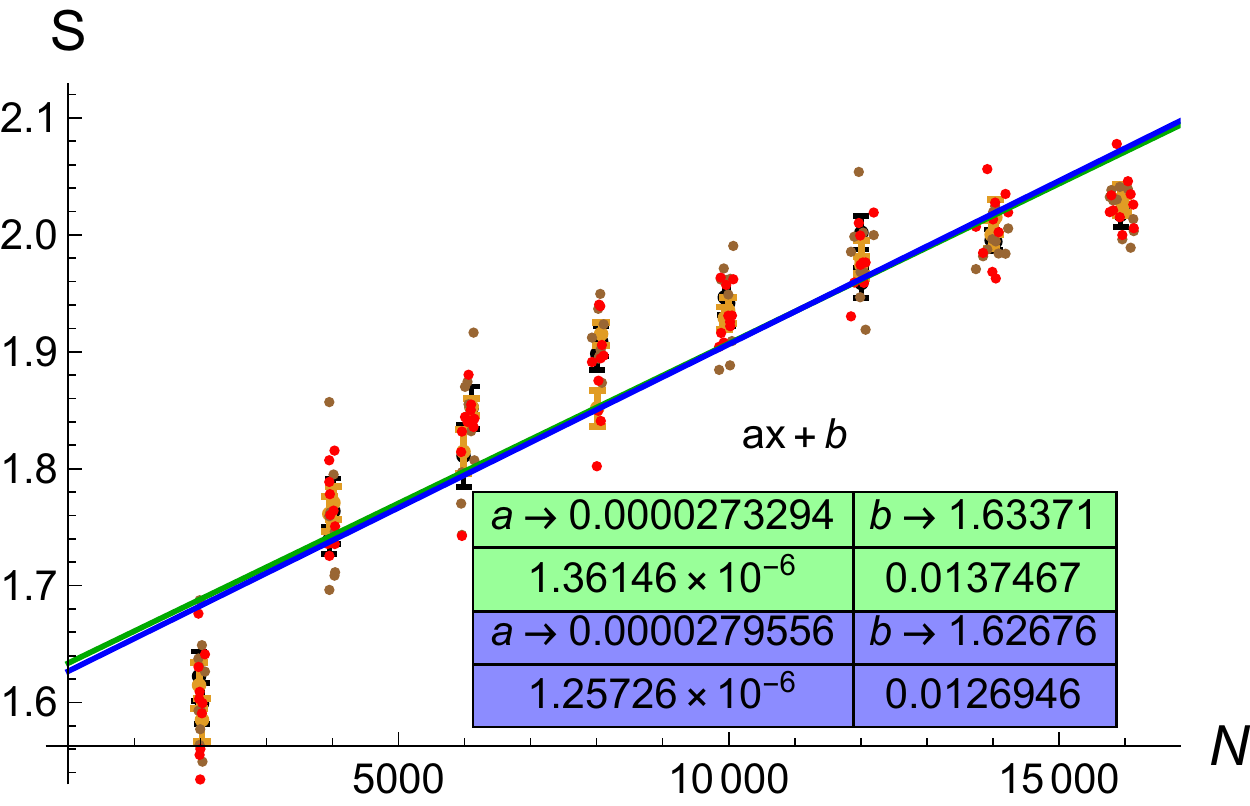}
  \caption{Number truncation with $n_\mx=N_s^{1/2}$}
  \label{ds2num}
  \end{subfigure}
\begin{subfigure}[b]{\textwidth}
	\includegraphics[width=0.5\textwidth]{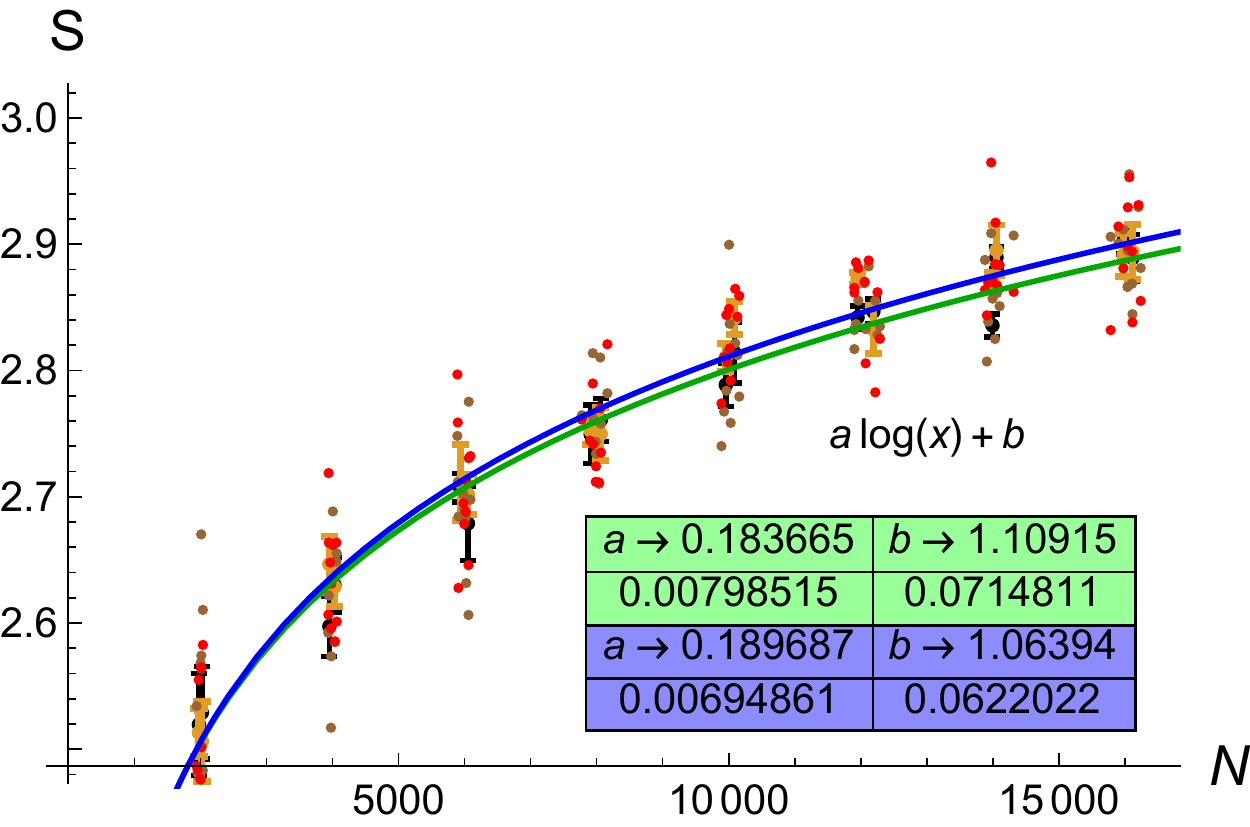}
	\includegraphics[width=0.5\textwidth]{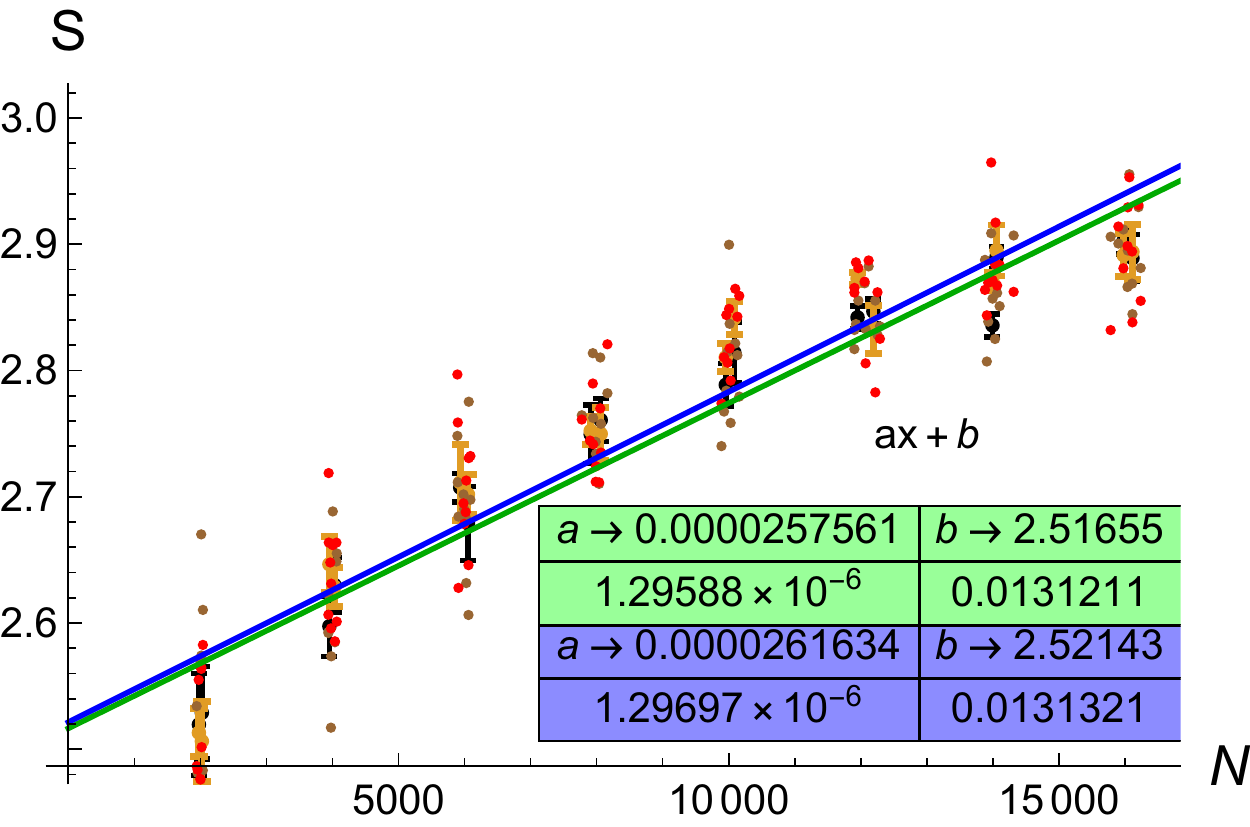}
	\caption{Number truncation with $n_\mx=2N_s^{1/2}$}
	\label{ds2fac2}
\end{subfigure}
  \begin{subfigure}[b]{\textwidth}
  \includegraphics[width=0.5\textwidth]{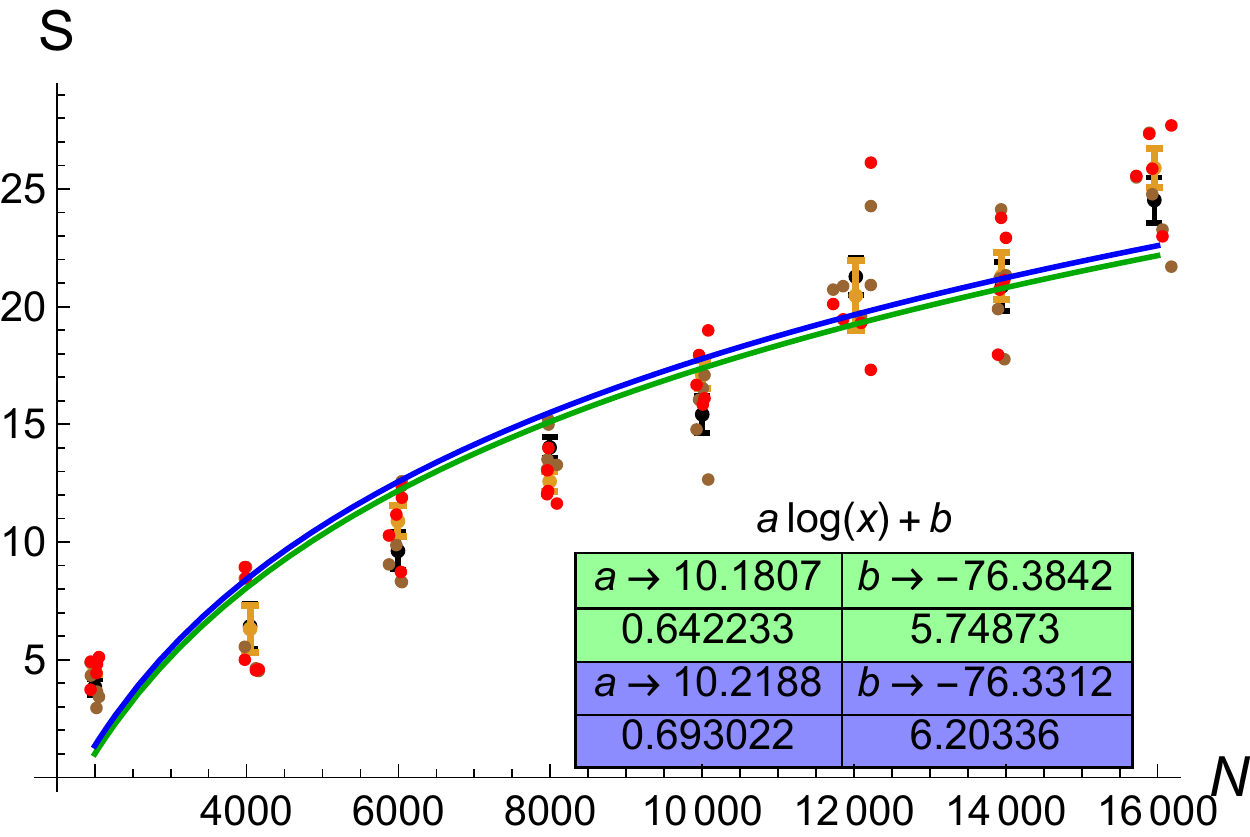}
  \includegraphics[width=0.5\textwidth]{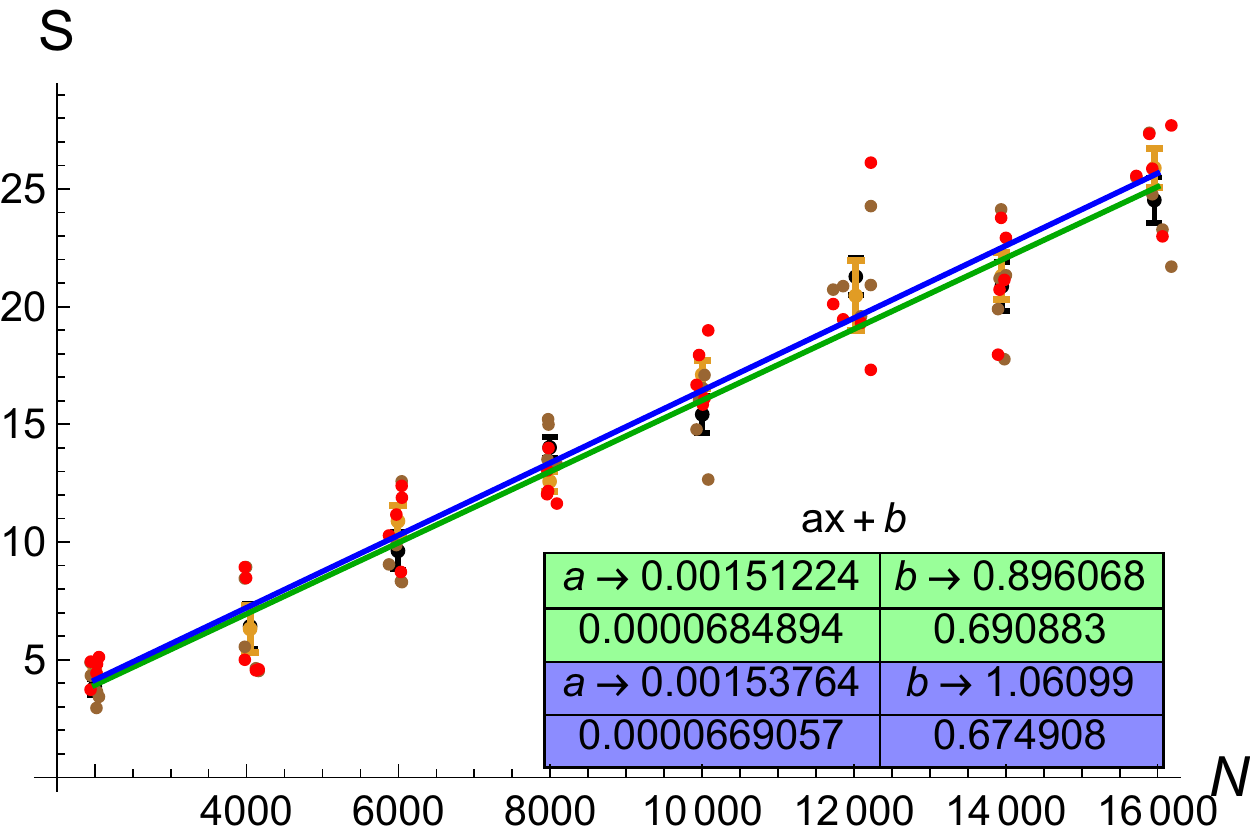}
  \caption{Linear truncation}
  \label{ds2linear}
  \end{subfigure}
  \caption{SSEE vs. $N$ with three different choices of truncation in $\deS_2$.  The green and blue represent the data for
    the two Rindler-like wedges. A comparison of the two fits $a\ln{x}+b$ and $ax+b$ is shown on the left and the right
    for each choice of truncation.}
  \label{ds2 entropy trunc}
\end{figure}

  Finally, in Figure \ref{ds2 entropy trunc} we show the SSEE calculated using the above three truncations for both
  $\cR_{o_S}$ and $\cR_{o_N}$.    For each truncation, on the left we show the fit to the logarithmic behaviour 
  \begin{equation}
    \sseec = a \ln N+ b,  
  \end{equation}
  (where the expected value of $a$ is  $1/6$) and on the right,  the fit to the volume behaviour $a N+ b$. {The
    errors in the best fit parameters are given below these values.}
  The fit and corresponding uncertainities are found using the least square method.
  We see in all three cases that the data has a high degree of scatter, which is also the case for the $d=2$ nested diamonds
\cite{Sorkin:2016pbz}  and seems to be a characteristic of $d=2$.  All cases are reasonably consistent with an area law,
but the linear truncation is  surprisingly more consistent with a volume law. All cases 
also satisfy complementarity up to Poisson fluctuations. 

From these results we conclude that the truncation that is closest to the expected EE values is the
choice $n_\mx=2\sqrt{N}$, with $a$ and $b$ values given in Figure \ref{ds2fac2}. This case gives  $a\sim0.18$ which is
closest to the expected value of $1/6$. 
 
\subsection{$\deS_4$}
 \label{ds4}
 
 The $\deS_4$ slab is again taken to have height $T=1.2$.  We consider causal set sprinklings with $\langle N \rangle$
 ranging from $2000$ to $16000$.
 
\begin{figure}[!h]
 	\centering
   \includegraphics[width=0.6\textwidth]{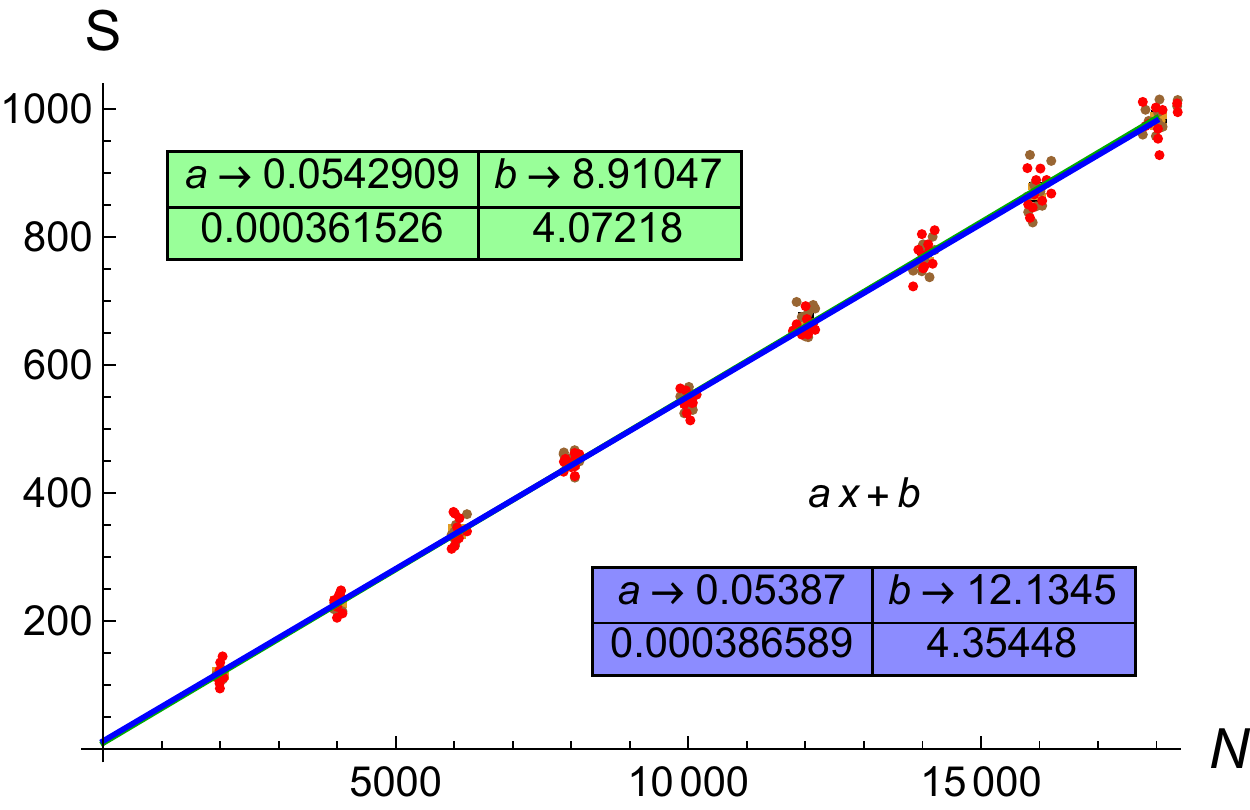}
  \caption{Untruncated SSEE vs. $N$ in a $\deS_4$ slab of height $T=1.2$ for the two
    Rindler-like wedges $\cR_{o_S}$ and $\cR_{o_N}$ (shown in green and blue).  The best fits are shown.}
   \label{ds4notrunc}
 \end{figure}

 In Figure \ref{ds4notrunc} we show the untruncated SSEE which again clearly scales linearly with $N$ and
 therefore obeys a spacetime volume law.

We present results for three choices of truncations, the number truncations $n_\mx=N^{\frac{3}{4}},\,2 N^{\frac{3}{4}}$ and the linear truncation.  We run $10$ simulations for each fixed $\langle N \rangle$ for both types of truncation. 
As in $\deS_2$,  for the  latter,  the estimation of the linear regime is done for the SJ  spectrum in the slab as well as for the SJ spectrum in the Rindler-like wedges. We {find that a choice of $\delta \simeq 0.15$ for both the slab and the Rindler-like wedge spectrum gives the best results. 

Note that since the knee is fairly sharp in the  log-log spectrum,  even a seemingly large tolerance does not lead to  very drastic
changes in the spectrum but does change the SSEE so obtained.  
\begin{figure}[!h]
	\begin{subfigure}[b]{0.5\textwidth}
		\includegraphics[width=\textwidth]{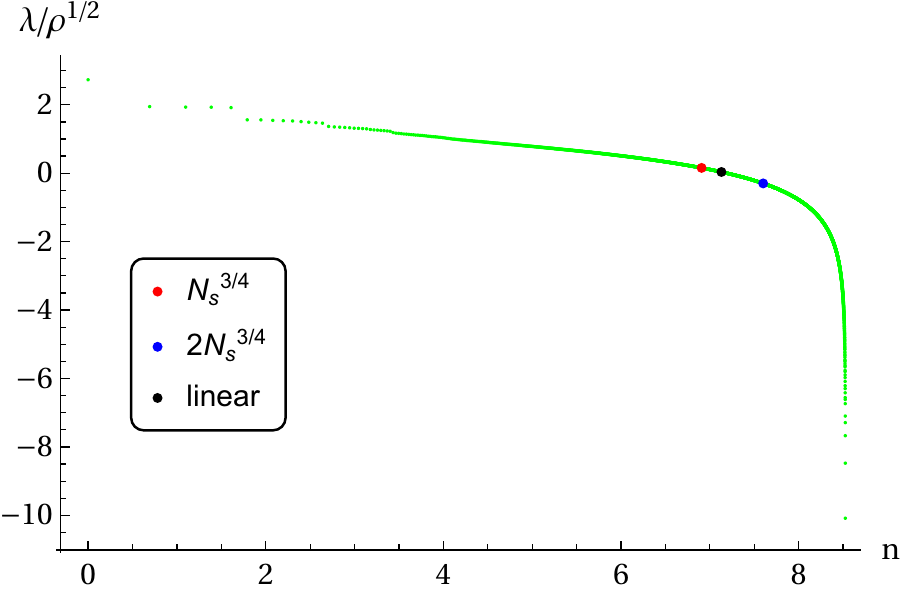}
		\caption{}
		\label{ds4spectrum}
	\end{subfigure}
	\begin{subfigure}[b]{0.5\textwidth}
		\includegraphics[width=\textwidth]{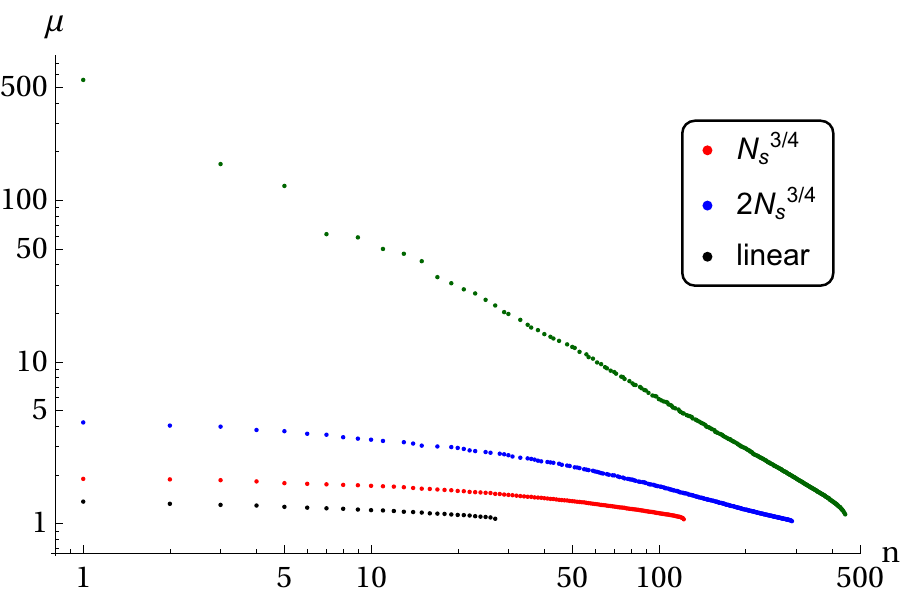}
		\caption{}
		\label{dS4 generalized spectrum}
	\end{subfigure}
	\caption{For $N=10k$ in $\deS_4$, (a) is the spectrum of $i\Delta$ with different truncations marked, and (b) is a plot of the
		solutions of the generalised equation \eqref{s4c} for these truncations.}
\end{figure}

In Figure \ref{ds4spectrum} we show the causal set SJ spectrum in the $\deS_4$ slab with the different truncations marked, and in Figure
 \ref{dS4 generalized spectrum} we show the generalised spectrum with and without these truncations. Again the broad features are the same -- the truncations lie in the linear regime of the SJ
 spectrum and drastically cut down both the magnitude and number of the generalised spectrum. However, the differences in the generalised spectrum post truncation are more
 marked in $d=4$.
 
  The area law for the SSEE for $\deS_4$ is given by
 \begin{equation}
 \ssee = a \sqrt{N} + b. 
 \end{equation} 
 With $T=1.2$, we expect $a \sim 0.17$ for the  Bekenstein-Hawking entropy for the $\deS$  horizon
 \eqref{4dcsentropy}.  In Figure \ref{ds4entropytrunc} we show the results for the SSEE. We note
 that interestingly, the scatter is far less than in $d=2$, which makes the results easier to interpret.
 
 In all  cases we see that an area law and complementarity are compatible with the data, but that the linear truncation scheme is also compatible with a volume law. The  number truncations  $n_\mx=N^{\frac{3}{4}}$, and  $n_\mx=2 N^{\frac{3}{4}}$ give a much more convincing area law.
 
 A comparison with the Bekenstein-Hawking formula however shows that {\it all} the values of $a$ in Figure \ref{ds4num} exceed
 the expected value of $a=0.17$. So even though an area law is obtained, it is one that contains more entropy than
 expected. For $n_\mx=N^{\frac{3}{4}}$ the SSEE  is about $5$ times larger, and the difference is even greater for $n_\mx=2 N^{\frac{3}{4}}$.
  \begin{figure}[!h]
 	\begin{subfigure}[b]{\textwidth}
 		\includegraphics[width=0.5\textwidth]{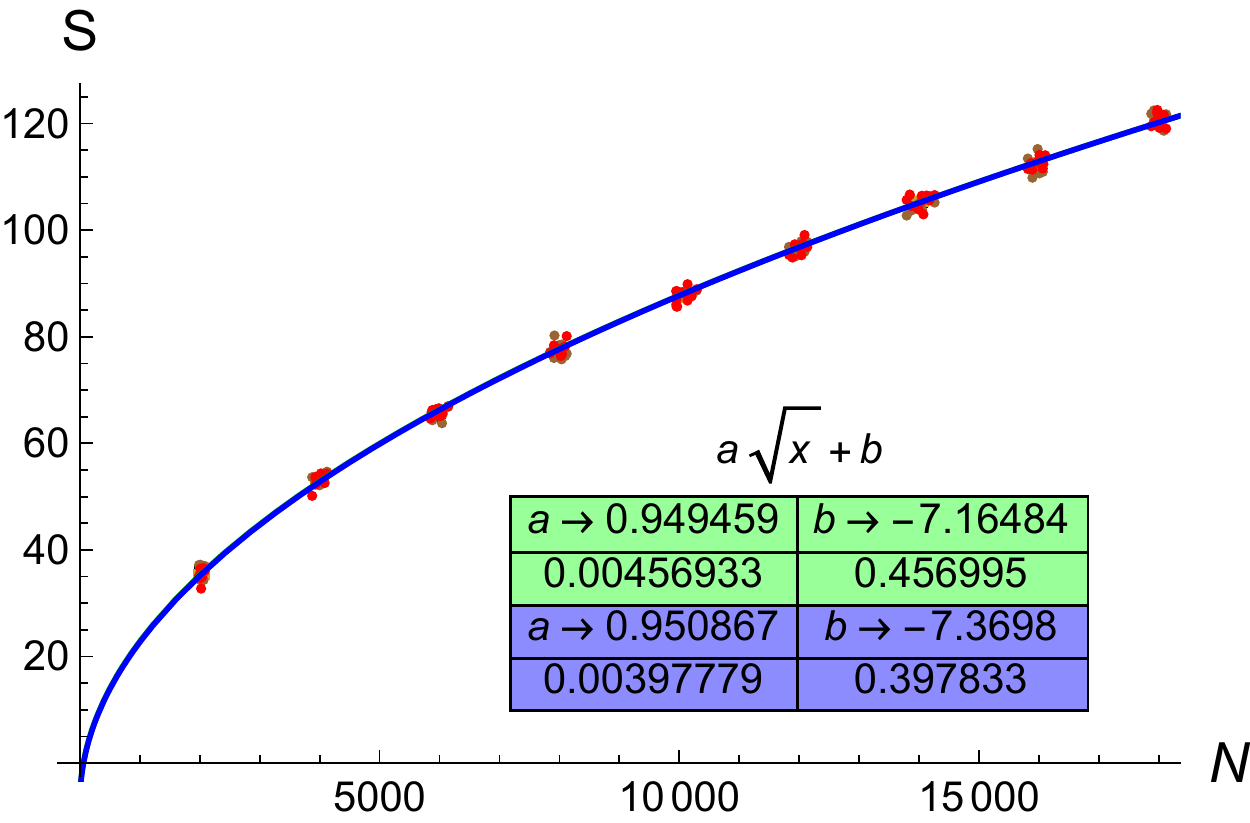}
 		\includegraphics[width=0.5\textwidth]{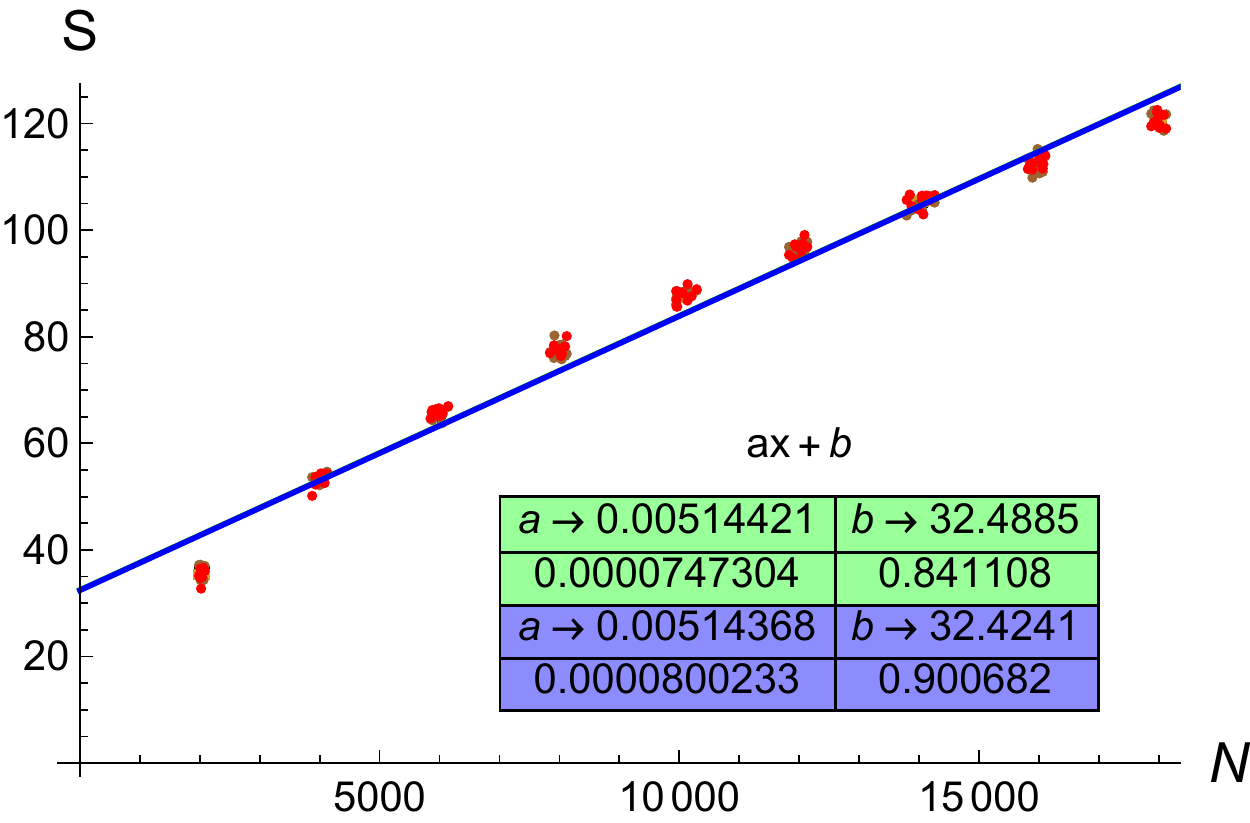}
 		\caption{Number truncation with $n_\mx=N_s^{3/4}$}
 		\label{ds4num}
 	\end{subfigure}
 	\begin{subfigure}[b]{\textwidth}
 		\includegraphics[width=0.5\textwidth]{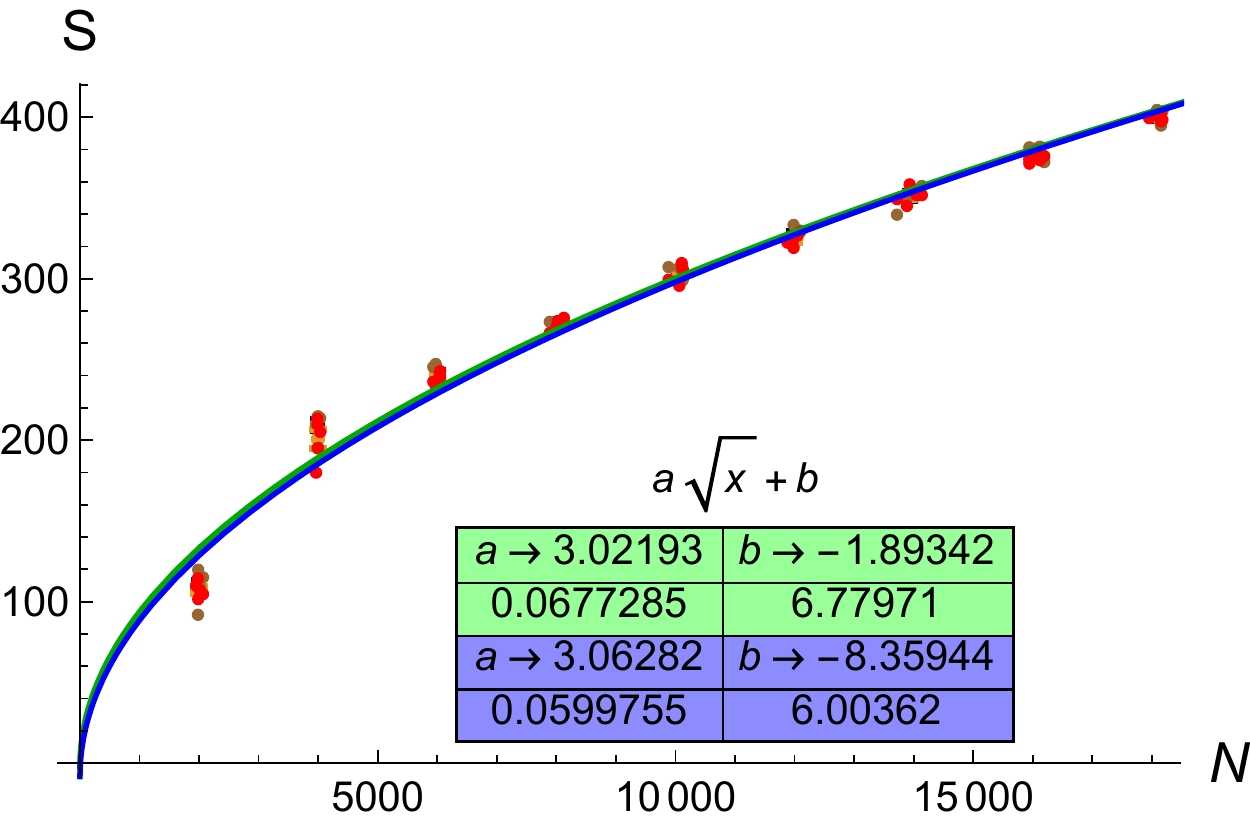}
 		\includegraphics[width=0.5\textwidth]{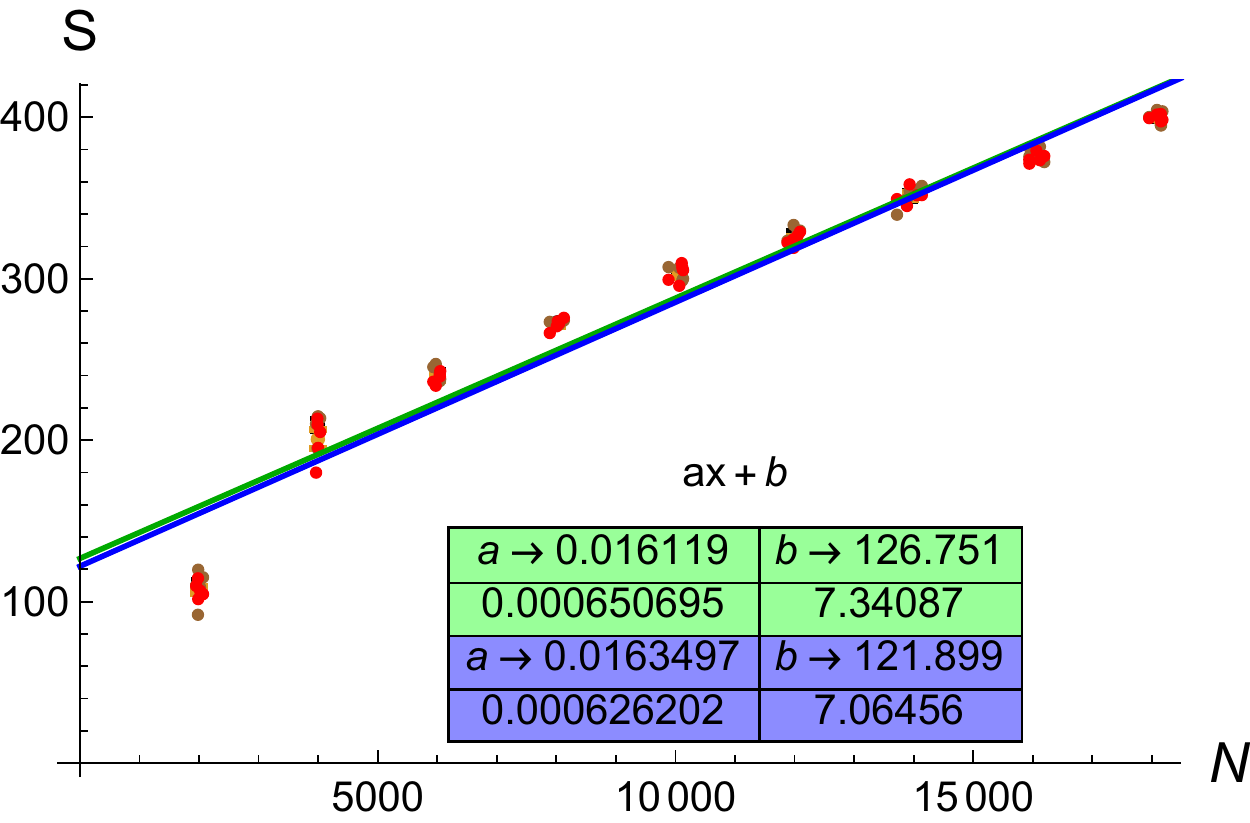}
 		\caption{Number truncation with $n_\mx=2 N_s^{3/4}$}
 		\label{ds4num2}
 	\end{subfigure}
  \end{figure}
\begin{figure}[!h]
\ContinuedFloat
 	\begin{subfigure}[b]{\textwidth}
 		\includegraphics[width=0.5\textwidth]{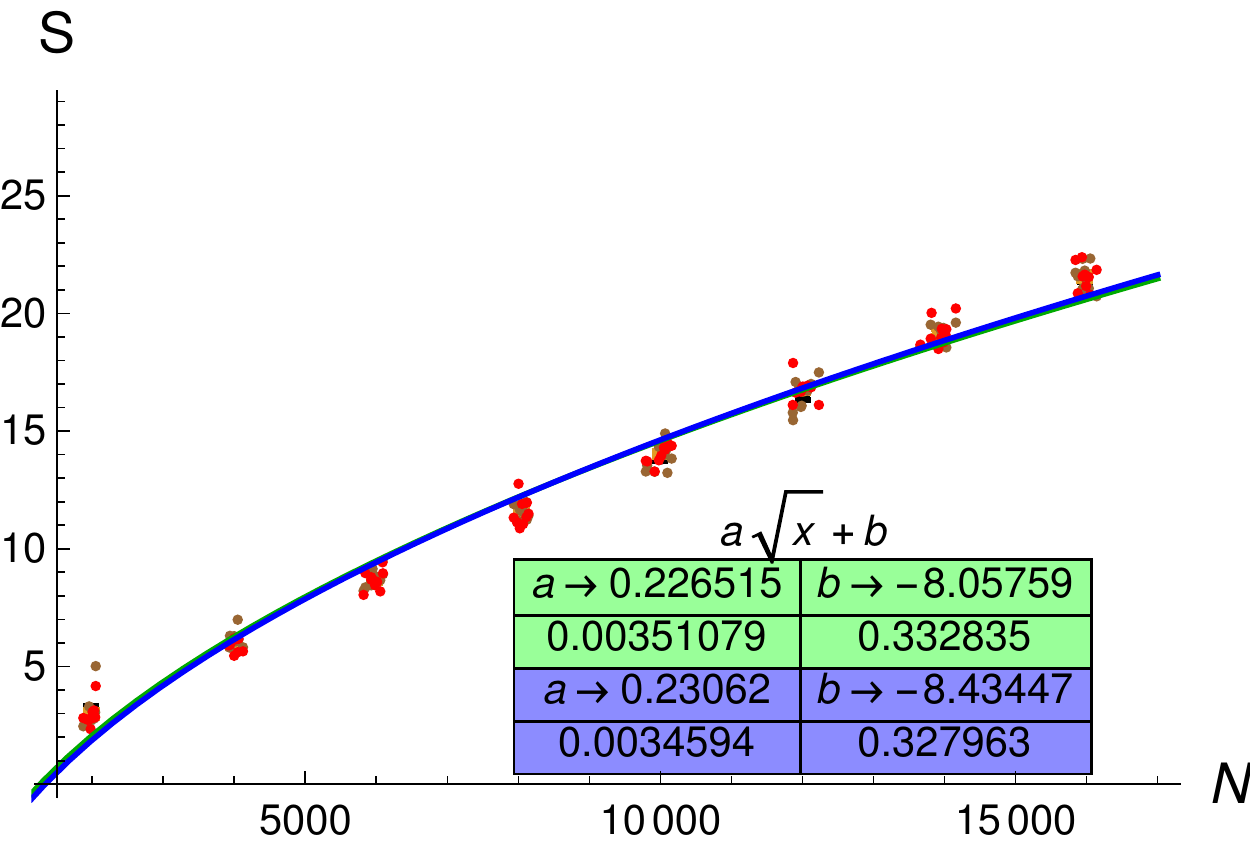}
 		\includegraphics[width=0.5\textwidth]{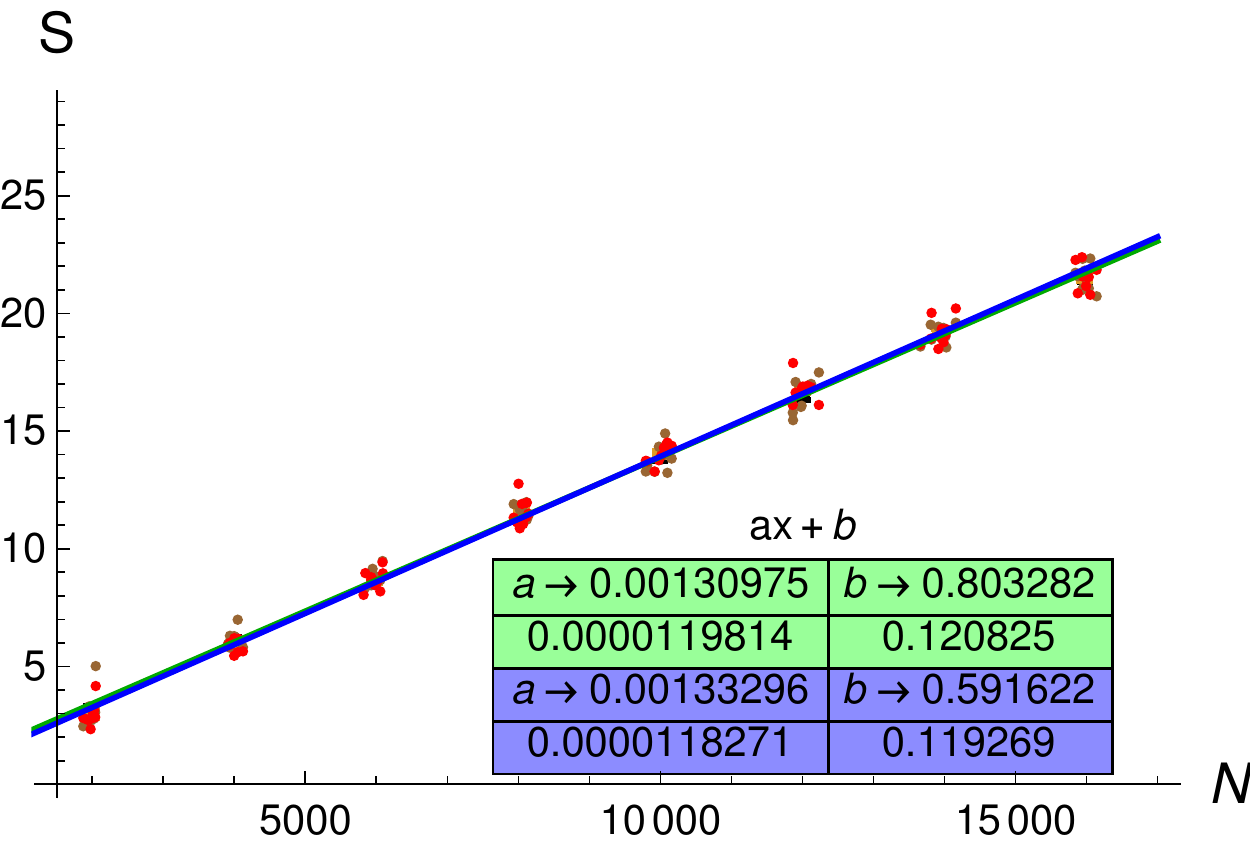}
 		\caption{Linear truncation}
 		\label{ds4linear}
 	\end{subfigure}
 	\caption{SSEE vs. $N$ with different choices of truncation in $\deS_4$.  The green and blue represent the data for
 		the two Rindler-like wedges. A comparison of the two fits $a\sqrt{x}+b$ and $ax+b$ is shown on the left and the right
 		for each choice of truncation.}
 	\label{ds4entropytrunc}
 \end{figure}   

Perhaps this is not surprising because we do not know the proportionality constant $\alpha$ in \eqref{eq:numtrunc}, although it is reasonable to expect an $\alpha$ of order $1$. The linear truncation gives an SSEE area law that is closer to the expectation, although the coefficient again is in excess.
	
\subsection{Nested Causal Diamonds in $\mink^4$}
	\label{mink.app}
	
	\begin{figure}[!h]
		\centering
		\includegraphics[width=0.7\textwidth]{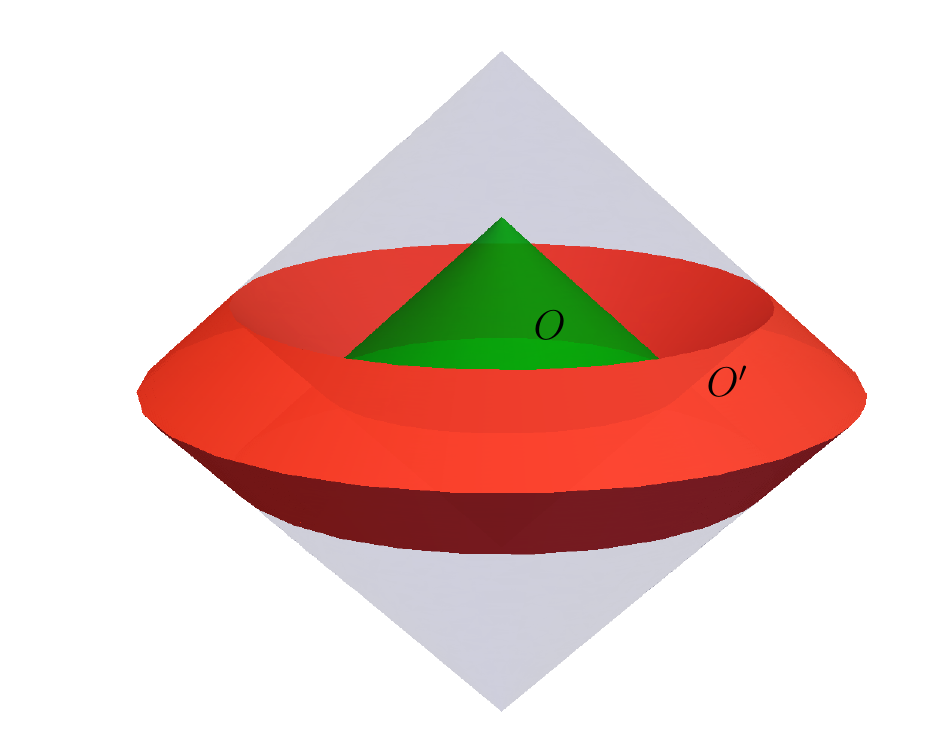}
		\caption{A nested causal diamond $O$ and its complement $O'$ in $3d$.}
		\label{fig:4ddiamond}
	\end{figure} 
	
	Here we present some results for nested causal diamonds in $\mink^4$ which is a non-trivial extension of the $\mink^2$ case. 
	
	We consider a similar set up to $\mink^2$, namely nested causal  diamonds $\diam^4_\ell \subset \diam_L^4 \subset \mink^4$. The causal complement $O' \subset \diam_L^4$ of $\diam_\ell^4$ is connected and is the domain of dependence of a $d=3$ open ball with a concentric spherical hole. We show this connectivity in Figure \ref{fig:4ddiamond}, where one of the dimensions has been suppressed.
	
	In $\diam^4$, the entropy-area relation $S=A/4$ of \eqref{sh} is  
	\begin{equation}
	S= \pi r^2 , 
	\end{equation}where $r$ is the radius of the smaller diamond.
	The expression in the causal set, \eqref{eq: causalsetentropy}, is
	\begin{equation}
	S^{(c)}= \sqrt{\frac{3 \pi}{2}} (r/R)^2 \sqrt{N} ,
	\label{4dcsentropymink} 
	\end{equation}
	where $R$ is the radius of the larger diamond. In our simulations we set $r/R=0.6$, therefore we expect $S \approx 0.78\sqrt{N}$. 
	
	We consider a number truncation with $n_\mx = N^{3/4}$ in all regions, as well as a number truncation with $n_\mx = N^{3/4}$ in $\diam^4_L$ and $\diam^4_\ell$ while $n'_\mx = 2 N^{3/4}$ in the complement of $\diam^4_\ell$. The motivation for the factor of $2$ in the latter number truncation is that the relative spatial volume of the subset of the $t=0$, time-symmetric Cauchy slice that lies in the complementary region is around twice as large as the subset that lies in $\diam^4_\ell$. We also consider the linear truncation with $m'=-0.25-|\epsilon|$ and $\epsilon=0.05$ (or $\delta=0.2$) in all regions.
	
	In the simulations, we consider $\langle N \rangle$ values ranging from $4000$ to $18000$. For each $\langle N \rangle$ we consider 5 realizations. 
	
	\begin{figure}[!h]
		\begin{subfigure}[b]{0.55\textwidth}
			\includegraphics[width=\textwidth]{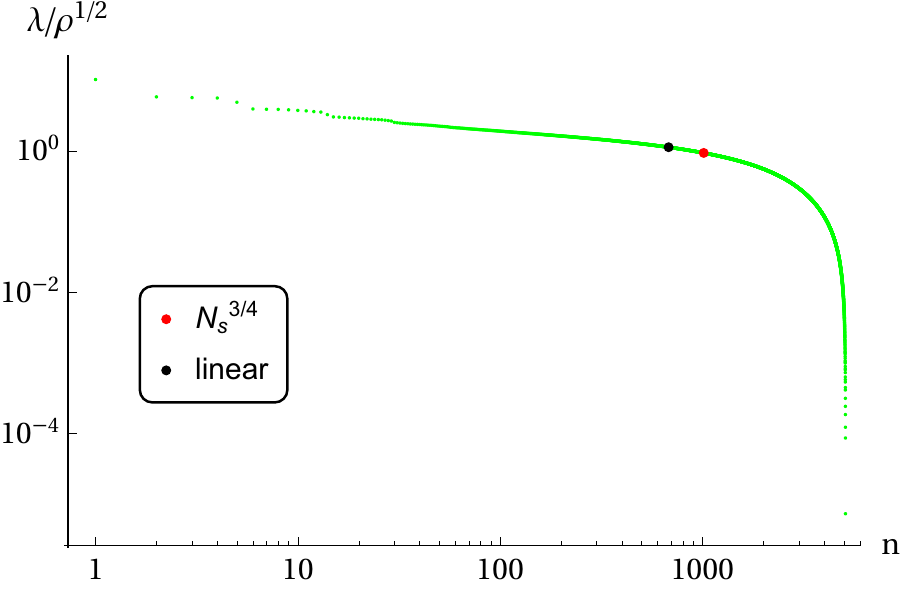}
			\caption{}
			\label{4dmink spectrum marked}
		\end{subfigure}
		\begin{subfigure}[b]{0.55\textwidth}
			\includegraphics[width=\textwidth]{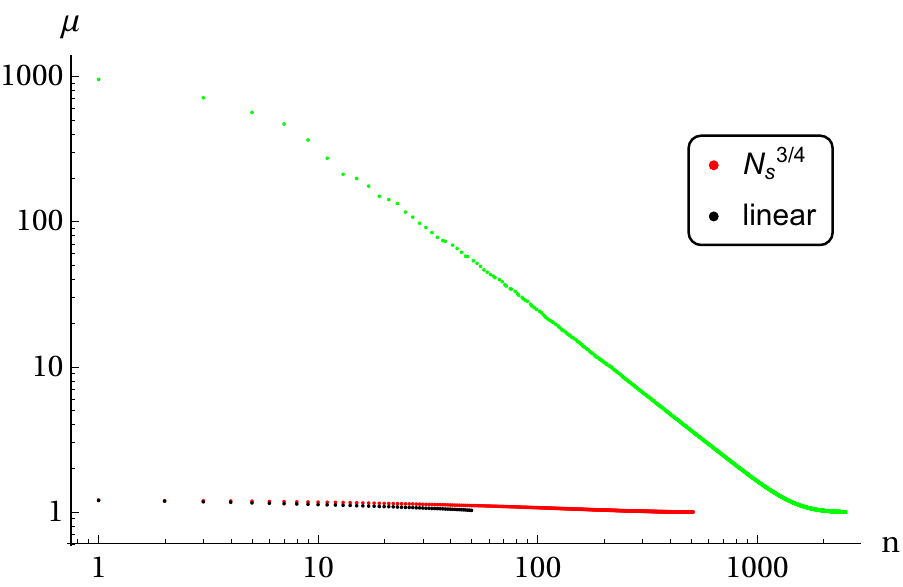}
			\caption{}
			\label{4dmink generalized spectrum}
		\end{subfigure}
		\caption{For $N=10k$, (a) is the spectrum of $i\Delta$ with different truncations marked, and (b) is a plot of the solutions of the generalised equation \eqref{s4c} for these truncation schemes.}
	\end{figure} 
	\begin{figure}[!h]
		\centering
		\includegraphics[width=0.6\textwidth]{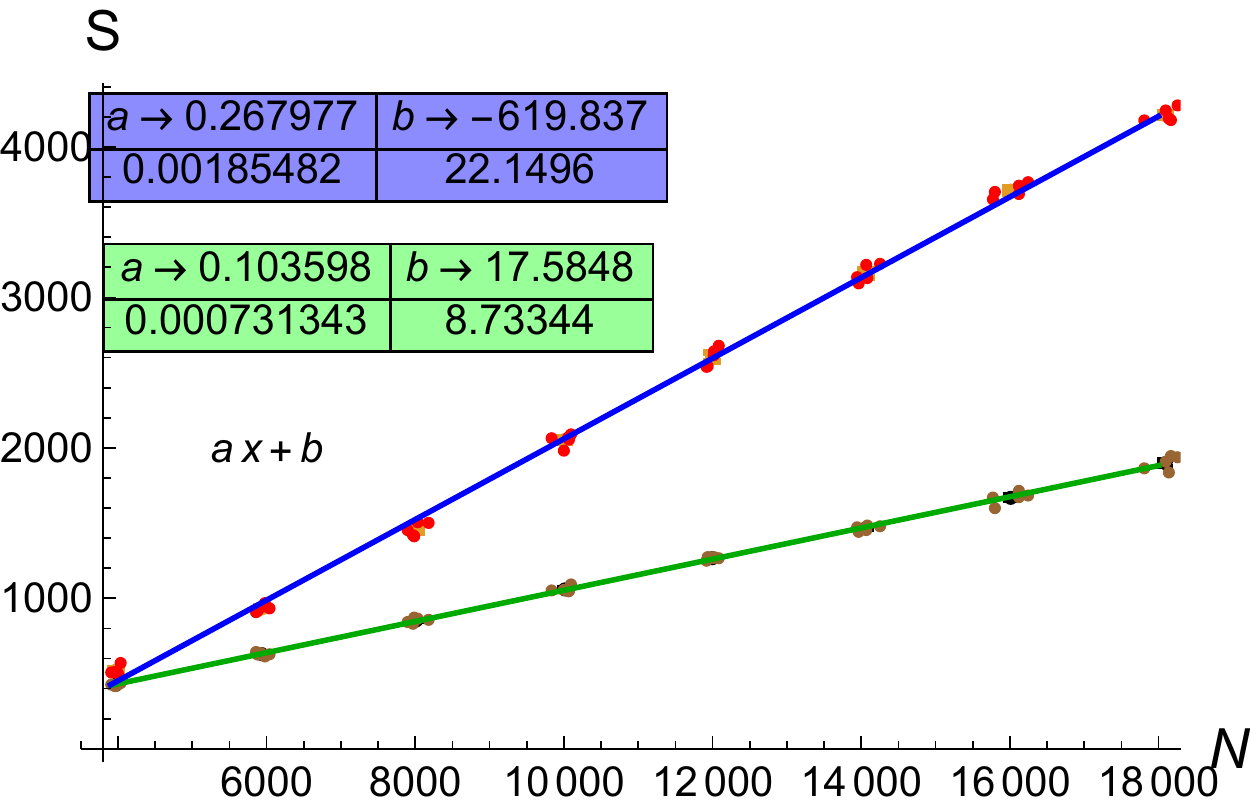}
		\caption{SSEE vs. $N$ without truncation. Green represents the data for $\diam^4_\ell$ and blue represents its complementary region. The best fits are shown. }
		\label{4dmink entropy notrunc}
	\end{figure}
	In Figure \ref{4dmink spectrum marked} we show the SJ spectrum for $\diam^4_L$, and where the truncations we consider lie. We also show the SSEE eigenvalues $\mu$ in Figure \ref{4dmink generalized spectrum} for one realization.
	The causal set SSEE without truncation is shown in Figure \ref{4dmink entropy notrunc} and can be seen to obey a spacetime volume law as anticipated. 
	\begin{figure}[!h]
		\begin{subfigure}[b]{\textwidth}
			\includegraphics[width=0.5\textwidth]{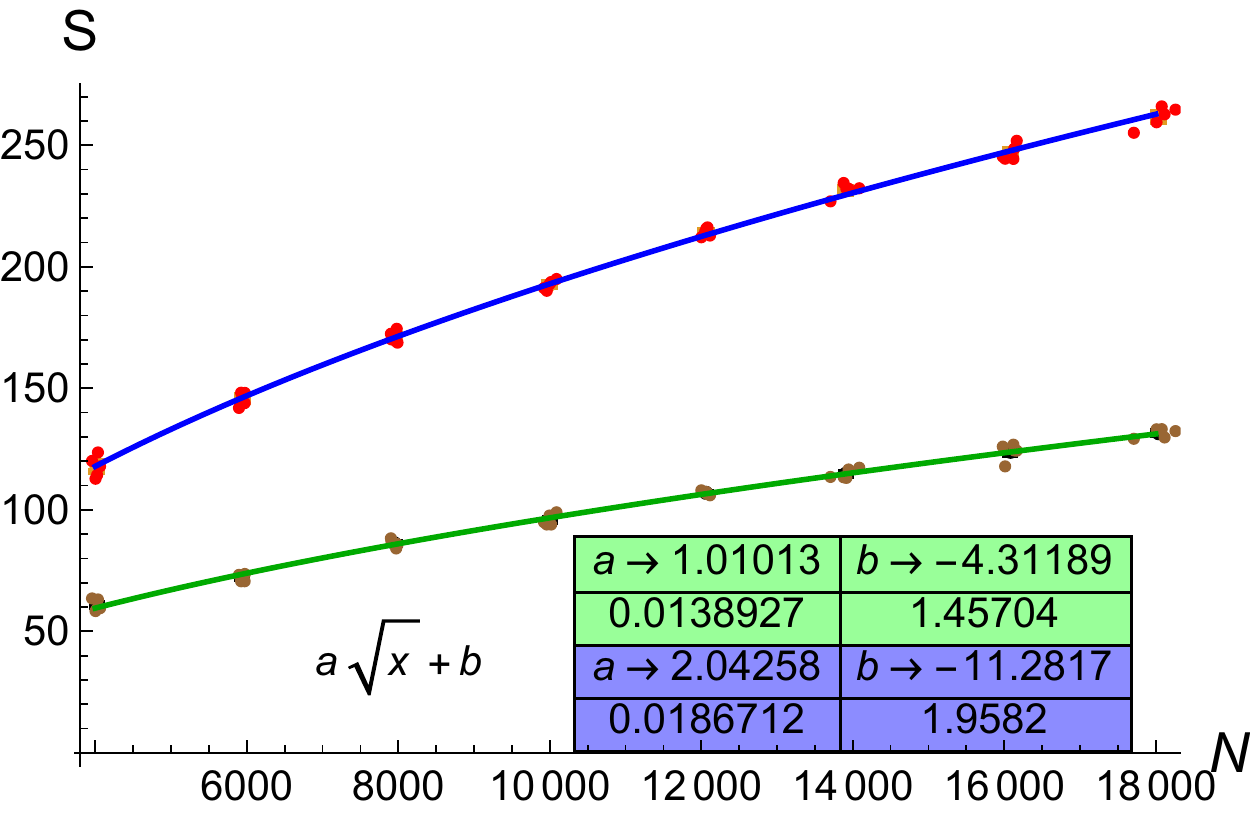}
			\includegraphics[width=0.5\textwidth]{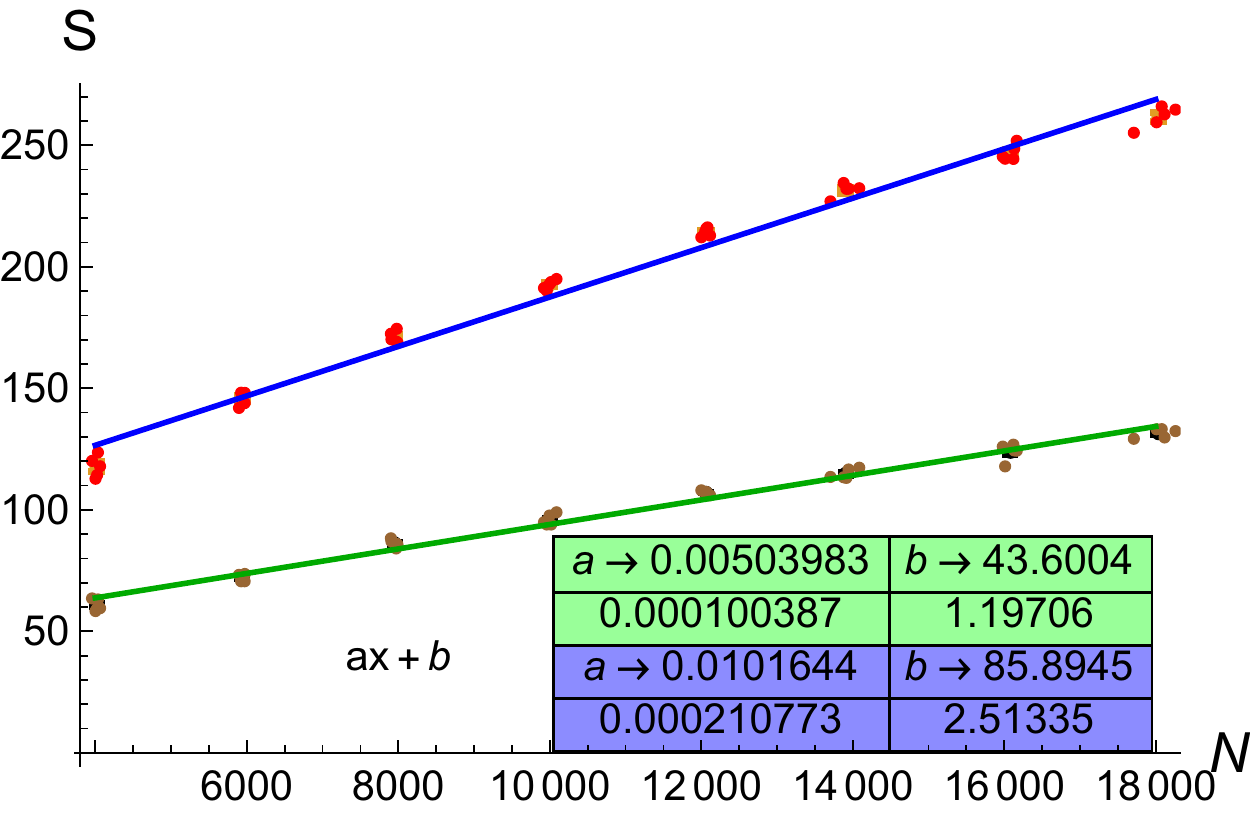}
			\caption{Number truncation with $n_\mx=N_s^{3/4}$}
			\label{4dmink entropy num nofactor}
		\end{subfigure}
		\begin{subfigure}[b]{\textwidth}
			\includegraphics[width=0.5\textwidth]{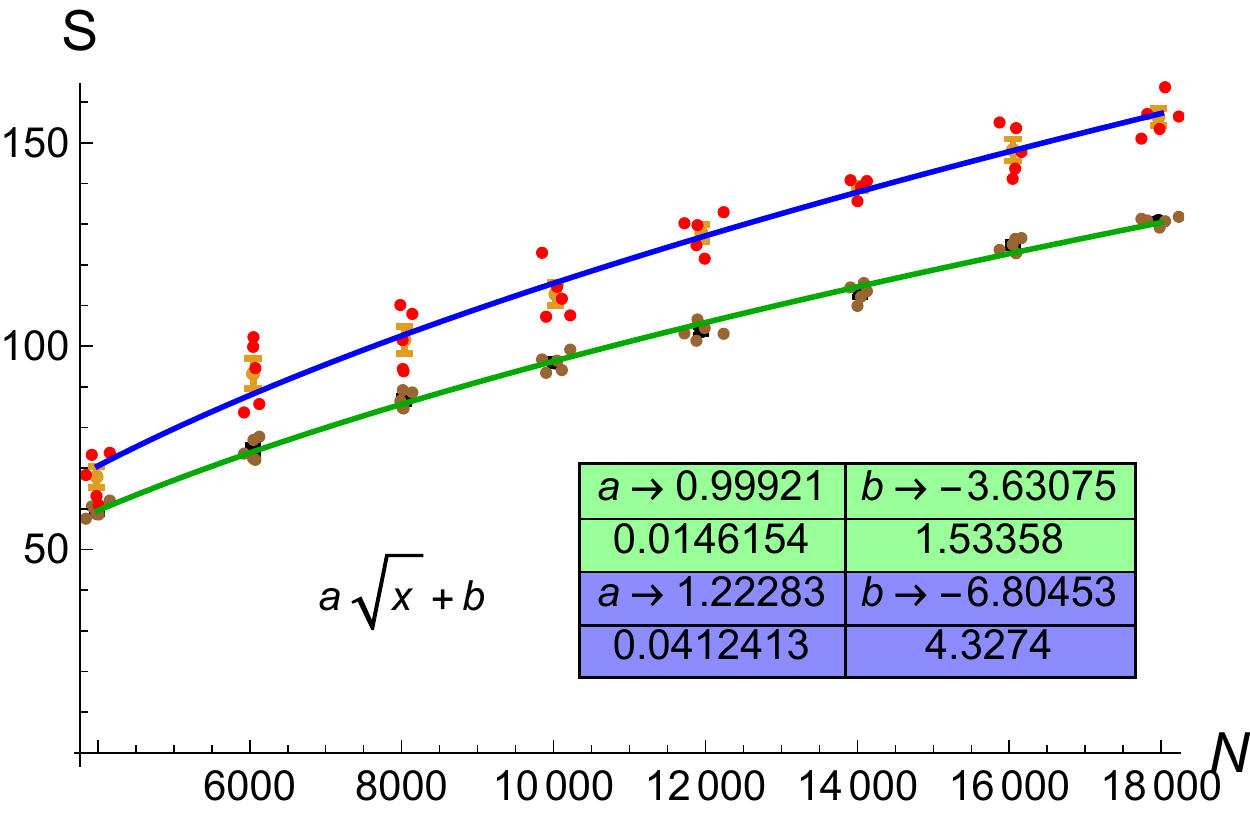}
			\includegraphics[width=0.5\textwidth]{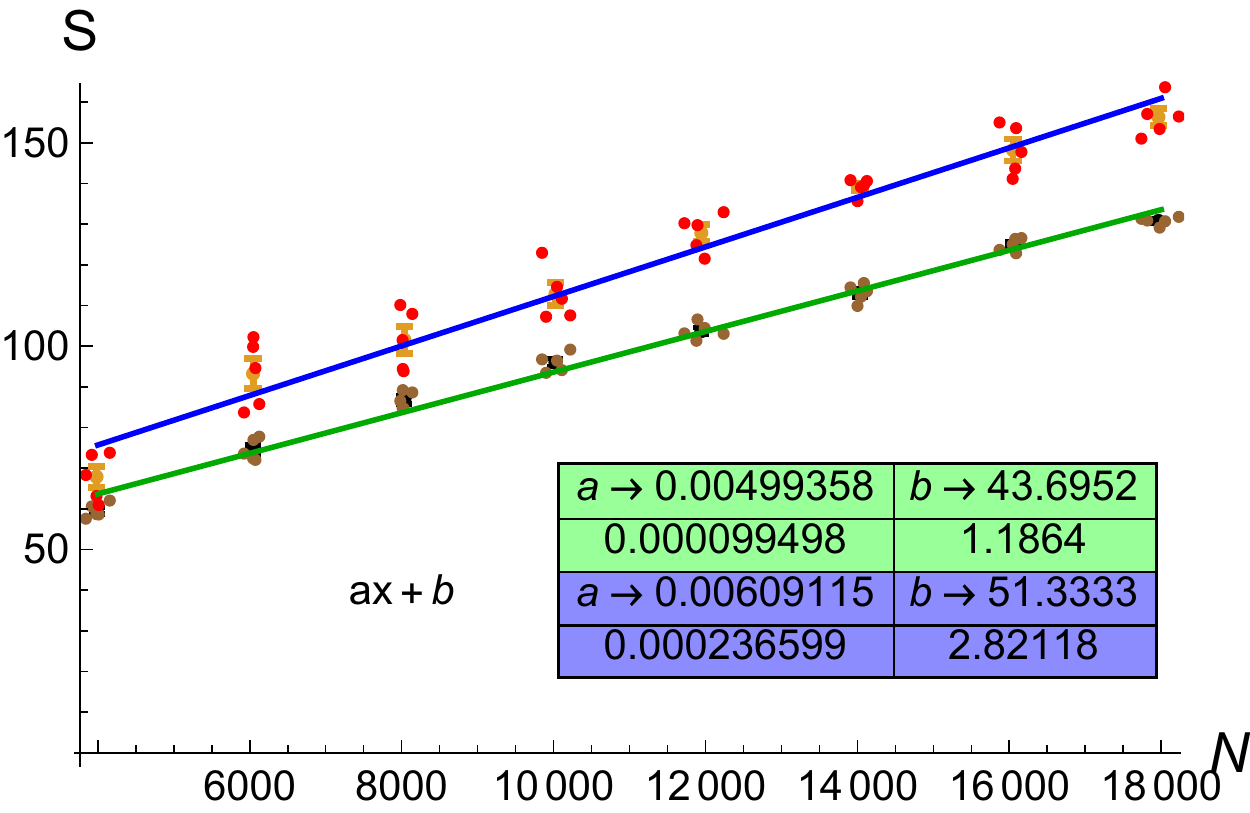}
			\caption{Number truncation with $n'_\mx=2 \,n_\mx=2N_s^{3/4}$}
			\label{4dmink entropy num factor2}
		\end{subfigure}
	\end{figure}
	\begin{figure}[!h]
		\ContinuedFloat
		\begin{subfigure}[b]{\textwidth}
			\includegraphics[width=0.5\textwidth]{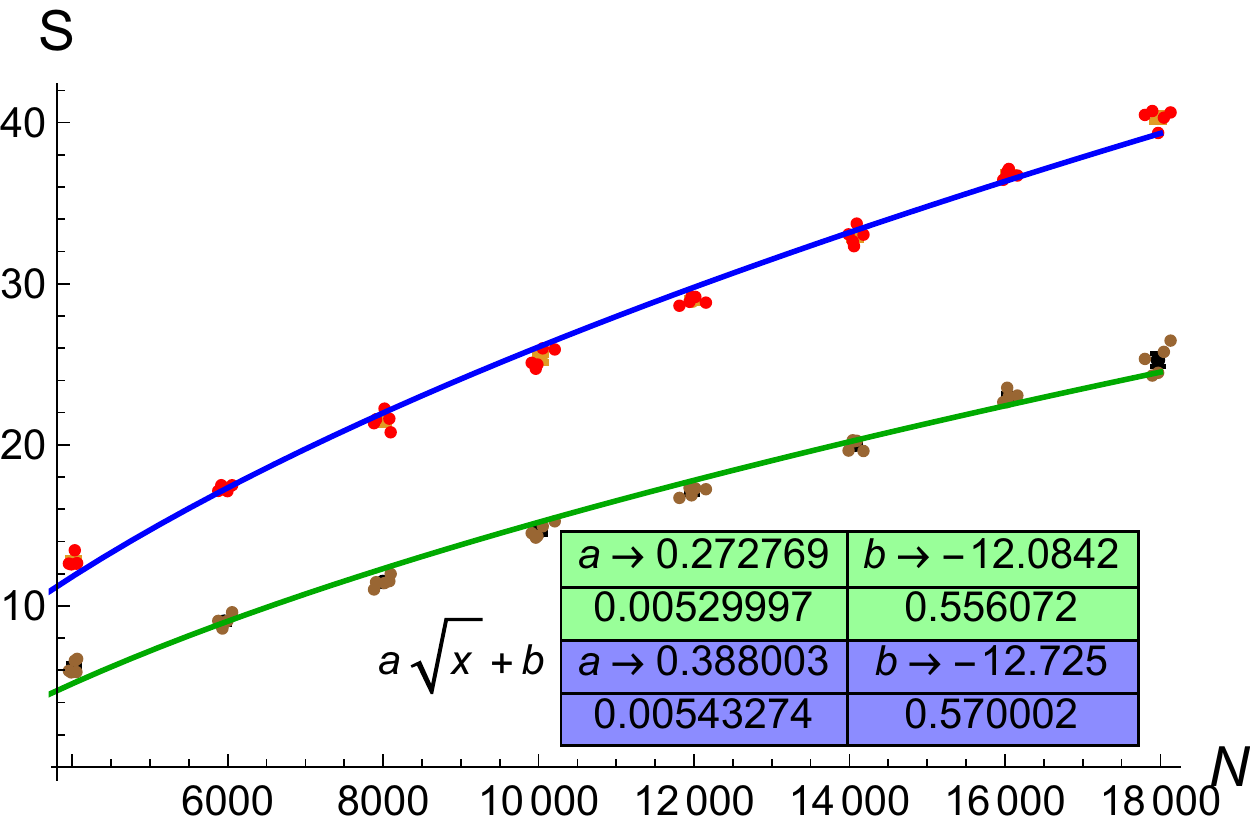}
			\includegraphics[width=0.5\textwidth]{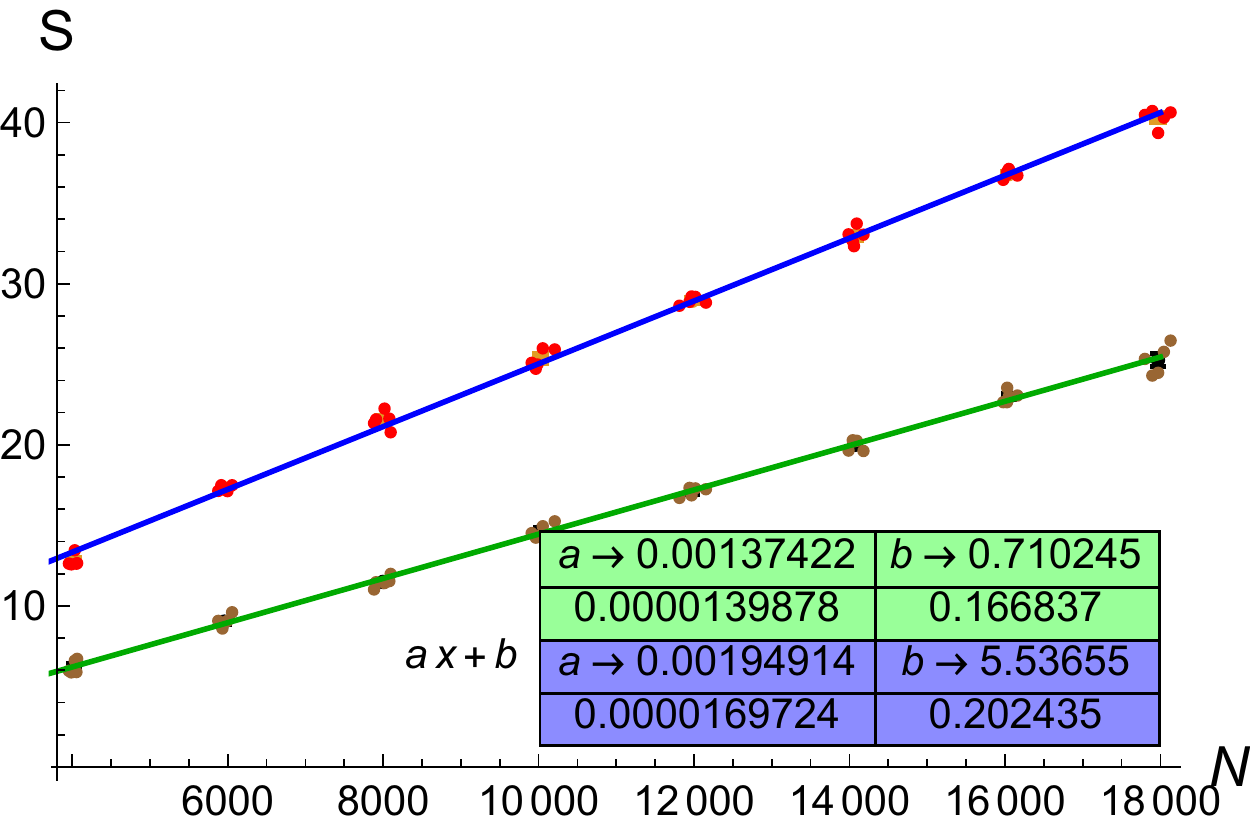}
			\caption{Linear truncation}
			\label{4dmink entropy dynamic linear}
		\end{subfigure}
		\caption{SSEE vs. $N$ with different truncations. Green represents the data for $\diam^4_\ell$ and blue represents its complementary region. A comparison of the two fits $a\sqrt{x}+b$ and $ax+b$ is also shown. Here $n'_\mx$ is the number truncation in the region complementary to $\diam^4_\ell$.} 
		\label{4dmink entropy trunc}
	\end{figure}
	
	Next, we show in Figure \ref{4dmink entropy trunc} how the  SSEE is modified with the application of the various truncations. An area law is recovered with the two number truncations $n_\mx \propto N^{3/4}$, while the linear truncation is more consistent with a volume law. This is similar to what we found in the dS cases we studied above. The area law coefficients in most cases are $\sim 1$ and are therefore close to the expected value $0.78$. What is more challenging and non-trivial here, compared to the dS cases, is achieving complementarity. The geometries of $\diam^4_\ell$ and its complement are very different (see Figure \ref{fig:4ddiamond}) and therefore the truncations ought to take this difference into account.  
	
	As we can see by comparing Figures \ref{4dmink entropy num nofactor} and \ref{4dmink entropy num factor2}, it is correct to truncate the complementary region more than $\diam^4_\ell$. The SSEE can get closer to satisfying both complementarity and the expected area law coefficient if one appropriately tunes the proportionality constant in $n_\mx \propto N^{3/4}$. As discussed in the main text, we do not have a covariant argument by which to uniquely set this proportionality constant. In the absence of such an argument, we do not pursue tuning the constant(s).

\section{Violation of Causality}
\label{acausality}

Before concluding, we mention another unavoidable consequence of truncation. The matrix\footnote{We ignore the $i$ for now as it is irrelevant for this discussion.} $\Delta$ is constructed from $G_R$ and $G_A$, this contains $\Delta_{xy}=1$ if $y\prec x$, $\Delta_{xy}=-1$ if $x\prec y$ and $0$ otherwise. When we truncate the eigenbasis of $i\Delta$ the new Pauli-Jordan matrix $\Delta^{\text{tr}}$ contains new entries which are non-zero even when $x$ and $y$ are unrelated i.e., there is a violation of causality. 

As an example consider a random causal set obtained by sprinkling 6 points in $\diam_\ell^2$. The eigenvalues of $i\Delta$ are $\{2.876,-2.876,0.851,-0.851,0,0\}$, as a sample truncation we retain only the higher value. Below we show what happens to $\Delta$ post truncation 
$$
\left(
\begin{array}{cccccc}
0 & 0 & 0 & 1 & 1 & 1 \\
0 & 0 & 0 & 1 & 0 & 1 \\
0 & 0 & 0 & 1 & 0 & 1 \\
-1 & -1 & -1 & 0 & 0 & 1 \\
-1 & 0 & 0 & 0 & 0 & 1 \\
-1 & -1 & -1 & -1 & -1 & 0 \\
\end{array}
\right)\xrightarrow[]{\text{truncation}}
\left(
\begin{array}{cccccc}
0 & 0.13 & 0.13 & 1.09 & 0.56 & 1.09 \\
-0.13 & 0 & 0 & 0.83 & 0.39 & 0.96 \\
-0.13 & 0 & 0 & 0.83 & 0.39 & 0.96 \\
-1.09 & -0.83 & -0.83 & 0 & -0.26 & 1.09 \\
-0.56 & -0.39 & -0.39 & 0.26 & 0 & 0.83 \\
-1.09 & -0.96 & -0.96 & -1.09 & -0.83 & 0 \\
\end{array}
\right)
$$
\bfig[h]
\centering
\includegraphics[scale=1]{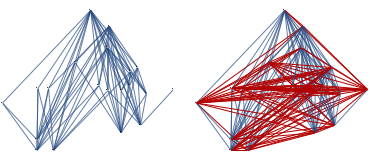}
\caption{A sample causal set with 15 elements along with the relations is shown. The red lines on the right are the extra relations that are introduced due to the truncation of $i\Delta$.}
\efig
The extra numbers that appear in $\Delta$ post truncation are small and come with $\pm$ signs. A possibility is to think of these as fluctuations arising from the truncation of an individual causal set. Since the continuum will correspond to an ensemble of causal sets this kind of fluctuation could ``average" out to zero. Of course we would have to specify what is meant by averaging. Each time we sprinkle, not only the number of points but the points themselves are different hence it cannot be a simple average of matrix entries. Even if such an averaging procedure could be devised and these extra relations be shown to average to zero it does not tell us why we need the truncation in the first place. As we have seen, truncation seems to be an essential ingredient in making sense of the SSEE.   

Finally, in the context of the equations of motion on the causal set, we can also interpret the truncation as augmenting the $\kr(i\Delta)$. The is motivated from the fact that in the continuum we expect the set of solutions of $\hB$ to be a very small subset of its domain. The remaining part of the domain is spanned by its image. From \eqref{eq:kerim} this would imply that the $\kr(i\Delta)$ is much larger than its image. Hence by truncating the spectrum of $i\Delta$ we are in fact increasing the size of its kernel. To quantify these ideas and put them on stronger footing would require a thorough study from all these directions.
	
\section{Discussion}
\label{discussion}  

Numerical studies such as the one we have carried out in this paper have shown
that an excess in the coefficient of the SSEE area law due to the small eigenvalues of the SJ spectrum is a common occurrence, and thus generically gives rise to a volume rather than an area law.
Here we have presented evidence that the causal set SSEE for $\deS_{2,4}$  horizons satisfies a volume rather than an area law, when the full causal set SJ spectrum is used.

On implementing {certain} double truncation schemes on the SJ spectrum, inspired by the $\diam_L^2$ case,  we show that area laws can be  obtained, {which also, as expected, satisfy complementarity}.  In this sense, the properties of the causal set SSEE obtained in  \cite{Sorkin:2016pbz} for the nested causal diamonds $\diam_\ell^2\subset \diam_L^2$ appear to be universal.

Out of the many truncation schemes explored in $\deS_2$, the number truncation $n_\mx=2 \sqrt{N}$ gave results most compatible with the Bekenstein-Hawking entropy, though even in this case, the coefficient of the SSEE area law is in excess by $\sim 8 \%$. In $\deS_4$, the linear truncation gave the best results, but the coefficient is in excess of the expected value for the $\deS_4$ horizon by $\sim 30 \%$. In $\mink^4$, the number truncation $n'_\mx=2 \,n_\mx=2N_s^{3/4}$ gave the best results, the larger concern in this case being the clear lack of complementarity. 

The following ideas need to be explored further - 
\begin{itemize}
	
\item An important question is whether this over-estimation of the area law coefficient is generic or whether it can be removed by fine tuning the values of $\con$ and $\delta$ and bettering our data. In our investigations, several choices of these parameters have been scanned, with the values presented here being the most optimal in terms of the area law and data compatible with complementarity. While the possibility always exists of further fine-tuning, or using  a different truncation scheme, this is not necessarily helpful without further physical understanding. One might of course also resort to the possibility that $N$ is not large enough and what we are seeing are finite size effects. In light of the fact that bigger and faster computers are always on the horizon, this can be checked, but as mentioned before, we are perhaps at the optimal values, given current limitations on available RAM. We believe these studies suggest a deeper origin to these questions which cannot be fully understood by further numerical studies alone.  

\item In all cases we find an over-estimation of the SSEE area law coefficient in the truncated case and the volume law in the untruncated case. A possible reason for this could lie in spacetime discreteness and the use of the causal set $\deS$ SJ vacuum. The SSEE \eqref{s4c} does not specify a choice of vacuum. However, the SJ vacuum is the only way we know how to define a vacuum in the causal set. Does this modified discrete vacuum then imply a profound change in our understanding of the Bekenstein-Hawking entropy? If so, how can this be compatible with effective field theory descriptions of horizon entropy? 

\item Even if a truncation scheme could be found that does not lead to an overestimation, the truncation procedure comes with its own additional questions. As discussed in section \ref{acausality}, in the nested causal diamonds in $\mink^2$, the truncation leads to a violation of causality. Such a violation needs to be explained and it points to the need for a deeper understanding of the truncation process.  

\item What also needs to be understood better is the untruncated volume law and the nature of the extra contributions that lead to it. The fact that the SSEE obeys a volume law without truncation seems to arise from the non-locality of the causal set. As discussed in \cite{Eisert:2008ur} systems with long-range order exhibit volume rather than area laws. The non-locality in a causal set, which enables an element near the past boundary to be linked to one near the future boundary, is fundamental to the discrete-continuum correspondence. Localising influences near sets of measure zero, even if they are genuine horizons, are not commensurate with this feature. Thus, a volume law seems particularly convincing in causal set theory. However, since area laws are a fundamental feature of General Relativity, which causal set theory must approximate, locality must be emergent, and with it, an area law for the SSEE. 

\item While the causal set offers a ready covariant spacetime cutoff, the recovery of an area law for the SSEE is highly non-trivial for the case of $\deS$. Suggestions in \cite{Belenchia:2017cex} for a deeper understanding from an Algebraic QFT perspective need to explored further. Recently the continuum $\deS$ SJ spectrum in the slab has been found analytically and offers us a possible route to calculating the continuum $\deS$ SSEE, and hence finding a more physically motivated truncation.

\end{itemize}

It is possible that the detailed nature of the UV physics cannot come from causal sets that are manifold-like at all scales and that the modification to $\deS$ on the smallest of scales can have an effect on the SJ spectrum and in turn on the SSEE. This  conjectured modified UV behaviour may be the missing ingredient, but we are far from an understanding of what this might be.

\chapter{Conclusions}

The overarching theme of this thesis is quantum field theory of a free scalar on fixed causal sets approximated by Minkowski and de Sitter spacetimes. We begin with the construction of causal set retarded Green functions on curved backgrounds. We extended previous results to a Riemann normal neighbourhood and to regions of de Sitter ($\deS$) and anti de Sitter (adS) spacetime. We then use the Sorkin-Johnston (SJ) construction to go from the retarded Green function to the Wightman function $W$. This construction does not require a choice of timelike Killing vector and hence is a covariant way to define the vacuum via $W$. We find that the vacuum so obtained in a symmetric slab of $\deS$ spacetime is not one of the well known Mottola-Allen $\alpha$-vacua and discuss why this might be the case. The analysis was done on a causal set in a finite region of $\deS$ and our results show asymptotic convergence. Finally we use Sorkin's covariant spacetime entanglement entropy formulation to compute the entanglement entropy associated with $\deS$ horizons. We find that the entropy satisfies a volume law which can be traced to the non-locality inherent in the causal set. However, we were also able to identify that this contrast with the continuum may be due the difference in the spectrum of the Pauli-Jordan (PJ) operator. By truncating this spectrum appropriately we recover an area law. Finally, we compare this entropy to the Bekenstein-Hawking entropy.

This work is an exploration of ideas that have been proposed in causal set quantum gravity over the last 3 decades. Specifically, ideas related to working with quantum fields on fixed background causal sets. Our focus here has been on causal sets approximated by de Sitter spacetime. Although we have been able to show several results starting from first principles, it has also thrown up other, deeper questions about the nature of fundamental discreteness and the discrete-continuum correspondence. We have discussed these questions in detail in the thesis. We also reiterate some of the broader questions here 

\begin{itemize}
	\item There are multiple candidates for a Green function in a causal set and it is impossible to choose one without an explicit comparison with the continuum. If we consider causal sets to be more fundamental then we must have an inherent choice based on some physical principle. Recent work on the construction of discretized wave operators based on the idea of a \textit{preferred past} may help resolve some ambiguities \cite{Dable-Heath:2019sej}. 
	\item In continuum $\deS$ spacetime, the $\alpha$-vacua are the complete set of available $\deS$ invariant vacua. A new vacuum that doesn't fall in this family may have phenomenological consequences for early universe cosmology. However, our result is obtained in a finite region and the comparison is only asymptotic. There are also basic differences in the construction of the SJ vacuum due to spacetime discreteness and we need a better analytic understanding.
	\item The SJ construction is entirely covariant and does not explicitly rely on the symmetries of spacetime. However when we work with causal sets in a finite region we may be breaking these symmetries, at least near the boundaries. As we have suggested, this might be the reason for the difference between the SJ vacuum and the $\alpha$-vacua. A study on the status of symmetries while working with causal sets in finite regions of spacetime is needed. One way to do this is to use non-symmetrical boundaries for the slab i.e., instead of the spacelike boundaries of the slab we could use slanted or boosted boundaries. Any boundary effects will reflect in the form of $W$. Other deformations of the boundaries can also be considered. 
	\item It is important to understand the thermal properties of the SJ vacuum. In the context of horizons such descriptions always reveal important geometric properties of the underlying region. In this context a comparison of the SJ vacuum with known thermal vacua should be helpful.
	\item The spacetime entanglement entropy arising from causal sets is a volume dependent quantity and, apart from the need to compare with the continuum, there is no fundamental reason why it must follow an area law. This disconnect between the causal set volume law and the continuum area law needs a satisfactory explanation.
	\item A truncation is needed in the spectrum of the PJ operator in order to get an area law for the entanglement entropy. This truncation can be traced back to the question of equations of motion on the causal set. Although we have tried several ways of implementing the truncation, there is no clarity on the need for it or a reason to pick one way over another. The choice of truncation as well as the initial value formulation on causal sets need to be studied further. 
	\item Numerical studies of Sorkin's spacetime entanglement entropy in $\deS_{2,4}$ and in $\mink^4$ give us area laws with coefficients larger than what we expect from known results. We expect that the entanglement entropy may account for a part of the horizon entropy, it may be an underestimate because we consider only the free scalar field. An overestimate signals a flaw in our understanding.          
\end{itemize} 

To conclude, we comment on the relevance of computational results in quantum gravity. Unlike many other areas of physics, the problem of quantum gravity is aggravated by the lack of phenomenology. In such a scenario, to check new hypothesis, we are dependent on indirect tools like mathematical consistency, low energy limits, analog models etc. Recently, emphasis has been on using computational methods in QG \cite{Glaser:2018kty} - numerical relativity toolkits, Monte Carlo for evaluating the path integral in various approaches, AI in pattern recognition and dynamics etc. These methods give us (1) a chance to test scenarios that are analytically intractable, (2) intuition about phenomenon like black hole entropy which are not directly observable. It is possible that major insights in quantum gravity will be along the directions indicated by numerical results.




\bibliography{refs_main}{}
\bibliographystyle{ieeetr} 

\end{document}